
\magnification \magstep1
\raggedbottom
\openup 4\jot
\headline={\ifnum\pageno=1\hfill\else
\hfill {\it Dirac Operator and Eigenvalues in Riemannian
Geometry} \hfill \fi}
\voffset6truemm
\rightline {DSF report 95/33}
\vskip 3.3cm
\leftline {\bf Giampiero Esposito}
\vskip 0.3cm
\hrule
\vskip 5cm
\centerline {\bf DIRAC OPERATOR AND}
\centerline {\bf EIGENVALUES IN RIEMANNIAN GEOMETRY}
\vskip 6cm
\leftline {Based on 5 graduate lectures given by the author
at SISSA, April 1994.}
\vskip 100cm
\leftline {\bf Author}
\leftline {\it Giampiero Esposito}
\leftline {\it Istituto Nazionale di Fisica Nucleare}
\leftline {\it Gruppo IV, Sezione di Napoli}
\leftline {\it Mostra d'Oltremare Padiglione 20}
\leftline {\it I-80125 Napoli, Italy}
\vskip 1cm
\leftline {\it Dipartimento di Scienze Fisiche}
\leftline {\it Mostra d'Oltremare Padiglione 19}
\leftline {\it I-80125 Napoli, Italy}
\vskip 100cm
\centerline {\bf PROGRAM}
\vskip 1cm
\noindent
{\bf Lecture 1.} Differential operators on manifolds.
Index of elliptic operators. Dirac operator. (Refs. [1,2,6])
\vskip 0.3cm
\noindent
{\bf Lecture 2.} Index problem for manifolds with a boundary.
Index of the Dirac operator and anomalies. (Refs. [1,3,6])
\vskip 0.3cm
\noindent
{\bf Lecture 3.} Spectral asymmetry and Riemannian geometry.
Heat equation and asymptotic heat kernel. The $\eta$- and
$\zeta$-functions. (Refs. [2,4,6])
\vskip 0.3cm
\noindent
{\bf Lecture 4.} Two-component spinor calculus. Dirac and Weyl
equations in two-component spinor form. Weyl equation with
spectral or local boundary conditions. Potentials for massless
spin-${3\over 2}$ fields. (Refs. [4,5,6])
\vskip 0.3cm
\noindent
{\bf Lecture 5.} Self-adjointness of the boundary-value problem
for the Dirac operator in a particular class of Riemannian
four-geometries with boundary. Asymptotic expansion of the
corresponding heat kernel. (Refs. [4,6])
\vskip 1cm
\centerline {\bf REFERENCES}
\vskip 1cm
\noindent
\item {[1]}
Atiyah M. F. (1988a) {\it Collected Works, Vol. 3 (Index Theory: 1)}
(Oxford: Clarendon Press).
\item {[2]}
Atiyah M. F. (1988b) {\it Collected Works, Vol. 4 (Index Theory: 2)}
(Oxford: Clarendon Press).
\item {[3]}
Atiyah M. F. (1988c) {\it Collected Works, Vol. 5 (Gauge Theories)}
(Oxford: Clarendon Press).
\item {[4]}
Esposito G. (1994a) {\it Quantum Gravity, Quantum Cosmology and
Lorentzian Geometries}, second corrected and enlarged edition
(Berlin: Springer-Verlag).
\item {[5]}
Esposito G. (1994b) {\it Complex General Relativity}
(DSF preprint 93/47).
\item {[6]}
Problems and notes for the students.
\vskip 100cm
\centerline {\bf INTRODUCTION}
\vskip 1cm
When the reader is not familiar with the topic he is
going to study, he may have to read twice a detailed
introduction, and these lecture notes are no exception
in this respect. First, the reader might want to have
an idea of main concepts and results he is expected to
learn. Second, after having taken the trouble to study
definitions, proofs and calculations, he might come back to
the introduction, to appreciate (or disagree with) the
choice of topics, the logical order, and how the various
subjects are intertwined.

The aim of the Lectures is to introduce first-year Ph.D.
students to the theory of the Dirac operator, spinor
techniques, and their relevance for the theory of eigenvalues
in Riemannian geometry. For this purpose, Lecture 1 begins
with a review of exterior algebra and Clifford algebra. These
algebras lead, by Fourier transform, to certain (standard)
differential operators on vector spaces. After introducing a
Riemannian metric, these can be globalized to manifolds.
The naturally occurring differential operators are exterior
differentiation, Hodge-Laplace operator, signature operator,
and the Dirac operator. The latter is first introduced by
looking at Euclidean space $R^{2k}$, with Clifford algebra as a
full matrix algebra acting on spin-space. Spin-space is given
by the direct sum of positive and negative spin-spaces, and the
Dirac operator interchanges these spaces. In the case of a
compact oriented manifold $X$ of even dimension $2k$, assuming
a spin-structure on such a manifold, and using a Levi-Civita
connection on the principal $SO(2k)$ bundle of $X$, the Dirac
operator can be defined as a first-order, elliptic, self-adjoint
differential operator. Harmonic spinors are also introduced.

Symbols of elliptic operators and local Bott theorem are used to
show that the symbol of the Dirac operator on a compact manifold
is a generator for all elliptic symbols. Moreover, from the
global Bott theorem one knows that every symbol is equivalent
to the symbol of the restricted Dirac operator. Thus, if one
can evaluate the index of the restricted Dirac operator, one
gets a formula for the index of all elliptic operators (under
suitable conditions). The index, defined as the difference
between the dimensions of the null-spaces of a given operator and
of its adjoint, plays a key role in the theory of eigenvalues in
Riemannian geometry. From the point of view of theoretical and
mathematical physics, index theorems and index formulae are
used to study elliptic boundary-value problems relevant for
quantum amplitudes, and anomalies in quantum field theory
(see below).

When one obtains index formulae for closed manifolds, it is
natural to ask for extensions to manifolds with boundary.
One has then to understand how to measure the topological
implications of elliptic boundary conditions. Such boundary
conditions have, of course, an effect on the index. For
example, if $b$ is a nowhere vanishing vector field on the
unit disk $X$ in the plane, and $D$ is the Laplace operator
on $X$, the index of the boundary-value problem $(D,b)$ is
found to be 2(1-winding number of $b$). The powerful techniques
of differential geometry can be used to derive general index
formulae for the elliptic boundary-value problems of mathematical
physics, as done in Lecture 2. Remarkably, the index is
unchanged by perturbation of the operator. Hence it can be given
by a topological formula, and if the gauge field $A$ is
continuously varied, the corresponding index $D_{A}$ does not
change. It is possible, however, to extract more topological
information from a continuous family of operators. This
analysis is motivated by the attempts to quantize gauge theories
of fundamental interactions. In fact, in quantizing gauge
theories one has to introduce determinants of operators,
and a regularization is necessary to make sense of the
(divergent) quantum amplitudes. For this purpose, the tool
used by physicists is the generalized Riemann $\zeta$-function.
It relies on the study of the heat equation, self-adjointness
theory, and the theory of complex powers of elliptic operators.

Since these Lectures are mainly concerned with the Dirac
operator, Lecture 3 begins by introducing instead the
$\eta$-function. This function, obtained from the positive
and negative eigenvalues of a self-adjoint elliptic operator,
relates spectral invariants to the properties of oriented
Riemannian four-manifolds with boundary. More precisely,
the $\eta$-function is regular at the origin, and such a value
is proportional to the value taken by a certain continuous
function depending only on the boundary (and related in turn
to the signature of the quadratic form and to the first
Pontrjagin class). A detailed calculation of the $\eta$-function
and of its value at the origin is then given in a particular case,
following the work of Atiyah, Patodi and Singer.
The basic steps in performing $\zeta$-function calculations are
studied already at the end of Lecture 2.

Lecture 4 presents two-component spinor calculus by relying
on the definition of unprimed and primed spin-spaces as
symplectic spaces. The basic isomorphisms in the Riemannian
case are the one between a spin-space and its dual,
and the one between the tangent space at a point, and the tensor
product of unprimed and primed spin-space. However, unprimed and
primed spin-space are not isomorphic (as in the Lorentzian case),
and this makes it possible to have a rich theory of self-duality
and anti-self-duality for gauge fields and gravitation.
Moreover, the only conjugation available is the so-called Euclidean
conjugation, which maps a given spin-space to itself, and is
anti-involutory on spinor fields with an odd number of indices.
The Weyl equation is then studied as part of the massless free-field
equations of classical field theory, and the corresponding Dirac
equation in the massive case is also given. The classical
boundary-value problem for solutions of the Weyl equation subject
to spectral or supersymmetric boundary conditions on a three-sphere
is then studied in detail. One has indeed a choice of spectral or
local boundary conditions for fermionic fields by virtue of the
first-order nature of the Dirac operator. The former are motivated
by the work of Atiyah, Patodi and Singer described in Lecture 3,
whereas the latter are motivated by supersymmetric gauge theories
of gravitation, combining bosonic and fermionic fields in suitable
matter supermultiplets. It has been a recent achievement of
theoretical and mathematical physicists to show that such local
boundary conditions also lead to well-posed elliptic problems
for the Dirac operator, besides being physically more relevant.
Last but not least, Penrose potentials for spin three-halves are
studied, since the gauge freedom for such potentials is (again)
expressed in terms of solutions of positive- and negative-helicity
Weyl equations. The corresponding calculations, motivated by the
attempt of Penrose to give a purely geometric characterization
of vacuum Einstein theory for complex or real Riemannian four-metrics,
also shed new light on a class of elliptic boundary-value problems
relevant for cosmology and fundamental physics [5].

Lecture 5 begins with a proof of the existence of self-adjoint
extensions of the Dirac operator subject to supersymmetric
boundary conditions on a three-sphere, following [4]. For
simplicity, the background is flat Riemannian four-space, with a
three-sphere boundary. The proof involves the generalization
of a method due to von Neumann, originally developed for complex
scalar fields for which complex conjugation can be defined.
Here, one deals instead with spinor fields for which only Euclidean
conjugation (see above) can be defined. The corresponding
asymptotic expansion of the heat kernel is then derived in
detail. Although the result is non-trivial, it relies on the
standard techniques of integration in the complex domain and
uniform asymptotic expansions of Bessel functions. While
some readers may not be interested in the relevance of this
analysis for quantum field theory in the presence of boundaries
(a new research topic entirely revolutionized in the last few years),
I hope they will appreciate there is a fertile interplay between
topological, analytic and geometrical ideas in modern mathematical
physics.
\vskip 1cm
\centerline {\bf LECTURE 1.}
\vskip 1cm
\centerline {\bf L1.1 Exterior Algebra and Clifford Algebra}
\vskip 0.3cm
For our purposes, we only need the basic definitions of these
algebras. For the sake of completeness, however, a few more
details are given following [1,2], whereas a more comprehensive
treatment can be found, again, in [1,2]. In the Lectures,
only part of this section will be used.

If $V$ is a complex $n$-dimensional vector space, we can form
the exterior powers $\Lambda^{q}(V)$. They are vector spaces
of dimension $\pmatrix {n \cr q \cr}$, and $GL(V)$ acts naturally
on them. Representations $\Lambda^{q}$ of $GL(n,C)$, and hence
of $U(n)$, are obtained by taking $V=C^{n}$. If $v \in V$ then
exterior multiplication by $v$ defines a linear transformation
$T_{v}:\Lambda^{q}(V) \rightarrow \Lambda^{q+1}(V)$. If $v$ does
not vanish, the sequence
$$
0 \rightarrow \Lambda^{0}(V) \rightarrow \Lambda^{1}(V)
\rightarrow ... \rightarrow \Lambda^{n}(V)
\rightarrow 0
\eqno (L1.1.1)
$$
is {\it exact}. In fact, if $T_{v}^{*}$ denotes the adjoint of
$T_{v}$ with respect to the natural inner product of $\Lambda^{*}(V)$
given by an inner product on $V$, then
$$
T_{v}^{*} \; T_{v}+T_{v} \; T_{v}^{*}
= {\| v \|}^{2} \; I
\; \; \; \; ,
\eqno (L1.1.2)
$$
where $I$ is the identity. Eq. (L1.1.2) is checked by using an
orthonormal base of $V$ for which $v=\lambda e_{1}$. It immediately
implies the exactness of the sequence (L1.1.1), since $T_{v}(x)=0$
yields $x=T_{v} \biggr(T_{v}^{*} {x \over {\| v \|}^{2}}\biggr)$.

Analogous results hold for real vector spaces. Thus, the exterior
powers $\Lambda^{q}(R^{n})$ give real representations of
$GL(n,R)$ and hence of $O(n)$. Note that $\Lambda^{q}(R^{2n})$ and
$\Lambda^{q}(C^{n})$ are sharply distinguished, since the first
has real dimension $\pmatrix {2n \cr q \cr}$ while the second
has complex dimension $\pmatrix {n \cr q \cr}$. Since, for a
complex vector space $V$, one has a natural isomorphism
$V \otimes_{R} C \cong V \oplus {\overline V}$, and since
$\Lambda^{*}(V \oplus W) \cong \Lambda^{*}(V) \otimes
\Lambda^{*}(W)$, it follows that
$$
\Lambda^{q}(V) \otimes_{R} C \cong
\Lambda^{q}(V \otimes_{R} C) \cong \Lambda^{*}(V)
\otimes \Lambda^{*}({\overline V})
\; \; \; \; .
\eqno (L1.1.3)
$$
This tells us how the exterior-algebra representation
of $O(2n)$ decomposes under restriction to $U(n)$.

For Euclidean space $E$ one has a natural duality
isomorphism $*:\Lambda^{q}(E) \rightarrow \Lambda^{n-q}(E)$,
where $n={\rm dim} E$. Choosing an orthonormal base
$\Bigr(e_{1},...,e_{n}\Bigr)$ of $E$ this is given by
$$
* \biggr(e_{i_{1}}\wedge e_{i_{2}}\wedge ... \wedge e_{i_{q}}
\biggr) = \epsilon \; e_{j_{1}} \wedge e_{j_{2}} \wedge ...
\wedge e_{j_{n-q}}
\; \; \; \; ,
\eqno (L1.1.4)
$$
where $\biggr(i_{1},i_{2},...,i_{q},j_{1},j_{2},...,j_{n-q}\biggr)$
is a permutation of $(1,2,...,n)$ of sign $\epsilon$. This
isomorphism depends on having an orientation of $E$. If $E$ is
even-dimensional: ${\rm dim}E=2l$, then on the middle dimension
$\Lambda^{l}(E)$ we have $*^{2}=(-1)^{l}$. If $l=2k$ then
$*^{2}=1$ and we can decompose $\Lambda^{l}(E)$ into subspaces
$\Lambda_{+}^{l}(E)$ and $\Lambda_{-}^{l}(E)$ given by the
eigenspaces of $*$. Thus, as representations of $SO(4k)$ we have
$$
\Lambda^{2k} \cong \Lambda_{+}^{2k} \oplus \Lambda_{-}^{2k}
\; \; \; \; .
\eqno (L1.1.5)
$$

If $Q$ is a quadratic form on a real vector space $V$, the
corresponding Clifford algebra $C(Q)$ is defined as the
quotient of the tensor algebra of $V$ by the ideal generated
by all elements of the form $x \otimes x -Q(x)1$ for
$x \in V$. If $Q=0$ we get the exterior algebra of $V$.
If $V=R^{n}$ and $Q(x)=-\sum_{i=1}^{n}x_{i}^{2}$ we get
the Clifford algebra $C_{n}$ which has generators
$e_{1},...,e_{n}$ obeying the relations
$$
e_{i}^{2}=-1
\; \; \; \; ,
\eqno (L1.1.6)
$$
$$
e_{i}e_{j}+e_{j}e_{i}=0
\; \; \; \;
i \not = j
\; \; \; \; .
\eqno (L1.1.7)
$$
Interestingly, ${\rm dim} C_{n}={\rm dim}
\Lambda^{*}(R^{n})=2^{n}$. An important result holds,
according to which if $n=2k$, the complexified
Clifford algebra ${\widetilde C}_{n} \equiv
C_{n} \otimes_{R}C$ is a full matrix algebra $M(2^{k})$. Moreover,
if $n=2k+1$, ${\widetilde C}_{n} \cong M(2^{k})
\oplus M(2^{k})$ (see problem L1.1).
\vskip 0.3cm
\centerline {\bf L1.2 Exterior Differentiation and
Hodge-Laplace Operator}
\vskip 0.3cm
Let $X$ be an $n$-dimensional $C^{\infty}$ manifold and
let $\Omega^{q}(X)$ denote the space of $C^{\infty}$
exterior differential forms of degree $q$ on $X$. One then
has a naturally occurring differential operator
$d: \Omega^{q}(X) \rightarrow \Omega^{q+1}(X)$ which extends
the differential of a function. The sequence
$$
0 \rightarrow \Omega^{0}(X) \rightarrow \Omega^{1}(X)
\rightarrow ... \rightarrow \Omega^{n}(X) \rightarrow 0
$$
is a complex in that $d^{2}=0$, and the main theorem of
de Rham states that the cohomology of this complex is
isomorphic to the cohomology of $X$, with real coefficients.
In other words, if $\gamma_{1},...,\gamma_{N}$ is a basis for
the homology in dimension $q$, then a $q$-form $\omega$
exists with $d\omega=0$ having prescribed periods
$\int_{\omega} \gamma_{i}$, and $\omega$ is unique modulo
the addition of forms $d\theta$. In particular, if $X=R^{n}$,
the corresponding $\Omega^{q}(X)$ can be identified with
$C^{\infty}$ maps $R^{n} \rightarrow
\Lambda^{q}({R^{n}}^{*})$,
and $d$ becomes a constant-coefficient operator whose
Fourier transform is exterior multiplication by $i\xi$,
for $\xi \in (R^{n})^{*}$.

If $X$ is compact oriented, and a Riemannian metric is
given on $X$, we can define a positive-definite inner
product on $\Omega^{q}(X)$ by
$$
<u,v> \equiv \int_{X}u_{\wedge} *v
\; \; \; \; ,
\eqno (L1.2.1)
$$
where $*: \Omega^{q}(X) \rightarrow \Omega^{n-q}(X)$
is the isomorphism given by the metric as in section L1.1.
By virtue of the identity
$$
d \Bigr(u_{\wedge} *v\Bigr)
=du_{\wedge} *v+(-1)^{q}u_{\wedge}d*v
\; \; \; \; ,
\eqno (L1.2.2)
$$
where $u \in \Omega^{q}, v \in \Omega^{q-1}$, one gets
$$
\int_{X}du_{\wedge} *v +(-1)^{q}\int_{X}u_{\wedge}
d*v=0
\; \; \; \; .
\eqno (L1.2.3)
$$
This implies
$$
<du,v>=(-1)^{q+1}<u,{*}^{-1}d*v>
\; \; \; \; .
\eqno (L1.2.4)
$$
Thus, the adjoint $d^{*}$ of $d$ on $\Omega^{q}$
is given by
$$
d^{*}=(-1)^{q+1}{*}^{-1}d^{*}
=\epsilon \; { }^{*}d^{*}
\; \; \; \; ,
\eqno (L1.2.5)
$$
where $\epsilon$ may take the values $\pm 1$ depending on
$n,q$.

The corresponding Hodge-Laplace operator on $\Omega^{q}$
is therefore defined as
$$
\bigtriangleup \equiv dd^{*}+d^{*}d
\; \; \; \; ,
\eqno (L1.2.6)
$$
and the harmonic forms ${\cal H}^{q}$ are the solutions
of $\bigtriangleup u=0$. The main theorem of Hodge theory
is that ${\cal H}^{q}$ is isomorphic to the $q$-th de Rham
group and hence to the $q$-th cohomology group of $X$.
\vskip 0.3cm
\centerline {\bf L1.3 Index of Elliptic Operators and
Signature Operator}
\vskip 0.3cm
It is convenient to consider the operator
$d+d^{*}$ acting on $\Omega^{*}(X) \equiv \oplus_{q}
\Omega^{q}(X)$. Since $d^{2}=(d^{*})^{2}=0$, one has
$\bigtriangleup=(d+d^{*})^{2}$ and the harmonic forms are
also the solutions of $(d+d^{*})u=0$. If we consider
$d+d^{*}$ as an operator $\Omega^{{\rm ev}} \rightarrow
\Omega^{{\rm odd}}$, its null-space is
$\oplus {\cal H}^{2q}$ while the null-space of its adjoint
is $\oplus {\cal H}^{2q+1}$. Hence its {\it index} is the
Euler characteristic of $X$. It is now appropriate to define
elliptic differential operators and their index.

Let us denote again by $X$ a compact oriented smooth manifold,
and by $E,F$ two smooth complex vector bundles over $X$. We
consider linear differential operators
$$
D:\Gamma(E) \rightarrow \Gamma(F)
\; \; \; \; ,
\eqno (L1.3.1)
$$
i.e. linear operators defined on the spaces of smooth
sections and expressible locally by a matrix of partial
derivatives. Note that the extra generality involved in
considering vector bundles presents no serious difficulties
and it is quite essential in a systematic treatment on
manifolds, since all geometrically interesting operators
operate on vector bundles.

Let $T^{*}(X)$ denote the cotangent vector bundle of $X$,
$S(X)$ the unit sphere-bundle in $T^{*}(X)$ (relative to
some Riemannian metric), $\pi: S(X) \rightarrow X$ the
projection. Then, associated with $D$ there is a vector-bundle
homomorphism
$$
\sigma(D):\pi^{*}E \rightarrow \pi^{*}F
\; \; \; \; ,
\eqno (L1.3.2)
$$
which is called the symbol of $D$. In terms of local coordinates
$\sigma(D)$ is obtained from $D$ replacing
${\partial \over \partial x_{j}}$ by $i\xi_{j}$ in the
highest-order terms of $D$ ($\xi_{j}$ is the $j$th coordinate
in the cotangent bundle). By definition, $D$ is elliptic if
$\sigma(D)$ is an isomorphism. Of course, this implies that
the complex vector bundles $E,F$ have the same dimension.

One of the basic properties of elliptic operators is that
${\rm Ker} \; D$ (i.e. the null-space) and ${\rm Coker} \; D
\equiv \Gamma(F)/D\Gamma(E)$ are both finite-dimensional. The
{\it index} $\gamma(D)$ is defined by
$$
\gamma(D) \equiv {\rm dim} \; {\rm Ker} \; D
- {\rm dim} \; {\rm Coker} \; D
\; \; \; \; .
\eqno (L1.3.3)
$$
If $D^{*}:\Gamma(F) \rightarrow \Gamma(E)$ denotes the
formal adjoint of $D$, relative to metrics in $E,F,X$,
then $D^{*}$ is also elliptic and
$$
{\rm Coker} \; D \cong {\rm Ker} \; D^{*}
\; \; \; \; ,
\eqno (L1.3.4)
$$
so that
$$
\gamma(D)={\rm dim} \; {\rm Ker} \; D
-{\rm dim} \; {\rm Ker} \; D^{*}
\; \; \; \; .
\eqno (L1.3.5)
$$

Getting back to the definition of symbol, we find it
helpful for the reader to say that, for the exterior
derivative $d$, its symbol is exterior multiplication by
$i \xi$, for $\bigtriangleup$ it is $-{\| \xi \|}^{2}$,
and for $d+d^{*}$ it is $iA_{\xi}$, where $A_{\xi}$ is
Clifford multiplication by $\xi$ [1,2].

For a given $\alpha \in \Lambda^{p}(R^{2l})
\otimes_{R}C$, let us denote by $\tau$ the involution
$$
\tau(\alpha) \equiv i^{p(p-1)+l} \; { }^{*}\alpha
\; \; \; \; .
\eqno (L1.3.6)
$$
If $n=2l$, $\tau$ can be expressed as $i^{l}\omega$, where
$\omega$ denotes Clifford multiplication by the volume
form ${ }^{*}1$. Remarkably, $(d+d^{*})$ and $\tau$
anti-commute. In the Clifford Algebra one has
$\xi \omega =-\omega \xi$ for $\xi \in T^{*}$. Thus, if
$\Omega^{\pm}$ denote the $\pm1$-eigenspaces of $\tau$,
$(d+d^{*})$ maps $\Omega^{+}$ into $\Omega^{-}$. The restricted
operator
$$
d+d^{*}:\Omega_{+} \rightarrow \Omega_{-}
\; \; \; \; ,
\eqno (L1.3.7)
$$
is called the {\it signature operator} and denoted by $A$.
If $l=2k$, the index of this operator is equal to
${\rm dim} \; {\cal H}_{+}-{\rm dim} \; {\cal H}_{-}$
where ${\cal H}_{\pm}$ are the spaces of harmonic forms in
$\Omega_{\pm}$. If $q \not = l$, the space
${\cal H}^{q} \oplus {\cal H}^{n-q}$ is stable under $\tau$,
which just interchanges the two factors. Hence one gets a
vanishing contribution to the index from such a space. For
$q=l=2k$, however, one has $\tau(\alpha)=i^{l(l-1)+l}
{ }^{*}\alpha={ }^{*}\alpha$ for $\alpha \in \Omega^{l}$.
One thus finds
$$
{\rm index} \; A = {\rm dim} \; {\cal H}_{+}^{l}
-{\rm dim} \; {\cal H}_{-}^{l}
\; \; \; \; ,
\eqno (L1.3.8)
$$
where ${\cal H}_{\pm}^{l}$ are the $\pm1$-eigenspaces
of $*$ on ${\cal H}^{l}$. Thus, since the inner product in
${\cal H}^{l}$ is given by $\int u_{\wedge}{ }^{*}v$, one
finds
$$
{\rm index} \; A={\rm Sign}(X)
\; \; \; \; ,
\eqno (L1.3.9)
$$
where ${\rm Sign}(X)$ is the signature of the quadratic
form on $H^{2k}(X;R)$ given by the cup-product. Hence the name
for the operator $A$. It is now worth recalling that the
cup-product is a very useful algebraic structure occurring in
cohomology theory. It works as follows. Given $[\omega] \in
H^{p}(M;R)$ and $[\nu] \in H^{q}(M;R)$, then we define the
cup-product of $[\omega]$ and $[\nu]$, written
$[\omega] \bigcup [\nu]$, by
$$
[\omega] \bigcup [\nu] \equiv [\omega \wedge \nu]
\; \; \; \; .
\eqno (L1.3.10)
$$
The right-hand-side of (L1.3.10) is a $(p+q)$-form so that
$[\omega \wedge \nu] \in H^{p+q}(M;R)$. One can check, using
the properties of closed and exact forms, that this is a
well-defined product of cohomology classes. The cup-product
$\bigcup$ is therefore a map of the form
$$
\bigcup : H^{p}(M;R) \times H^{q}(M;R) \rightarrow
H^{p+q}(M;R)
\; \; \; \; .
\eqno (L1.3.11)
$$
If we define the sum of all cohomology groups $H^{*}(M;R)$ by
$$
H^{*}(M;R) \equiv \oplus_{p>0}H^{p}(M;R)
\; \; \; \; ,
\eqno (L1.3.12)
$$
the cup-product has the neater looking form
$$
\bigcup : H^{*}(M;R) \times H^{*}(M;R)
\rightarrow H^{*}(M;R)
\; \; \; \; .
\eqno (L1.3.13)
$$
The product on $H^{*}(M;R)$ defined in (L1.3.13) makes
$H^{*}(M;R)$ into a ring. It can happen that two spaces
$M$ and $N$ have the same cohomology groups and yet are
not topologically the same. This can be proved by evaluating
the cup-product $\bigcup$ for $H^{*}(M;R)$ and $H^{*}(N;R)$,
and showing that the resulting rings are different. An example
is provided by choosing $M \equiv S^{2} \times S^{4}$
and $N \equiv CP^{3}$.
\vskip 10cm
\centerline {\bf L1.4 Dirac Operator}
\vskip 0.3cm
We begin our analysis by considering Euclidean space
$R^{2k}$. The Clifford algebra ${\widetilde C}_{2k}$
of section 1 is a full matrix algebra acting on the
$2^{k}$-dimensional spin-space $S$. Moreover, $S$ is
given by the direct sum of $S^{+}$ and $S^{-}$, i.e. the
eigenspaces of $\omega=e_{1}e_{2}...e_{2k}$, and
$\omega=\pm i^{k}$ on $S^{\pm}$. Let us now denote by
$E_{1},...,E_{2k}$ the linear transformations on $S$
representing $e_{1},...,e_{2k}$. The Dirac operator is
then defined as the first-order differential operator
$$
D \equiv \sum_{i=1}^{2k}E_{i}{\partial \over \partial
x_{i}}
\; \; \; \; ,
\eqno (L1.4.1)
$$
and satisfies $D^{2}=-\bigtriangleup \cdot I$, where $I$
is the identity of spin-space. Note that, since the $e_{i}$
anti-commute with $\omega$, the Dirac operator interchanges
the positive and negative spin-spaces, $S^{+}$ and $S^{-}$.
Restriction to $S^{+}$ yields the operator
$$
B : C^{\infty}\Bigr(R^{2k},S^{+}\Bigr) \rightarrow
C^{\infty} \Bigr(R^{2k},S^{-}\Bigr)
\; \; \; \; .
\eqno (L1.4.2)
$$
Since $E_{i}^{2}=-1$, in the standard metric the $E_{i}$
are unitary and hence skew-adjoint. Moreover, since
${\partial \over \partial x_{i}}$ is also formally
skew-adjoint, it follows that $D$ is formally self-adjoint,
i.e. the formal adjoint of $B$ is the restriction of
$D$ to $S^{-}$.

The next step is the global situation of a compact oriented
$2k$-dimensional manifold $X$. First, we assume that a
spin-structure exists on $X$. By this we mean that the
principal $SO(2k)$ bundle $P$ of $X$, consisting of oriented
orthonormal frames, lifts to a principal Spin($2k$) bundle
$Q$, i.e. the map $Q \rightarrow P$ is a double covering
inducing the standard covering Spin($2k$) $\rightarrow
SO(2k)$ on each fibre. If $S^{\pm}$ are the two half-spin
representations of Spin($2k$), we consider the associated
vector bundles on $X$: $E^{\pm} \equiv Q
\times_{{\rm Spin}(2k)}S^{\pm}$. Sections of these vector
bundles are the spinor fields on $X$. The {\it total} Dirac
operator on $X$ is a differential operator acting on
$E=E^{+} \oplus E^{-}$ and switching factors as above. To
define $D$ we have to use the Levi-Civita connection on $P$,
which lifts to one on $Q$. This enables one to define the
covariant derivative
$$
\nabla : C^{\infty}\Bigr(X,E\Bigr) \rightarrow C^{\infty}
\Bigr(X,E \otimes T^{*}\Bigr)
\; \; \; \; ,
\eqno (L1.4.3)
$$
and $D$ is defined as the composition of $\nabla$ with
the map $C^{\infty}(X,E \otimes T^{*}\Bigr)
\rightarrow C^{\infty}(X,E)$ induced by Clifford multiplication.
By using an orthonormal base $e_{i}$ of $T$ at any point we
can write
$$
Ds \equiv \sum_{i=1}^{2k} e_{i} \nabla_{i}s
\; \; \; \; ,
\eqno (L1.4.4)
$$
where $\nabla_{i}s$ is the covariant derivative in the
direction $e_{i}$, and $e_{i}( \; )$ denotes Clifford
multiplication. By looking at symbols, $D$ is skew-adjoint.
Hence $D-D^{*}$ is an algebraic invariant of the metric and
it only involves the first derivatives of $g$. Using normal
coordinates one finds that any such invariant vanishes,
which implies $D=D^{*}$, i.e. $D$ is self-adjoint.

As in Euclidean space, the restriction of $D$ to the
half-spinors $E^{+}$ is denoted by
$$
B: C^{\infty}\Bigr(X,E^{+}\Bigr) \rightarrow
C^{\infty} \Bigr(X,E^{-}\Bigr)
\; \; \; \; .
\eqno (L1.4.5)
$$
The index of the restricted Dirac operator is given by
$$
{\rm index} \; B={\rm dim} \; {\cal H}^{+}
-{\rm dim} \; {\cal H}^{-}
\; \; \; \; ,
\eqno (L1.4.6)
$$
where ${\cal H}^{\pm}$ are the spaces of solutions of
$Du=0$, for $u$ a section of $E^{\pm}$. Since $D$ is
elliptic and self-adjoint, these spaces are also the solutions
of $D^{2}u=0$, and are called {\it harmonic spinors}.
\vskip 100cm
\centerline {\bf LECTURE II.}
\vskip 1cm
\centerline {\bf L2.1 Index Problem for Manifolds with Boundary}
\vskip 0.3cm
Following [1,2], this section is devoted to the extension of
the index formula for closed manifolds, to manifolds with
boundary. The naturally occurring question is then how to
measure the topological implications of elliptic boundary
conditions, since boundary conditions have of course a
definite effect on the index.

For example, let $X$ be the unit disk in the plane, let $Y$
be the boundary of $X$, and let $b$ be a nowhere-vanishing
vector field on $Y$. Denoting by $D$ the Laplacian on $X$,
we consider the operator
$$
(D,b):C^{\infty}(X) \rightarrow C^{\infty}(X) \oplus
C^{\infty}(Y)
\; \; \; \; .
\eqno (L2.1.1)
$$
By definition, $(D,b)$ sends $f$ into $Df \oplus (bf \mid Y)$,
where $bf$ is the directional derivative of $f$ along $b$.
Since $D$ is elliptic and $b$ is non-vanishing, kernel and
cokernel of $(D,b)$ are both finite-dimensional, hence the
index of the boundary-value problem is finite. One would now
like it to express the index in terms of the topological
data given by $D$ and the boundary conditions.
The solution of this problem is expressed by a formula
derived by Vekua [1,2], according to which
$$
{\rm index} \; (D,b)=2 \Bigr(1-{\rm winding}
\; {\rm number} \; {\rm of} \; b \Bigr)
\; \; \; \; .
\eqno (L2.1.2)
$$

It is now necessary to recall the index formula for closed
manifolds, and this is here done in the case of vector
bundles. From now on, $X$ is the base manifold, $Y$ its
boundary, $T(X)$ the tangent bundle of $X$, $B=B(X)$
the ball-bundle (consisting of vectors in $T(X)$ of
length $\leq 1$), $S=S(X)$ the sphere-bundle of unit vectors
in $T(X)$. Of course, if $X$ is a closed manifold
(hence without boundary), then $S(X)$ is just the boundary of
$B(X)$, i.e.
$$
\partial B(X)=S(X)
\; \; \; \; {\rm if} \; \partial X = \emptyset
\; \; \; \; .
\eqno (L2.1.3)
$$
However, if $X$ has a boundary $Y$, then
$$
\partial B(X)=S(X) \cup B(X) \mid Y
\; \; \; \; ,
\eqno (L2.1.4)
$$
where $B(X) \mid Y$ is the subspace of $B(X)$ lying
over $Y$ under the natural map $\pi: B(X) \rightarrow X$.

We now focus on a system of $k$ linear partial differential
operators, i.e.
$$
Df_{i} \equiv \sum_{j=1}^{k} A_{ij} f_{j}
\; \; \; \; ,
\eqno (L2.1.5)
$$
defined on the $n$-dimensional manifold $X$. As we know from
Lecture 1, the symbol of $D$ is a function $\sigma(D)$ on
$T^{*}(X)$ attaching to each cotangent vector $\lambda$ of
$X$, the matrix $\sigma(D: \lambda)$ obtained from the
highest terms of $A_{ij}$ by replacing ${\partial^{\alpha}
\over \partial x^{\alpha}}$ with $(i \lambda)^{\alpha}$.
Moreover, the system $D$ is {\it elliptic} if and only if
the function $\sigma(D)$ maps $S(X)$ into the group
$GL(k,C)$ of non-singular $k \times k$ matrices with complex
coefficients. Thus, defining
$$
GL \equiv {\rm lim}_{m \to \infty}GL(m,C)
\; \; \; \; ,
\eqno (L2.1.6)
$$
the symbol defines a map
$$
\sigma(D):S(X) \rightarrow GL
\; \; \; \; .
\eqno (L2.1.7)
$$
Since the index of an elliptic system is invariant under
deformations, on a closed manifold the index of $D$ only
depends on the homotopy class of the map $\sigma(D)$ [1,2].

To write down the index formula we are looking for, one
constructs a definite differential form $ch \equiv
\sum_{i}{ch}^{i}$ on $GL$. Note that, strictly, we
define a differential form ${ch}(m)$ on each
$GL(m,C)$. Moreover, by using a universal expression in the
curvature of $X$, one constructs a definite form
${\cal F}(X) \equiv \sum_{i}{\cal F}^{i}(X)$ on $X$. The index
formula for a closed manifold then yields the index of
$D$ as an integral [1,2]
$$
{\rm index} (D)=
\int_{S(X)} \sigma(D)^{*} \; {ch}
\wedge \pi^{*} {\cal F}(X)
\; \; \; \; .
\eqno (L2.1.8)
$$
With our notation, $\sigma(D)^{*}ch$ is the form on $GL$
pulled back to $S(X)$ via $\sigma(D)$. In light of
(L2.1.3), the integral (L2.1.8) may be re-written as
$$
{\rm index}(D)=
\int_{\partial B(X)} \sigma(D)^{*} \; {ch}
\wedge \pi^{*} {\cal F}(X)
\; \; \; \; .
\eqno (L2.1.9)
$$
In this form the index formula is also meaningful
for a manifold with boundary, providing that $\sigma(D)$,
originally defined on $S(X)$, is extended to
$\partial B(X)$. It is indeed in this extension that the
topological data of a set of elliptic boundary conditions
manifest themselves. Following [1,2], we may in fact state
the following theorem:
\vskip 0.3cm
\noindent
{\bf Theorem L2.1.1} A set of elliptic boundary conditions,
$B$, on the elliptic system $D$, defines a definite map
$\sigma(D,B):\partial B(X) \rightarrow GL$, which extends
the map $\sigma(D): S(X) \rightarrow GL$ to the whole of
$\partial B(X)$.
\vskip 0.3cm
\noindent
One thus finds an index theorem formally analogous to the
original one, in that the index of $D$ subject to the
elliptic boundary condition $B$ is given by
$$
{\rm index}(D,B)=\int_{\partial B(X)}
\sigma(D,B)^{*} {ch} \wedge \pi^{*} {\cal F}(X)
\; \; \; \; .
\eqno (L2.1.10)
$$
\vskip 0.3cm
\centerline {\bf L2.2 Elliptic Boundary Conditions}
\vskip 0.3cm
Although this section is (a bit) technical, it is necessary
to include it for the sake of completeness. Relevant examples
of elliptic boundary-value problems will be given in Lectures
3, 4 and 5.

Following [1-2], let $D$ be a $k \times k$ elliptic system of
differential operators on $X$, and let $r$ denote the order
of $D$. The symbol $\sigma(D)$ is a function on the cotangent
vector bundle $T^{*}(X)$ whose values are $(k \times k)$-matrices,
and its restriction to the unit sphere-bundle $S(X)$ takes
non-singular values. For our purposes, it is now necessary to
consider a system $B$ of boundary operators given by an
$l \times k$ matrix with rows $b_{1},...,b_{l}$ of orders
$r_{1},...,r_{l}$ respectively. We now denote by $\sigma(b_{i})$
the symbol of $b_{i}$, and by $\sigma(B)$ the matrix with
$\sigma(b_{i})$ as $i$-th row. At a point of the boundary
$Y$ of $X$, let $\nu$ be the unit inward normal and let $y$
denote any unit tangent vector to $Y$. One puts
$$
\sigma_{y}(D)(t)=\sigma(D)(y+t\nu)
\; \; \; \; ,
\eqno (L2.2.1)
$$
$$
\sigma_{y}(B)(t)=\sigma(B)(y+t\nu)
\; \; \; \; ,
\eqno (L2.2.2)
$$
so that $\sigma_{y}(D)$ and $\sigma_{y}(B)$ are polynomials
in $t$. It then makes sense to consider the system of
ordinary linear equations
$$
\sigma_{y}(D)\biggr(-i{d\over dt}\biggr)u=0
\; \; \; \; ,
\eqno (L2.2.3)
$$
whose space of solutions is denoted by ${\cal M}_{y}$.
The ellipticity of the system $D$ means, by definition,
that ${\cal M}_{y}$ can be decomposed as
$$
{\cal M}_{y}={\cal M}_{y}^{+} \oplus
{\cal M}_{y}^{-}
\; \; \; \; ,
\eqno (L2.2.4)
$$
where ${\cal M}_{y}^{+}$ consists only of exponential
polynomials involving $e^{i\lambda t}$ with
${\rm Im}(\lambda)>0$, and ${\cal M}_{y}^{-}$ involves
those with ${\rm Im}(\lambda)<0$. The ellipticity
condition for the system of boundary operators, relative
to the elliptic system of differential operators, is that
the equations (L2.2.3) should have a unique solution
$u \in {\cal M}_{y}^{+}$ satisfying the boundary
condition
$$
\sigma_{y}(B) \biggr(-i{d\over dt}\biggr)
u {\left.  \right |}_{t=0}=V
\; \; \; \; ,
\eqno (L2.2.5)
$$
for any given $V \in C^{l}$.

Although your lecturer is not quite algebraically oriented,
he has to use a few (more) algebraic concepts at this point [1-2].
Let $\Lambda \subset C(t)$ be the ring of all rational functions
of $t$ with no poles in the half-plane ${\rm Im}(t)>0$. One may
then regard $\sigma_{y}(D)(t)$ as defining a homomorphism of free
$\Lambda$-modules of rank $k$, and its cokernel is denoted by
$M_{y}^{+}$. One thus has the exact sequence of $\Lambda$-modules
$$
0 \rightarrow \Lambda^{k} \rightarrow \Lambda^{k}
\rightarrow M_{y}^{+} \rightarrow 0
\; \; \; \; .
\eqno (L2.2.6)
$$
Now a lemma can be proved, according to which a natural
isomorphism of vector spaces exists
$$
M_{y}^{+} \cong {\cal M}_{y}^{+}
\; \; \; \; .
\eqno (L2.2.7)
$$
Hence the elliptic boundary condition yields an isomorphism
$$
\beta_{y}^{+}: M_{y}^{+} \rightarrow C^{l}
\; \; \; \; .
\eqno (L2.2.8)
$$
The set of all $M_{y}^{+}$, for $y \in S(Y)$, forms a vector
bundle $M^{+}$ over $S(Y)$, and (L2.2.8) defines an isomorphism
$\beta^{+}$ of $M^{+}$ with the trivial bundle $S(Y) \times
C^{l}$.

Possible boundary conditions are differential or integro-differential.
If only differential boundary conditions are imposed then the
map $\beta_{y}^{+}$, regarded as a function of $y$, cannot be an
arbitrary continuous function. To obtain {\it all} continuous
functions one has to enlarge the problem and consider
integro-differential boundary conditions as well [1-2]. This is,
indeed, an important topological simplification. Thus, an
elliptic problem $(D,B)$ has associated with it $\sigma(D),M^{+},
\beta^{+}$, where $\beta^{+}$ can be any vector-bundle isomorphism
of $M^{+}$ with the trivial bundle.
\vskip 0.3cm
\centerline {\bf L2.3 Index Theorem and Anomalies}
\vskip 0.3cm
Anomalies in gauge theories, and their relation to the index
theory of the Dirac operator, are at the core of modern attempts
to quantize the gauge theories of fundamental interactions.
The aim of the second part of this Lecture is to introduce the
reader to ideas and problems in this field of research,
following [3].

The index theorem is concerned with any elliptic differential
operator, but for physical applications one only needs the
special case of the Dirac operator. From Lecture 1, we know
this is globally defined on any Riemannian manifold
$(M,g)$ providing $M$ is oriented and has a spin-structure.
If $M$ is compact, the Dirac operator is elliptic, self-adjoint,
and has a discrete spectrum of eigenvalues. Of particular
interest is the $0$-eigenspace (or null-space). By definition,
the Dirac operator acts on spinor fields, and it should be
emphasized there is a basic algebraic difference between
spinors in even and odd dimensions. In fact in odd dimensions
spinors belong to an irreducible representation of the
spin-group, whereas in even dimensions spinors break up
into two irreducible pieces, here denoted by $S^{+}$ and
$S^{-}$. The Dirac operator interchanges these two, hence it
consists essentially of an operator
$$
D: S^{+} \rightarrow S^{-}
\; \; \; \; ,
\eqno (L2.3.1)
$$
and its adjoint
$$
D^{*}:S^{-} \rightarrow S^{+}
\; \; \; \; .
\eqno (L2.3.2)
$$
The dimensions of the null-spaces of $D$ and $D^{*}$ are denoted
by $N^{+}$ and $N^{-}$ respectively, and the index of $D$
(cf. Lecture 1) is defined by
$$
{\rm index}(D) \equiv N^{+}-N^{-}
\; \; \; \; .
\eqno (L2.3.3)
$$
Note that, although a formal symmetry exists between
positive and negative spinors (in fact they are interchanged
by reversing the orientation of $M$), the numbers
$N^{+}$ and $N^{-}$ need not be equal, due to topological
effects. Hence the origin of the {\it chiral anomaly}.

The index theorem provides an explicit formula for
${\rm index}(D)$ in terms of topological invariants
of $M$. Moreover, by using the Riemannian metric $g$
and its curvature tensor $R(g)$, one can write an explicit
integral formula
$$
{\rm index}(D)=\int_{M}\Omega(R(g))
\; \; \; \; ,
\eqno (L2.3.4)
$$
where $\Omega$ is a formal expression obtained purely
algebraically out of $R(g)$. For example, when $M$ is
four-dimensional one finds
$$
\Omega={{\rm Tr} \; \omega^{2} \over 96 \pi^{2}}
\; \; \; \; ,
\eqno (L2.3.5)
$$
where $\omega$ is regarded as a matrix of two-forms.
Interestingly, the index formula (L2.3.4) is, from the
physical point of view, purely gravitational since it
only involves the metric $g$. Moreover, the index vanishes
for the sphere or torus, and one needs more complicated
manifolds to exhibit a non-vanishing index. However,
(L2.3.4) can be generalized to gauge theories. This means
one is given a complex vector bundle $V$ over $M$ with a
unitary connection $A$. If the fibre of $V$ is isomorphic
with $C^{N}$, then $A$ is a $U(N)$ gauge field over $M$.
The extended Dirac operator now acts on vector-bundle-valued
spinors
$$
D_{A}:S^{+} \otimes V \rightarrow S^{-} \otimes V
\; \; \; \; ,
\eqno (L2.3.6)
$$
and one defines again the index of $D_{A}$  by a formula
like (L2.3.3). The index formula (L2.3.4) is then replaced
by
$$
{\rm index}\Bigr(D_{A}\Bigr)=\int_{M}
\Omega(R(g),F(A))
\; \; \; \; ,
\eqno (L2.3.7)
$$
where $F(A)$ is the curvature of $A$, and $\Omega(R,F)$ is a
certain algebraic expression in $R$ and $F$. For example,
if $M$ is four-dimensional, (L2.3.5) is generalized by
$$
\Omega={{\rm Tr} \; \omega^{2}\over 96 \pi^{2}}
-{{\rm Tr} \; F^{2} \over 8\pi^{2}}
\; \; \; \; .
\eqno (L2.3.8)
$$
\vskip 5cm
\centerline {\bf L2.4 Index of Two-Parameter Families}
\vskip 0.3cm
A very important property of the index is that it is
unchanged by perturbation of the operator, hence it can
be given by a topological formula. This is why, when we
vary the gauge field $A$ continuously, the index of the
extended Dirac operator $D_{A}$ does not change. It is
however possible to extract more topological information from
a continuous family of operators, and we here focus on the
two-parameter case [3].

For this purpose we consider a two-dimensional surface $X$
compact, connected and oriented, and suppose a fibre bundle
$Y$ over $X$ is given with fibre $M$. The twist involved in
making a non-trivial bundle is an essential topological
feature. By giving $Y$ a Riemannian metric, we get metrics
$g_{x}$ on all fibres $M_{x}$, hence we have metrics on
$M$ parametrized by $X$. The corresponding family of Dirac
operators on $M$ is denoted by $D_{x}$. We can now define
two topological invariants, i.e. ${\rm index}(D_{x})$ and
the {\it degree} of the family of Dirac operators. To define
such a degree, we restrict ourselves to the particular case
when the index vanishes.

Since ${\rm index}(D_{x})=0$ by hypothesis, the generic
situation (one can achieve this by perturbing the metric
on $Y$) is that $D_{x}$ is invertible $\forall x \in X$
except at a finite number of points $x_{1},...,x_{k}$.
Moreover, a multiplicity $\nu_{i}$ can be assigned to
each $x_{i}$, which is generically $\pm 1$. To obtain
this one can, in the neighbourhood of each $x_{i}$, replace
$D_{x}$ by a finite matrix $T_{x}$. Zeros of ${\rm det}(T_{x})$
have multiplicity $\nu_{i}$ at $x_{i}$. We are actually dealing
with the phase-variation of ${\rm det}(T_{x})$ as $x$
traverses positively a small circle centred at $x_{i}$.

The degree $\nu$ of the family of Dirac operators $D_{x}$
is defined by
$$
\nu \equiv \sum_{i=1}^{k}\nu_{i}
\; \; \; \; .
\eqno (L2.4.1)
$$
This is a topological invariant of the family in that it is
invariant under perturbation. Hence it only depends on the
topology of the fibre bundle $Y$ and vanishes for the
product $M \times X$. In the case of gauge fields, we can
fix the metric $g$ on $X$ and take $Y= M \times X$, but we
also take a vector bundle $V$ on $Y$ with a connection.
This gives a family $A_{x}$ of gauge fields, and the
corresponding Dirac family has again a degree which now
depends on the topology of $V$. Such a degree vanishes if
the vector bundle $V$ comes from a bundle on $M$.
\vskip 0.3cm
\centerline {\bf L2.5 Zeta-Function Regularization of Path
Integrals}
\vskip 0.3cm
In the path-integral approach to quantum field theories,
one deals with determinants of elliptic operators, and it
may not be obvious how to regularize and to retain all the
required properties (e.g. gauge-invariance,
supersymmetry) at the same time.
The formal expressions for one-loop quantum
amplitudes clearly diverge [4] since the eigenvalues
$\lambda_{n}$ increase without bound, and a regularization
is indeed necessary. For this purpose, the following technique
has been described and applied by many authors.

Bearing in mind that Riemann's zeta-function $\zeta_{R}(s)$ is
defined as
$$
\zeta_{R}(s) \equiv \sum_{n=1}^{\infty}n^{-s}
\; \; \; \; ,
\eqno (L2.5.1)
$$
one first defines a generalized zeta-function $\zeta(s)$
obtained from the (positive) eigenvalues of the second-order,
self-adjoint operator ${\cal B}$.
Such $\zeta(s)$ can be defined as
$$
\zeta(s) \equiv \sum_{n=n_{0}}^{\infty}
\sum_{m=m_{0}}^{\infty} d_{m}(n) \lambda_{n,m}^{-s}
\; \; \; \; .
\eqno (L2.5.2)
$$
This means that all the eigenvalues are completely
characterized by two integer labels $n$ and $m$, while
their degeneracy $d_{m}$ only depends on $n$. This is the
case studied in the following lectures.
Note that formal differentiation of (L2.5.2) at the origin
yields
$$
{\rm det} \Bigr({\cal B}\Bigr) = e^{-\zeta'(0)}
\; \; \; \; .
\eqno (L2.5.3)
$$
This result can be given a sensible meaning since, in four
dimensions, $\zeta(s)$ converges for $Re(s)>2$, and one can
perform its analytic extension to a meromorphic function of
$s$ which only has poles at $s={1\over 2},1,{3\over 2},2$.
Since ${\rm det} \Bigr(\mu {\cal B}\Bigr)=\mu^{\zeta(0)}
{\rm det} \Bigr({\cal B}\Bigr)$, one finds the useful formula
for the quantum amplitude [4]
$$
\log \Bigr({\widetilde A}_{\phi}\Bigr)={1\over 2}\zeta'(0)
+{1\over 2} \log \Bigr(2\pi \mu^{2}\Bigr) \zeta(0)
\; \; \; \; .
\eqno (L2.5.4)
$$
It may happen quite often that
the eigenvalues appearing in (L2.5.2) are unknown, since the
eigenvalue condition, i.e. the equation leading to the
eigenvalues by virtue of the boundary conditions,
is a complicated equation which cannot
be solved exactly for the eigenvalues. However, since the
scaling properties of the one-loop amplitude are still
given by $\zeta(0)$ (and $\zeta'(0)$) as shown in (L2.5.4),
efforts have been made to compute the regularized $\zeta(0)$
also in this case. The various steps of this program are as
follows [4].

(1) One first studies the heat equation for the operator
${\cal B}$
$$
{\partial \over \partial \tau}F(x,y,\tau) + {\cal B}
F(x,y,\tau)=0 \; \; \; \; ,
\eqno (L2.5.5)
$$
where the Green's function $F$ satisfies the initial
condition $F(x,y,0)=\delta(x,y)$.

(2) Assuming completeness of the set $\Bigr \{ \phi_{n}
\Bigr \}$ of eigenfunctions of ${\cal B}$, the field
$\phi$ can be expanded as
$$
\phi=\sum_{n=n_{i}}^{\infty}a_{n}\phi_{n}
\; \; \; \; .
$$

(3) The Green's function $F(x,y,\tau)$ is then given by
$$
F(x,y,\tau)=\sum_{n=n_{0}}^{\infty}\sum_{m=m_{0}}^{\infty}
e^{-\lambda_{n,m}\tau}
\phi_{n,m}(x) \otimes \phi_{n,m}(y)
\; \; \; \; .
\eqno (L2.5.6)
$$

(4) The corresponding (integrated) heat kernel is then
$$
G(\tau) = \int_{M}Tr \; F(x,x,\tau)\sqrt{g} \; d^{4}x
=\sum_{n=n_{0}}^{\infty}\sum_{m=m_{0}}^{\infty}
e^{-\lambda_{n,m}\tau}
\; \; \; \; .
\eqno (L2.5.7)
$$

(5) In light of (L2.5.2) and (L2.5.7), the generalized
zeta-function can be also obtained as an integral transform
of the integrated heat kernel
$$
\zeta(s)={1\over \Gamma(s)} \int_{0}^{\infty}
\tau^{s-1}G(\tau) \; d\tau
\; \; \; \; .
\eqno (L2.5.8)
$$

(6) The hard part of the analysis is now to prove that
$G(\tau)$ has an asymptotic expansion as
$\tau \rightarrow 0^{+}$. This property has
been proved for all boundary conditions such that the
Laplace operator is self-adjoint [4]. The corresponding
asymptotic expansion of $G(\tau)$ can be written as
$$
G(\tau) \sim b_{1}\tau^{-2}+b_{2}\tau^{-{3\over 2}}
+b_{3}\tau^{-1}+b_{4}\tau^{-{1\over 2}}+b_{5}+
{\rm O} \Bigr(\sqrt{\tau} \Bigr)
\; \; \; \; ,
\eqno (L2.5.9)
$$
which implies
$$
\zeta(0)=b_{5} \; \; \; \; .
\eqno (L2.5.10)
$$
The result (L2.5.10) is proved by splitting the integral in (L2.5.8)
into an integral from $0$ to $1$ and an integral from
$1$ to $\infty$. The asymptotic expansion of
$\int_{0}^{1} \tau^{s-1}G(\tau) \; d\tau$ is then obtained
by using (L2.5.9).

In other words, for a given second-order self-adjoint elliptic
operator, we study the corresponding heat equation, and the
integrated heat kernel $G(\tau)$. The regularized $\zeta(0)$
value is then given by the constant term appearing in the
asymptotic expansion of $G(\tau)$ as
$\tau \rightarrow 0^{+}$. The
regularized $\zeta(0)$ value also yields the one-loop
divergences of the theory for bosonic and fermionic fields [4].
\vskip 0.3cm
\centerline {\bf L2.6 Determinants of Dirac Operators}
\vskip 0.3cm
For fermionic fields one has to define determinants of
Dirac operators, and these are not positive-definite by
virtue of their first-order nature. Consider as before
an even-dimensional Riemannian manifold $(M,g)$ with Dirac
operators $D_{g}:S^{+} \rightarrow S^{-}$, depending on the
metric $g$. We look for a regularized complex-valued
determinant ${\rm det}(D_{g})$ which should have the
following properties [3]:

(1) ${\rm det}(D_{g})$ is a differentiable function
of $g$.

(2) ${\rm det}(D_{g})$ is gauge-invariant, i.e.
${\rm det}\Bigr(D_{f_{g}}\Bigr)={\rm det}(D_{g})$ for any
diffeomorphism $f$ of $M$.

(3) ${\rm det}(D_{g})=0$ to first-order if and only if
$D_{g}^{*} \; D_{g}$ has exactly one zero-eigenvalue.
\vskip 0.3cm
\noindent
The third property is the characteristic property of determinants
in finite dimensions and it is a minimum requirement for
any regularized determinant. Note also that, since
$D_{g}^{*} \; D_{g}$ is a Laplace-type operator, there is no
difficulty in defining its determinant, so that one can also
define $\left | {\rm det}(D_{g}) \right |$. The problem is to
define the {\it phase} of ${\rm det}(D_{g})$.

Suppose now one can find a fibration over a surface $X$ with
fibre $M$, so that the Dirac operators $D_{x}$ all have
vanishing index whereas the {\it degree} of the family does
not vanish. This implies there is no function ${\rm det}(D_{g})$
with the 3 properties just listed. Suppose in fact such a
${\rm det}(D_{g})$ exists. Then (1) and (2) imply that
$h(x) \equiv {\rm det} \Bigr(D_{g_{x}}\Bigr)$ is a differentiable
complex-valued function on $X$. Moreover, (3) implies that
$\nu$ is the number of zeros of $h$ counted with multiplicities
and signs, i.e. that $\nu$ is the degree of $h:X \rightarrow C$.
But topological arguments may be used to show that the degree
of $h$ has to vanish, hence $\nu=0$, contradicting the
assumption [3].

We should now bear in mind that the local multiplicities
$\nu_{i}$ at points $x_{i}$ where $D_{x}$ is not invertible,
are defined as local phase variations obtained by using an
eigenvalue cut-off. Remarkably, if it were possible to use
such a cut-off consistently at all points, the total phase
variation $\nu$ would have to vanish. Thus, a fibration with
non-vanishing degree implies a behaviour of the eigenvalues
which prevents a good cut-off regularization. Such a remark is
not of purely academic interest, since two-dimensional examples
can be found, involving surfaces $M$ and $X$ of fairly high
genus, where all these things happen. What is essentially
involved is the variation in the conformal structure of
$M$ [3].

Other examples of non-vanishing degree, relevant for gauge theory,
are obtained when $M$ is the $2n$-sphere $S^{2n}$, $X=S^{2}$
and the vector bundle $V$ is $(n+1)$-dimensional, so that the
gauge group $G$ is $U(n+1)$ or $SU(n+1)$.
If one represents $S^{2}$ as $R^{2} \cup \infty$ then removing the
point at infinity, one obtains a two-parameter family of gauge
fields on $M$ parametrized by $R^{2}$. Since this family is
defined by a bundle over $M \times S^{2}$, gauge fields are all
gauge-equivalent as one goes to $\infty$ in $R^{2}$. What are
we really up to ? By taking a large circle in $R^{2}$, one gets
a closed path in the group ${\cal G}$ of gauge transformations
of the bundle on $M$. The degree $\nu$ is essentially a
homomorphism $\pi_{1}({\cal G}) \rightarrow Z$, and if
$\nu$ does not vanish, the variation in phase of
${\rm det}(D_{g})$ going round this path in ${\cal G}$ proves
the lack of gauge-invariance. For $M=S^{2n}$ one has
$$
\pi_{1}({\cal G})=\pi_{2n+1}(G)
\; \; \; \; ,
\eqno (L2.6.1)
$$
$$
\pi_{2n+1} \Bigr(SU(n+1)\Bigr)=Z
\; \; \; \; ,
\eqno (L2.6.2)
$$
and the isomorphism is given by $\nu$, by virtue of the general
index theorem for families [3].
\vskip 100cm
\centerline {\bf LECTURE 3.}
\vskip 1cm
\centerline {\bf L3.1 Spectral Asymmetry and Riemannian Geometry}
\vskip 0.3cm
If $X$ is a closed, oriented, four-dimensional Riemannian
manifold, in addition to the Gauss-Bonnet formula there is
another formula which relates cohomological invariants with
curvature. In fact, it is known that the signature (i.e.
number of positive eigenvalues minus number of negative
eigenvalues) of the quadratic form on $H^{2}(X;R)$ given by
the cup-product is expressed by
$$
{\rm sign}(X)={1\over 3} \int_{X}p_{1}
\; \; \; \; .
\eqno (L3.1.1)
$$
In (L3.1.1), $p_{1}$ is the differential four-form which
represents the first Pontrjagin class [1-2,4], and is equal
to $(2\pi)^{-2} \; {\rm Tr}(R^{2})$, where $R$ is the curvature
matrix. However, (L3.1.1) does not hold in general for
manifolds with boundary, so that one has
$$
{\rm sign}(X)-{1\over 3}\int_{X}p_{1}=f(Y) \not = 0
\; \; \; \; ,
\eqno (L3.1.2)
$$
where $Y \equiv \partial X$. Thus, if $X'$ is another manifold
with the same boundary, i.e. such that $Y = \partial X'$, one
has
$$
{\rm sign}(X)-{1\over 3}\int_{X}p_{1}
={\rm sign}(X')-{1\over 3}\int_{X'}p_{1}'
\; \; \; \; .
\eqno (L3.1.3)
$$
Hence one is looking for a continuous function $f$ of the
metric on $Y$ such that $f(-Y)=-f(Y)$. Atiyah et al. [1-2]
were able to prove that $f(Y)$ is a spectral invariant,
evaluated as follows. One looks at the Laplace operator
$\triangle$ acting on forms as well as on scalar functions.
This operator $\triangle$ is the square of the self-adjoint
first-order operator $B \equiv \pm \Bigr(d*-*d\Bigr)$, where
$d$ is the exterior-derivative operator, and $*$ is the
Hodge-star operator mapping $p$-forms to $(l-p)$-forms in
$l$ dimensions. Thus, if $\lambda$ is an eigenvalue of $B$,
the eigenvalues of $\triangle$ are of the form $\lambda^{2}$.
However, the eigenvalues of $B$ can be both negative and positive.
One takes this property into account by defining the
$\eta$-function
$$
\eta(s) \equiv \sum_{\lambda \not = 0}d(\lambda)
({\rm sign}(\lambda))
{\mid \lambda \mid}^{-s}
\; \; \; \; ,
\eqno (L3.1.4)
$$
where $d(\lambda)$ is the multiplicity of the eigenvalue
$\lambda$. Note that, since $B$ involves the $*$ operator,
in reversing the orientation of the boundary $Y$ we change
$B$ into $-B$, and hence $\eta(s)$ into $-\eta(s)$. The main
result of Atiyah et al., in its simplest form, states therefore
that [1-2,4]
$$
f(Y)={1\over 2}\eta(0)
\; \; \; \; .
\eqno (L3.1.5)
$$
Now, for a manifold $X$ with boundary $Y$, if one tries to
set up an elliptic boundary-value problem for the signature
operator of Lecture 1, one finds there is no local boundary
condition for this operator. For global boundary conditions,
however, expressed by the vanishing of a given integral
evaluated on $Y$, one has a good elliptic theory and a finite
index. Thus, one has to consider the theorem expressed by
(L3.1.5) within the framework of index theorems for global
boundary conditions. Atiyah et al. [1-2] were also able
to derive the relation between the index of the Dirac operator
on $X$ with a global boundary condition and $\eta(0)$,
where $\eta$ is the $\eta$-function of the Dirac operator
on $Y$.
\vskip 10cm
\centerline {\bf L3.2 $\eta(0)$ Calculation}
\vskip 0.3cm
A non-trivial result is regularity at the origin of the
$\eta$-function defined in (L3.1.4) for first-order
elliptic operators. Following Atiyah et al. [1-2]
we now evaluate $\eta(0)$ in a specific example.

Let $Y$ be a closed manifold, $E$ a vector bundle over $Y$
and $A: C^{\infty}(Y,E) \rightarrow C^{\infty}(Y,E)$ a
self-adjoint, elliptic, first-order differential operator.
By virtue of this hypothesis, $A$ has a discrete spectrum
with real eigenvalues $\lambda$ and eigenfunctions
$\phi_{\lambda}$. Let $P$ denote the projection of
$C^{\infty}(Y,E)$ onto the space spanned by the $\phi_{\lambda}$
for $\lambda \geq 0$. We now form the product
$Y \times R^{+}$ of $Y$ with the half-line $u \geq 0$ and
consider the operator
$$
D \equiv {\partial \over \partial u}+A
\; \; \; \; ,
\eqno (L3.2.1)
$$
acting on sections $f(y,u)$ of $E$ lifted to $Y \times R^{+}$
(still denoted by $E$). Clearly $D$ is elliptic and its
{\it formal} adjoint is
$$
D^{*} \equiv -{\partial \over \partial u}+A
\; \; \; \; .
\eqno (L3.2.2)
$$
The following boundary condition is imposed for $D$:
$$
Pf(\cdot,0)=0
\; \; \; \; .
\eqno (L3.2.3)
$$
This is a {\it global} condition for the boundary value
$f(\cdot,0)$ in that it is equivalent to
$$
\int_{Y}\Bigr(f(y,0),\phi_{\lambda}(y)\Bigr)=0
\; \; \; \; {\rm for} \; {\rm all} \;
\lambda \geq 0
\; \; \; \; .
\eqno (L3.2.4)
$$
Of course, the adjoint boundary condition to (L3.2.3) is
$$
(1-P)f(\cdot,0)=0
\; \; \; \; .
\eqno (L3.2.5)
$$
The naturally occurring second-order self-adjoint operators
obtained out of $D$ are
$$
\triangle_{1} \equiv {\cal D}^{*}{\cal D}
\; \; \; \; ,
\eqno (L3.2.6)
$$
$$
\triangle_{2} \equiv {\cal D}{\cal D}^{*}
\; \; \; \; ,
\eqno (L3.2.7)
$$
where ${\cal D}$ is the closure of the operator $D$ on
$L^{2}$ with domain given by (L3.2.3). For $t >0$, we can then
consider the bounded operators $e^{-t \triangle_{1}}$ and
$e^{-t \triangle_{2}}$. The explicit kernels of these operators
will be given in terms of the eigenfunctions $\phi_{\lambda}$ of
$A$. For this purpose, consider first $\triangle_{1}$, i.e. the
operator given by
$$
-{\partial^{2}\over \partial u^{2}}+A^{2}
$$
with the boundary condition
$$
Pf(\cdot,0)=0
\; \; \; \; ,
\eqno (L3.2.8a)
$$
and
$$
(1-P){\left({\partial f \over \partial u}+Af \right)}_{u=0}
=0
\; \; \; \; .
\eqno (L3.2.8b)
$$
Expansion in terms of the $\phi_{\lambda}$, so that
$f(y,u)=\sum f_{\lambda}(u)\phi_{\lambda}(y)$, shows
that for each $\lambda$ one has to study the operator
$$
-{d^{2}\over du^{2}}+\lambda^{2}
$$
on $u \geq 0$ with the boundary conditions
$$
f_{\lambda}(0)=0
\; \; \; \; {\rm if} \; \; \; \;
\lambda \geq 0
\; \; \; \; ,
\eqno (L3.2.9)
$$
$$
{\left({df_{\lambda}\over du}+\lambda f_{\lambda}\right)}_{u=0}
=0 \; \; \; \; {\rm if} \; \; \; \;
\lambda <0
\; \; \; \; .
\eqno (L3.2.10)
$$
The fundamental solution for
$$
{\partial \over \partial t}-{\partial^{2}\over \partial u^{2}}
+\lambda^{2}
$$
with the boundary condition (L3.2.9) is found to be
$$
w_{A}=
{e^{-\lambda^{2}t}\over \sqrt{4\pi t}}
\left [{\rm exp}\left({-(u-v)^{2}\over 4t}\right)
-{\rm exp}\left({-(u+v)^{2}\over 4t}\right) \right]
\; \; \; \; ,
\eqno (L3.2.11)
$$
while for the boundary condition (L3.2.10), Laplace-transform
methods yield (Carslaw and Jaeger 1959)
$$
w_{B}=
{e^{-\lambda^{2}t}\over \sqrt{4\pi t}}
\left [{\rm exp}\left({-(u-v)^{2}\over 4t}\right)
+{\rm exp}\left({-(u+v)^{2}\over 4t}\right) \right]
+\lambda e^{-\lambda(u+v)}\; {\rm erfc}
\left[{(u+v)\over 2\sqrt{t}}-\lambda \sqrt{t}\right]
\; ,
\eqno (L3.2.12)
$$
where ${\rm erfc}$ is the (complementary) error function
defined by
$$
{\rm erfc}(x) \equiv {2\over \sqrt{\pi}}
\int_{x}^{\infty}e^{-\xi^{2}} \; d\xi
\; \; \; \; .
\eqno (L3.2.13)
$$
Thus, the kernel $K_{1}$ of $e^{-t\triangle_{1}}$ at a point
$(t;y,u;z,v)$ is obtained as
$$
K_{1}(t;y,u;z,v)
=\sum_{\lambda}w_{A}
\phi_{\lambda}(y)
{\overline {\phi_{\lambda}(z)}}
\; \; \; \; {\rm if} \; \lambda \geq 0
\; \; \; \; ,
\eqno (L3.2.14)
$$
$$
K_{1}(t;y,u;z,v)
=\sum_{\lambda}w_{B}
\phi_{\lambda}(y)
{\overline {\phi_{\lambda}(z)}}
\; \; \; \; {\rm if} \; \lambda <0
\; \; \; \; .
\eqno (L3.2.15)
$$

For the operator $\triangle_{2}$, the boundary conditions
for each $\lambda$ are
$$
f_{\lambda}(0)=0
\; \; \; \; {\rm if} \; \; \; \;
\lambda < 0
\; \; \; \; ,
\eqno (L3.2.16)
$$
$$
{\left(-{df_{\lambda}\over du}+\lambda f_{\lambda}\right)}_{u=0}
=0 \; \; \; \; {\rm if} \; \; \; \;
\lambda \geq 0
\; \; \; \; .
\eqno (L3.2.17)
$$
The fundamental solution for
$$
{\partial \over \partial t}-{\partial^{2}\over \partial u^{2}}
+\lambda^{2}
$$
subject to (L3.2.16) is ${\widetilde w}_{A}=w_{A}$, while for
the boundary conditions (L3.2.17) one finds
${\widetilde w}_{B}=w_{B}(-\lambda)$. Moreover, by virtue
of the inequality
$$
\int_{x}^{\infty}e^{-\xi^{2}} \; d\xi < e^{-x^{2}}
\; \; \; \; ,
\eqno (L3.2.18)
$$
$w_{A}$ and $w_{B}$ are both bounded by
$$
F_{\lambda}(t;u,v) \equiv
\left[{e^{-\lambda^{2}t} \over \sqrt{\pi t}}
+{2 \mid \lambda \mid \over \sqrt{\pi}}e^{-\lambda^{2}t}
\right] \;
{\rm exp}\left({-(u-v)^{2}\over 4t}\right)
\; \; \; \; .
\eqno (L3.2.19)
$$
If one now multiplies and divides by $\sqrt{t}$ the second
term in square brackets in (L3.2.19), application of the
inequality
$$
x \leq e^{x^{2}/2}
\; \; \; \; ,
\eqno (L3.2.20)
$$
to the resulting term $\mid \lambda \mid \sqrt{t}$ shows that
the kernel $K_{1}(t;y,u;z,v)$ of $e^{-t \triangle_{1}}$ is
bounded by
$$
G(t;y,u;z,v) \equiv
{3\over 2 \sqrt{\pi t}}
\left[\sum_{\lambda}e^{-\lambda^{2}t/2}
\left({\mid \phi_{\lambda}(y) \mid}^{2}
+{\mid \phi_{\lambda}(z) \mid}^{2}\right)\right]
{\rm exp}\left({-(u-v)^{2}\over 4t}\right)
\; \; \; \; .
\eqno (L3.2.21)
$$
Moreover, since the kernel of $e^{-t \triangle_{2}}$ on the
diagonal of $Y \times Y$ is bounded by
$C t^{-n/2}$, one finds that the kernels of $e^{-t \triangle_{1}}$
and $e^{-t \triangle_{2}}$ are exponentially small in $t$ as
$t \rightarrow 0$ for $u \not = v$, in that they are bounded by
$$
C \; t^{-(n+1)/2} \;
{\rm exp}\left({-(u-v)^{2}\over 4t}\right)
\; \; \; \; ,
$$
where $C$ is some constant, as $t \rightarrow 0$.

Thus, the contribution outside the diagonal is asymptotically
negligible, so that we are mainly interested in the contribution
from the diagonal. For this purpose, we study $K(t,y,u)$,
i.e. the kernel of $e^{-t \triangle_{1}}-e^{-t \triangle_{2}}$ at
the point $(y,u;y,u)$ of $(Y \times R^{+}) \times
(Y \times R^{+})$. Defining ${\rm sign}(\lambda)
\equiv +1 \; \forall
\lambda \geq 0$, ${\rm sign}(\lambda)\equiv -1
\; \forall \lambda <0$,
one thus finds
$$ \eqalignno{
K(t,y,u) & \equiv \Bigr[w_{A}(t;y,u;y,u)-{\widetilde w}_{B}
(t;y,u;y,u)\Bigr]
+\Bigr[w_{B}(t;y,u;y,u)-{\widetilde w}_{A}(t;y,u;y,u)\Bigr]\cr
&=\sum_{\lambda \geq 0}{\mid \phi_{\lambda}(y)\mid}^{2}
\left[{e^{-\lambda^{2}t} \over \sqrt{4\pi t}}
\left(1-e^{-u^{2}/t}\right)
-{e^{-\lambda^{2}t} \over \sqrt{4\pi t}}
\left(1+e^{-u^{2}/t}\right) \right. \cr
&\left. -(-\lambda)e^{2\lambda u}
{\rm erfc} \; \left({u\over \sqrt{t}}+\lambda \sqrt{t}\right)
\right]\cr
&+\sum_{\lambda<0}{\mid \phi_{\lambda}(y)\mid}^{2}
\left[{e^{-\lambda^{2}t}\over \sqrt{4\pi t}}
\left(1+e^{-u^{2}/t}\right)
+\lambda e^{-2\lambda u}{\rm erfc}
\left({u\over \sqrt{t}}-\lambda \sqrt{t}\right) \right. \cr
&\left. -{e^{-\lambda^{2}t}\over \sqrt{4 \pi t}}
\left(1-e^{-u^{2}/t}\right)\right]\cr
&=\sum_{\lambda}
{\mid \phi_{\lambda}(y)\mid}^{2}
{\rm sign}(\lambda)
\left[-{e^{-\lambda^{2}t}e^{-u^{2}/t}\over \sqrt{\pi t}}
+\mid \lambda \mid e^{2\mid \lambda \mid u}
{\rm erfc} \left({u\over \sqrt{t}}+\mid \lambda \mid
\sqrt{t}\right)\right]\cr
&=\sum_{\lambda}
{\mid \phi_{\lambda}(y)\mid}^{2}
{\rm sign}(\lambda){\partial \over \partial u}
\left[{1\over 2}e^{2 \mid \lambda \mid u}
{\rm erfc} \left({u\over \sqrt{t}}+\mid \lambda \mid
\sqrt{t}\right)\right]
\; \; \; \; .
&(L3.2.22)\cr}
$$
Thus, integration on $Y \times R^{+}$ and elemantary rules
for taking limits yield
$$
K(t) \equiv \int_{0}^{\infty}\int_{Y}K(t,y,u)dy \; du
=-{1\over 2}\sum_{\lambda}{\rm sign}(\lambda) \;
{\rm erfc}(\mid \lambda \mid \sqrt{t})
\; \; \; \; ,
\eqno (L3.2.23)
$$
which implies (on differentiating the error function
and using the signature of eigenvalues)
$$
K'(t)={1\over \sqrt{4\pi t}}\sum_{\lambda}\lambda
e^{-\lambda^{2}t}
\; \; \; \; .
\eqno (L3.2.24)
$$
It is now necessary to derive the limiting behaviour of the
(integrated) kernel (L3.2.23) as $t \rightarrow \infty$
and as $t \rightarrow 0$. Indeed, denoting by $h$ the
degeneracy of the zero-eigenvalue, one finds
$$
\lim_{t \to \infty}K(t)=-{h\over 2}
\; \; \; \; ,
\eqno (L3.2.25)
$$
whereas, as $t \rightarrow 0$, the following bound holds:
$$
\mid K(t) \mid \leq {1\over 2}\sum_{\lambda}
{\rm erfc}(\mid \lambda \mid \sqrt{t})
\leq {1\over \sqrt{\pi}}\sum_{\lambda}e^{-\lambda^{2}t}
< C \; t^{-n/2}
\; \; \; \; ,
\eqno (L3.2.26)
$$
where $C$ is a constant. Moreover, the result (L3.2.25) may be
supplemented by saying that $K(t)+{h\over 2}$ tends to $0$
exponentially as $t \rightarrow \infty$. Thus, combining
(L3.2.25)-(L3.2.26) and this property, one finds that, for
$Re(s)$ sufficiently large, the integral
$$
I(s) \equiv \int_{0}^{\infty}\Bigr(K(t)+{h\over 2}\Bigr)
t^{s-1} \; dt
\; \; \; \; ,
\eqno (L3.2.27)
$$
converges. Integration by parts, definition of the $\Gamma$
function
$$
\Gamma(z) \equiv \int_{0}^{\infty}t^{z-1}e^{-t} \; dt
=k^{z}\int_{0}^{\infty}t^{z-1}e^{-kt} \; dt
\; \; \; \; ,
\eqno (L3.2.28)
$$
and careful consideration of positive and negative eigenvalues
with their signatures then lead to
$$
I(s)
=-{\Gamma \left(s+{1\over 2}\right) \over 2s\sqrt{\pi}}
\sum_{\lambda}{{\rm sign}(\lambda)\over {{\mid \lambda \mid}^{2s}}}
=-{\Gamma \left(s+{1\over 2}\right) \over 2s\sqrt{\pi}}
\eta(2s)
\; \; \; \; .
\eqno (L3.2.29)
$$
The following analysis relies entirely on the {\it assumption}
than an asymptotic expansion of the integrated kernel $K(t)$
exists as $t \rightarrow 0$ [1-2].
By writing such an expansion in the form
$$
K(t) \sim \sum_{k \geq -n} a_{k}t^{k/2}
\; \; \; \; ,
\eqno (L3.2.30)
$$
equations (L3.2.27)-(L3.2.30) yield (on splitting the integral
(L3.2.27) into an integral from $0$ to $1$ plus an integral
from $1$ to $\infty$)
$$
\eta(2s) \sim -{2s \sqrt{\pi} \over
\Gamma \left(s+{1\over 2}\right)}
\left[{h\over 2s}+
\sum_{k=-n}^{N}{a_{k}\over \left({k\over 2}+s\right)}
+\theta_{N}(s)\right]
\; \; \; \; .
\eqno (L3.2.31)
$$
This is the analytic continuation of $\eta(2s)$ to the
whole $s$-plane. Hence one finds
$$
\eta(0)=-\Bigr(2a_{0}+h\Bigr)
\; \; \; \; .
\eqno (L3.2.32)
$$
Remarkably, regularity at the origin of the $\eta$- and
$\zeta$-functions results from the existence of the
asymptotic expansion of the integrated kernel as
$t \rightarrow 0^{+}$ [1-2,4].
\vskip 100cm
\centerline {\bf Further References}
\vskip 1cm
\item {[1]}
Carslaw H. S. and Jaeger J. C. (1959) {\it Conduction of
Heat in Solids} (Oxford: Clarendon Press).
\vskip 100cm
\def\cstok#1{\leavevmode\thinspace\hbox{\vrule\vtop{\vbox{\hrule\kern1pt
\hbox{\vphantom{\tt/}\thinspace{\tt#1}\thinspace}}
\kern1pt\hrule}\vrule}\thinspace}
\centerline {\bf LECTURE 4.}
\vskip 1cm
\centerline {\bf L4.1 Two-Component Spinor Calculus}
\vskip 0.3cm
Two-component spinor calculus is a powerful tool for
studying classical field theories in four-dimensional
space-time models. Within this framework,
the basic object is spin-space,
a two-dimensional complex vector space $S$ with a
symplectic form $\epsilon$, i.e. an antisymmetric
complex bilinear form. Unprimed spinor indices
$A,B,...$ take the values $0,1$ whereas primed spinor
indices $A',B',...$ take the values $0',1'$, since there
are actually two such spaces: unprimed spin-space
$(S,\epsilon)$ and primed spin-space $(S',\epsilon')$.
The whole two-spinor calculus in {\it Lorentzian}
four-manifolds relies on three fundamental isomorphisms:

(i) The isomorphism between $\Bigr(S,\epsilon_{AB}\Bigr)$ and
its dual $\Bigr(S^{*},\epsilon^{AB}\Bigr)$. This is provided
by the symplectic form $\epsilon$, which raises and
lowers indices according to the rules
$$
\epsilon^{AB} \; \varphi_{B}=\varphi^{A} \; \in \; S
\; \; \; \; ,
\eqno (L4.1.1)
$$
$$
\varphi^{B} \; \epsilon_{BA}=\varphi_{A} \; \in \; S^{*}
\; \; \; \; .
\eqno (L4.1.2)
$$
Thus, since
$$
\epsilon_{AB}=\epsilon^{AB}=\pmatrix {0&1\cr -1&0 \cr}
\; \; \; \; ,
\eqno (L4.1.3)
$$
one finds in components $\varphi^{0}=\varphi_{1},
\varphi^{1}=-\varphi_{0}$.

Similarly, one has the
isomorphism $\Bigr(S',\epsilon_{A'B'}\Bigr)
\cong \Bigr((S')^{*},\epsilon^{A'B'}\Bigr)$, which implies
$$
\epsilon^{A'B'} \; \varphi_{B'}=\varphi^{A'} \; \in \; S'
\; \; \; \; ,
\eqno (L4.1.4)
$$
$$
\varphi^{B'} \; \epsilon_{B'A'}=\varphi_{A'} \; \in
\; (S')^{*}
\; \; \; \; ,
\eqno (L4.1.5)
$$
where
$$
\epsilon_{A'B'}=\epsilon^{A'B'}=\pmatrix
{0'&1'\cr -1'&0'\cr}
\; \; \; \; .
\eqno (L4.1.6)
$$

(ii) The (anti)-isomorphism between $\Bigr(S,\epsilon_{AB}\Bigr)$
and $\Bigr(S',\epsilon_{A'B'}\Bigr)$, called complex conjugation,
and denoted by an overbar. According to a standard convention,
one has
$$
{\overline {\psi^{A}}} \equiv {\overline \psi}^{A'}
\; \in \; S'
\; \; \; \; ,
\eqno (L4.1.7)
$$
$$
{\overline {\psi^{A'}}} \equiv {\overline \psi}^{A}
\; \in \; S
\; \; \; \; .
\eqno (L4.1.8)
$$
In components, if $w^{A}$ is thought as
$w^{A}=\pmatrix {\alpha \cr \beta \cr}$, the action of (L4.1.7)
leads to
$$
{\overline {w^{A}}} \equiv {\overline w}^{A'}
\equiv \pmatrix {{\overline \alpha} \cr {\overline \beta}\cr}
\; \; \; \; ,
\eqno (L4.1.9)
$$
whereas, if $z^{A'}=\pmatrix {\gamma \cr \delta \cr}$, then
(L4.1.8) leads to
$$
{\overline {z^{A'}}} \equiv {\overline z}^{A}
=\pmatrix {{\overline \gamma}\cr {\overline \delta}\cr}
\; \; \; \; .
\eqno (L4.1.10)
$$
With our notation, $\overline \alpha$ denotes complex
conjugation of the function $\alpha$, and so on. Note that
the symplectic structure is preserved by complex conjugation,
since ${\overline \epsilon}_{A'B'}=\epsilon_{A'B'}$.

(iii) The isomorphism between the tangent space $T$ at a
point of space-time and the tensor product of the
unprimed spin-space $\Bigr(S,\epsilon_{AB}\Bigr)$ and the
primed spin-space $\Bigr(S',\epsilon_{A'B'}\Bigr)$:
$$
T \cong \Bigr(S,\epsilon_{AB}\Bigr) \otimes
\Bigr(S',\epsilon_{A'B'}\Bigr)
\; \; \; \; .
\eqno (L4.1.11)
$$
The Infeld-van der
Waerden symbols $\sigma_{\; \; AA'}^{a}$ and
$\sigma_{a}^{\; \; AA'}$ express this isomorphism, and the
correspondence between a vector $v^{a}$ and a spinor
$v^{AA'}$ is given by
$$
v^{AA'} \equiv v^{a} \; \sigma_{a}^{\; \; AA'}
\; \; \; \; ,
\eqno (L4.1.12)
$$
$$
v^{a} \equiv v^{AA'} \; \sigma_{\; \; AA'}^{a}
\; \; \; \; .
\eqno (L4.1.13)
$$
These mixed spinor-tensor symbols obey the identities
$$
{\overline \sigma}_{a}^{\; \; AA'}=\sigma_{a}^{\; \; AA'}
\; \; \; \; ,
\eqno (L4.1.14)
$$
$$
\sigma_{a}^{\; \; AA'} \; \sigma_{\; \; AA'}^{b}
=\delta_{a}^{\; \; b}
\; \; \; \; ,
\eqno (L4.1.15)
$$
$$
\sigma_{a}^{\; \; AA'} \; \sigma_{\; \; BB'}^{a}
=\epsilon_{B}^{\; \; A} \; \epsilon_{B'}^{\; \; \; A'}
\; \; \; \; ,
\eqno (L4.1.16)
$$
$$
\sigma_{[a}^{\; \; AA'} \; \sigma_{b]A}^{\; \; \; \; \; B'}
=-{i\over 2} \; \epsilon_{abcd} \; \sigma^{cAA'}
\; \sigma_{\; \; A}^{d \; \; B'}
\; \; \; \; .
\eqno (L4.1.17)
$$
Similarly, a one-form $\omega_{a}$ has a spinor equivalent
$$
\omega_{AA'} \equiv \omega_{a} \; \sigma_{\; \; AA'}^{a}
\; \; \; \; ,
\eqno (L4.1.18)
$$
whereas the spinor equivalent of the metric is
$$
\eta_{ab} \; \sigma_{\; \; AA'}^{a}
\; \sigma_{\; \; BB'}^{b} \equiv
\epsilon_{AB} \; \epsilon_{A'B'}
\; \; \; \; .
\eqno (L4.1.19)
$$
In particular, in Minkowski space-time Eqs. (L4.1.12)
and (L4.1.17) enable one to write down a coordinate system
in $2 \times 2$ matrix form
$$
x^{AA'}={1\over \sqrt{2}}
\pmatrix {{x^{0}+x^{3}}&{x^{1}-ix^{2}}\cr
{x^{1}+ix^{2}}&{x^{0}-x^{3}}\cr}
\; \; \; \; .
\eqno (L4.1.20)
$$

In the Lorentzian-signature case, the Maxwell curvature
two-form $F \equiv F_{ab}dx^{a} \wedge dx^{b}$ can be
written spinorially as
$$
F_{AA'BB'}={1\over 2}\Bigr(F_{AA'BB'}-F_{BB'AA'}\Bigr)
=\varphi_{AB} \; \epsilon_{A'B'}
+\varphi_{A'B'} \; \epsilon_{AB}
\; \; \; \; ,
\eqno (L4.1.21)
$$
where
$$
\varphi_{AB} \equiv {1\over 2}
F_{AC'B}^{\; \; \; \; \; \; \; \; \; C'}
=\varphi_{(AB)}
\; \; \; \; ,
\eqno (L4.1.22)
$$
$$
\varphi_{A'B'} \equiv {1\over 2}
F_{CB' \; \; A'}^{\; \; \; \; \; \; C}
=\varphi_{(A'B')}
\; \; \; \; .
\eqno (L4.1.23)
$$
These formulae are obtained by applying the identity
$$
T_{AB}-T_{BA}=\epsilon_{AB} \; T_{C}^{\; \; C}
\eqno (L4.1.24)
$$
to express ${1\over 2}\Bigr(F_{AA'BB'}-F_{AB'BA'}\Bigr)$
and ${1\over 2}\Bigr(F_{AB'BA'}-F_{BB'AA'}\Bigr)$.
Note also that round brackets $(AB)$ denote (as usual)
symmetrization over the spinor indices $A$ and $B$, and that
the antisymmetric part of $\varphi_{AB}$ vanishes by virtue
of the antisymmetry of $F_{ab}$, since
$\varphi_{[AB]}={1\over 4}\epsilon_{AB} \;
F_{CC'}^{\; \; \; \; \; CC'}={1\over 2}\epsilon_{AB}
\; \eta^{cd} \; F_{cd}=0$. Last but not least, in the
Lorentzian case
$$
{\overline {\varphi_{AB}}} \equiv {\overline \varphi}_{A'B'}
=\varphi_{A'B'}
\; \; \; \; .
\eqno (L4.1.25)
$$
The symmetric spinor fields $\varphi_{AB}$ and
$\varphi_{A'B'}$ are the anti-self-dual and self-dual parts
of the curvature two-form, respectively.

Similarly, the Weyl curvature $C_{\; \; bcd}^{a}$, i.e. the
part of the Riemann curvature tensor invariant under conformal
rescalings of the metric, may be expressed spinorially,
omitting soldering forms (see below)
for simplicity of notation, as
$$
C_{abcd}=\psi_{ABCD} \; \epsilon_{A'B'} \;
\epsilon_{C'D'}
+{\overline \psi}_{A'B'C'D'} \;
\epsilon_{AB} \; \epsilon_{CD}
\; \; \; \; .
\eqno (L4.1.26)
$$

In canonical gravity two-component spinors
lead to a considerable simplification of calculations. Denoting
by $n^{\mu}$ the future-pointing unit timelike normal to a
spacelike three-surface, its spinor version obeys the relations
$$
n_{AA'} \; e_{\; \; \; \; \; i}^{AA'}=0
\; \; \; \; ,
\eqno (L4.1.27)
$$
$$
n_{AA'} \; n^{AA'}=1
\; \; \; \; ,
\eqno (L4.1.28)
$$
where $e_{\; \; \; \; \; \mu}^{AA'} \equiv e_{\; \; \mu}^{a}
\; \sigma_{a}^{\; \; AA'}$ is the two-spinor version of the tetrad,
often referred to in the literature as soldering form.
Denoting by $h$ the induced Riemannian
metric on the three-surface, other
useful relations are
$$
h_{ij}=-e_{AA'i} \; e_{\; \; \; \; \; j}^{AA'}
\; \; \; \; ,
\eqno (L4.1.29)
$$
$$
e_{\; \; \; \; \; 0}^{AA'}=N \; n^{AA'}
+N^{i} \; e_{\; \; \; \; \; i}^{AA'}
\; \; \; \; ,
\eqno (L4.1.30)
$$
$$
n_{AA'} \; n^{BA'}={1\over 2} \epsilon_{A}^{\; \; B}
\; \; \; \; ,
\eqno (L4.1.31)
$$
$$
n_{AA'} \; n^{AB'}={1\over 2} \epsilon_{A'}^{\; \; \; B'}
\; \; \; \; ,
\eqno (L4.1.32)
$$
$$
n_{[EB'} \; n_{A]A'}={1\over 4}\epsilon_{EA} \;
\epsilon_{B'A'}
\; \; \; \; ,
\eqno (L4.1.33)
$$
$$
e_{AA'j} \; e_{\; \; \; \; \; k}^{AB'}
=-{1\over 2}h_{jk} \; \epsilon_{A'}^{\; \; \; B'}
-i \epsilon_{jkl}\sqrt{{\rm det} \; h} \;
n_{AA'} \; e^{AB'l}
\; \; \; \; .
\eqno (L4.1.34)
$$
In Eq. (L4.1.30), $N$ and $N^{i}$ are the lapse and shift
functions respectively [4].

The space-time curvature is obtained after defining the
spinor covariant derivative $\nabla_{a}=\nabla_{AA'}$.
If $\theta,\phi,\psi$ are spinor fields, $\nabla_{AA'}$
is a map such that
\vskip 0.3cm
\noindent
(1) $\nabla_{AA'}(\theta+\phi)=\nabla_{AA'}\theta
+\nabla_{AA'}\phi$ (i.e. linearity).
\vskip 0.3cm
\noindent
(2) $\nabla_{AA'}(\theta \psi)=\Bigr(\nabla_{AA'}\theta\Bigr)\psi
+\theta \Bigr(\nabla_{AA'}\psi\Bigr)$ (i.e. Leibniz rule).
\vskip 0.3cm
\noindent
(3) $\psi=\nabla_{AA'}\theta$ implies
${\overline \psi}=\nabla_{AA'}{\overline \theta}$
(i.e. reality condition).
\vskip 0.3cm
\noindent
(4) $\nabla_{AA'}\epsilon_{BC}=\nabla_{AA'}\epsilon^{BC}=0$,
i.e. the symplectic form may be used to raise or lower indices
within spinor expressions acted upon by $\nabla_{AA'}$, in
addition to the usual metricity condition
$\nabla g=0$, which involves instead the product of two
$\epsilon$-symbols.
\vskip 0.3cm
\noindent
(5) $\nabla_{AA'}$ commutes with any index substitution
not involving $A,A'$.
\vskip 0.3cm
\noindent
(6) For any function $f$, one finds
$\Bigr(\nabla_{a}\nabla_{b}-\nabla_{b}\nabla_{a}\Bigr)f
=2S_{ab}^{\; \; \; c} \; \nabla_{c}f$, where
$S_{ab}^{\; \; \; c}$ is the torsion tensor.
\vskip 0.3cm
\noindent
(7) For any derivation $D$ acting on spinor fields, a spinor
field $\xi^{AA'}$ exists such that $D \psi=\xi^{AA'}
\; \nabla_{AA'} \psi, \forall \psi$.
\vskip 0.3cm
\noindent
Such a spinor covariant
derivative exists and is unique.

If Lorentzian space-time is replaced by a complex or
real Riemannian four-manifold, an important modification
should be made, since the (anti)-isomorphism between
unprimed and primed spin-space no longer exists. This
means that primed spinors can no longer be regarded as
complex conjugates of unprimed spinors, or viceversa,
as in (L4.1.7)-(L4.1.8). In particular, Eqs. (L4.1.21)
and (L4.1.26) should be re-written as
$$
F_{AA'BB'}=\varphi_{AB} \; \epsilon_{A'B'}
+{\widetilde \varphi}_{A'B'} \; \epsilon_{AB}
\; \; \; \; ,
\eqno (L4.1.35)
$$
$$
C_{abcd}=\psi_{ABCD} \; \epsilon_{A'B'}
\; \epsilon_{C'D'}
+{\widetilde \psi}_{A'B'C'D'} \;
\epsilon_{AB} \; \epsilon_{CD}
\; \; \; \; .
\eqno (L4.1.36)
$$
With our notation, $\varphi_{AB},{\widetilde \varphi}_{A'B'}$,
as well as $\psi_{ABCD},{\widetilde \psi}_{A'B'C'D'}$
are {\it completely independent} symmetric spinor fields,
not related by any conjugation.

Indeed, a conjugation can still be defined in the real
Riemannian case, but it no longer relates $\Bigr(S,\epsilon_{AB})$
and $\Bigr(S',\epsilon_{A'B'}\Bigr)$. It is instead an
anti-involutory operation which maps elements of a spin-space
(either unprimed or primed) to elements of the {\it same}
spin-space. By anti-involutory we mean that, when applied twice
to a spinor with an odd number of indices,
it yields the same spinor with the opposite
sign, i.e. its square is minus the identity, whereas the square
of complex conjugation as defined in (L4.1.9)-(L4.1.10) equals
the identity.
Euclidean conjugation, denoted by a {\it dagger}, is defined as follows.
$$
{\Bigr(w^{A}\Bigr)}^{\dagger} \equiv
\pmatrix {{\overline \beta}\cr -{\overline \alpha}\cr}
\; \; \; \; ,
\eqno (L4.1.37)
$$
$$
{\Bigr(z^{A'}\Bigr)}^{\dagger} \equiv
\pmatrix {-{\overline \delta}\cr {\overline \gamma}\cr}
\; \; \; \; .
\eqno (L4.1.38)
$$
This means that, in flat Euclidean four-space, a unit
$2 \times 2$ matrix $\delta_{BA'}$ exists such that
$$
{\Bigr(w^{A}\Bigr)}^{\dagger} \equiv
\epsilon^{AB} \; \delta_{BA'} \;
{\overline w}^{A'}
\; \; \; \; .
\eqno (L4.1.39)
$$
We are here using the freedom to regard $w^{A}$ either as an
$SL(2,C)$ spinor for which complex conjugation can be defined,
or as an $SU(2)$ spinor for which Euclidean conjugation is
instead available. The soldering forms for $SU(2)$ spinors
only involve spinor indices of the same spin-space, i.e.
${\widetilde e}_{i}^{\; \; AB}$ and
${\widetilde e}_{i}^{\; \; A'B'}$.
More precisely, denoting by $E_{a}^{i}$
a real {\it triad}, where $i=1,2,3$, and by
$\tau_{\; \; A}^{a \; \; \; B}$ the three Pauli matrices
obeying the identity
$$
\tau_{\; \; A}^{a \; \; \; B} \;
\tau_{\; \; B}^{b \; \; \; D}
=i \; \epsilon^{abc} \; \tau_{cA}^{\; \; \; \; D}
+\delta^{ab} \; \delta_{A}^{\; \; D}
\; \; \; \; ,
\eqno (L4.1.40)
$$
the $SU(2)$ soldering forms are defined by
$$
{\widetilde e}_{\; \; A}^{j \; \; \; B} \equiv
-{i \over \sqrt{2}} \; E_{a}^{j} \; \tau_{\; \; A}^{a \; \; \; B}
\; \; \; \; .
\eqno (L4.1.41)
$$
The soldering form in (L4.1.41)
provides an isomorphism between three-real-dimensional tangent
space at each point of $\Sigma$, and the three-real-dimensional
vector space of $2 \times 2$ trace-free Hermitian matrices.
The Riemannian three-metric on $\Sigma$ is then given by
$$
h^{ij}=-{\widetilde e}_{\; \; A}^{i \; \; \; B} \;
{\widetilde e}_{\; \; B}^{j \; \; \; A}
\; \; \; \; .
\eqno (L4.1.42)
$$
\vskip 0.3cm
\centerline {\bf L4.2 Dirac and Weyl Equations in
Two-Component Spinor Form}
\vskip 0.3cm
Dirac's theory of massive and massless spin-${1\over 2}$
particles is still a key element of modern particle physics
and field theory. From the point of view of theoretical
physics, the description of such particles motivates indeed
the whole theory of Dirac operators. We are here concerned
with a two-component spinor analysis of the corresponding
spin-${1\over 2}$ fields in Riemannian four-geometries
$(M,g)$ with boundary. A massive spin-${1\over 2}$ Dirac
field is then described by the four independent spinor
fields $\phi^{A},\chi^{A},{\widetilde \phi}^{A'},
{\widetilde \chi}^{A'}$, and the action functional takes
the form
$$
I \equiv I_{V}+I_{B}
\; \; \; \; ,
\eqno (L4.2.1)
$$
where
$$ \eqalignno{
I_{V} &\equiv
{i\over 2}\int_{M}\Bigr[{\widetilde \phi}^{A'}
\Bigr(\nabla_{AA'}\phi^{A}\Bigr)-\Bigr(\nabla_{AA'}
{\widetilde \phi}^{A'}\Bigr)\phi^{A}\Bigr]
\sqrt{{\rm det} \; g} \; d^{4}x \cr
&+{i\over 2}\int_{M}\Bigr[{\widetilde \chi}^{A'}
\Bigr(\nabla_{AA'}\chi^{A}\Bigr)-\Bigr(\nabla_{AA'}
{\widetilde \chi}^{A'}\Bigr)\chi^{A}\Bigr]
\sqrt{{\rm det} \; g} \; d^{4}x \cr
&+{m\over \sqrt{2}}\int_{M}\Bigr[\chi_{A}\phi^{A}
+{\widetilde \phi}^{A'}{\widetilde \chi}_{A'}\Bigr]
\sqrt{{\rm det} \; g} \; d^{4}x
\; \; \; \; ,
&(L4.2.2)\cr}
$$
and $I_{B}$ is a suitable boundary term, necessary to
obtain a well-posed variational problem. Its form is
determined once one knows which spinor fields are
fixed on the boundary (e.g. section L4.3). With our
notation, the occurrence of $i$ depends on conventions
for Infeld-van der Waerden symbols (see section L4.3).
One thus finds the field equations
$$
\nabla_{AA'}\phi^{A}={im\over \sqrt{2}} \;
{\widetilde \chi}_{A'}
\; \; \; \; ,
\eqno (L4.2.3)
$$
$$
\nabla_{AA'}\chi^{A}={im \over \sqrt{2}} \;
{\widetilde \phi}_{A'}
\; \; \; \; ,
\eqno (L4.2.4)
$$
$$
\nabla_{AA'}{\widetilde \phi}^{A'}=
-{im\over \sqrt{2}} \; \chi_{A}
\; \; \; \; ,
\eqno (L4.2.5)
$$
$$
\nabla_{AA'}{\widetilde \chi}^{A'}=
-{im\over \sqrt{2}} \; \phi_{A}
\; \; \; \; .
\eqno (L4.2.6)
$$
Note that this is a coupled system of first-order
differential equations, obtained after applying
differentiation rules for anti-commuting spinor
fields. This means the spinor field acted upon by
the $\nabla_{AA'}$ operator should be always brought
to the left, hence leading to a minus sign if such a
field was not already on the left. Integration by
parts and careful use of boundary terms are also
necessary. The equations (L4.2.3)-(L4.2.6) reproduce
the familiar form of the Dirac equation expressed in
terms of $\gamma$-matrices. In particular, for massless
fermionic fields one obtains the independent Weyl
equations
$$
\nabla^{AA'}\phi_{A}=0
\; \; \; \; ,
\eqno (L4.2.7)
$$
$$
\nabla^{AA'}{\widetilde \phi}_{A'}=0
\; \; \; \; ,
\eqno (L4.2.8)
$$
not related by any conjugation.

Explicit mode-by-mode solutions of these equations with
spectral boundary conditions on a three-sphere are studied
in the section {\bf Problems for Students} at the end of
these Lecture Notes, as well as the tensor equivalent
of Eqs. (L4.2.3)-(L4.2.8).
\vskip 0.3cm
\centerline {\bf L4.3 New Variational Problems for Massless
Fermionic Fields}
\vskip 0.3cm
Locally supersymmetric boundary conditions have been recently
studied in quantum gravity to understand its one-loop
properties in the presence of boundaries.
They involve the normal to the boundary and the field
for spin ${1\over 2}$, the normal to the boundary and the
spin-${3\over 2}$ potential for gravitinos, Dirichlet conditions
for real scalar fields, magnetic or electric field for
electromagnetism, mixed boundary conditions for the four-metric
of the gravitational field (and in particular Dirichlet conditions
on the perturbed three-metric). The aim of this section is to describe
the corresponding classical properties in the case of massless
spin-${1\over 2}$ fields.

For this purpose, we consider flat Euclidean four-space bounded
by a three-sphere of radius $a$. The spin-${1\over 2}$ field,
represented by a pair of independent spinor fields $\psi^{A}$
and ${\widetilde \psi}^{A'}$, is expanded on a family of
three-spheres centred on the origin as
$$
\psi^{A}={\tau^{-{3\over 2}}\over 2\pi}
\sum_{n=0}^{\infty}\sum_{p=1}^{(n+1)(n+2)}
\sum_{q=1}^{(n+1)(n+2)} \alpha_{n}^{pq}
\Bigr[m_{np}(\tau)\rho^{nqA}+{\widetilde r}_{np}(\tau)
{\overline \sigma}^{nqA}\Bigr]
\; \; \; \; ,
\eqno (L4.3.1)
$$
$$
{\widetilde \psi}^{A'}={\tau^{-{3\over 2}}\over 2\pi}
\sum_{n=0}^{\infty}\sum_{p=1}^{(n+1)(n+2)}
\sum_{q=1}^{(n+1)(n+2)} \alpha_{n}^{pq}
\Bigr[{\widetilde m}_{np}(\tau){\overline \rho}^{nqA'}
+r_{np}(\tau)\sigma^{nqA'}\Bigr]
\; \; \; \; .
\eqno (L4.3.2)
$$
With our notation, $\tau$ is the Euclidean-time coordinate,
the $\alpha_{n}^{pq}$ are block-diagonal matrices with
blocks $\pmatrix {1&1\cr 1&-1\cr}$, the $\rho-$ and
$\sigma$-harmonics obey the identities described in [4].
Last but not least, the modes $m_{np}$ and $r_{np}$ are
regular at $\tau=0$, whereas the modes ${\widetilde m}_{np}$
and ${\widetilde r}_{np}$ are singular at $\tau=0$ if the
spin-${1\over 2}$ field is massless. Bearing in mind that the
harmonics $\rho^{nqA}$ and $\sigma^{nqA'}$ have positive
eigenvalues ${1\over 2}\Bigr(n+{3\over 2}\Bigr)$ for the
three-dimensional Dirac operator on the bounding $S^3$ [4], the
decomposition (L4.3.1)-(L4.3.2) can be re-expressed as
$$
\psi^{A}=\psi_{(+)}^{A}+\psi_{(-)}^{A}
\; \; \; \; ,
\eqno (L4.3.3)
$$
$$
{\widetilde \psi}^{A'}={\widetilde \psi}_{(+)}^{A'}
+{\widetilde \psi}_{(-)}^{A'}
\; \; \; \; .
\eqno (L4.3.4)
$$
In (L4.3.3)-(L4.3.4),
the $(+)$ parts correspond to the modes $m_{np}$
and $r_{np}$, whereas the $(-)$ parts correspond to the singular
modes ${\widetilde m}_{np}$ and ${\widetilde r}_{np}$,
which multiply harmonics having negative eigenvalues
$-{1\over 2}\Bigr(n+{3\over 2}\Bigr)$ for the three-dimensional
Dirac operator on $S^3$. If one wants to find a classical
solution of the Weyl equation which is regular
$\forall \tau \in [0,a]$, one is thus forced to set to zero
the modes ${\widetilde m}_{np}$ and ${\widetilde r}_{np}$
$\forall \tau \in [0,a]$ [4]. This is why, if one requires the
local boundary conditions [4]
$$
\sqrt{2} \; {_{e}n_{A}^{\; \; A'}} \; \psi^{A}
\mp {\widetilde \psi}^{A'}=\Phi^{A'}
\; {\rm on} \; S^{3}
\; \; \; \; ,
\eqno (L4.3.5)
$$
such a condition can be expressed as [4]
$$
\sqrt{2} \; {_{e}n_{A}^{\; \; A'}} \; \psi_{(+)}^{A}
=\Phi_{1}^{A'}
\; {\rm on} \; S^{3}
\; \; \; \; ,
\eqno (L4.3.6)
$$
$$
\mp {\widetilde \psi}_{(+)}^{A'}=\Phi_{2}^{A'}
\; {\rm on} \; S^{3}
\; \; \; \; ,
\eqno (L4.3.7)
$$
where $\Phi_{1}^{A'}$ and $\Phi_{2}^{A'}$ are the parts of the
spinor field $\Phi^{A'}$ related to the ${\overline \rho}$-
and $\sigma$-harmonics respectively. In particular, if
$\Phi_{1}^{A'}=\Phi_{2}^{A'}=0$
on $S^3$ as in [4], one finds
$$
\sum_{n=0}^{\infty}\sum_{p=1}^{(n+1)(n+2)}
\sum_{q=1}^{(n+1)(n+2)}\alpha_{n}^{pq}
\; m_{np}(a) \; {_{e}n_{A}^{\; \; A'}}
\; \rho_{nq}^{A}=0
\; \; \; \; ,
\eqno (L4.3.8)
$$
$$
\sum_{n=0}^{\infty}\sum_{p=1}^{(n+1)(n+2)}
\sum_{q=1}^{(n+1)(n+2)}\alpha_{n}^{pq}
\; r_{np}(a)
\; \sigma_{nq}^{A'}=0
\; \; \; \; ,
\eqno (L4.3.9)
$$
where $a$ is the three-sphere radius. Since the harmonics
appearing in (L4.3.8)-(L4.3.9)
are linearly independent, these
relations lead to $m_{np}(a)=r_{np}(a)=0$ $\forall n,p$.
Remarkably, this simple calculation shows that the classical
boundary-value problems for regular solutions of the Weyl
equation subject to local or spectral conditions on $S^3$
share the same property provided $\Phi^{A'}$ is set to zero
in (L4.3.5): the regular modes $m_{np}$ and $r_{np}$ should vanish
on the bounding $S^3$.

To study the corresponding variational problem for a massless
fermionic field, we should now bear in mind that the
spin-${1\over 2}$ action
functional in a Riemannian 4-geometry takes the form
(cf. (L4.2.2))
$$
I_{E}={i\over 2}\int_{M}\Bigr[{\widetilde \psi}^{A'}
\Bigr(\nabla_{AA'}\psi^{A}\Bigr)
-\Bigr(\nabla_{AA'}{\widetilde \psi}^{A'}\Bigr)
\psi^{A}\Bigr]\sqrt{{\rm det} \; g} \; d^{4}x+{\widehat I}_{B}
\; \; \; \; .
\eqno (L4.3.10)
$$
This action is {\it real}, and the factor $i$ occurs
by virtue of the convention for Infeld-van der Waerden symbols
used in [4].
In (L4.3.10) ${\widehat I}_{B}$ is a suitable boundary term, to be
added to ensure that $I_{E}$ is stationary under the boundary
conditions chosen at the various components of the boundary
(e.g. initial and final surfaces). Of course, the
variation $\delta I_{E}$ of $I_{E}$ is linear in the
variations $\delta \psi^{A}$ and $\delta {\widetilde \psi}^{A'}$.
Defining $\kappa
\equiv {2\over i}$ and $\kappa {\widehat I}_{B} \equiv I_{B}$,
variational rules for anticommuting spinor fields lead to
$$ \eqalignno{
\kappa \Bigr(\delta I_{E}\Bigr)&=
\int_{M}\Bigr[2 \delta {\widetilde \psi}^{A'}
\Bigr(\nabla_{AA'}\psi^{A}\Bigr)\Bigr]
\sqrt{{\rm det} \; g} \; d^{4}x
-\int_{M}\Bigr[\Bigr(\nabla_{AA'}{\widetilde \psi}^{A'}
\Bigr)2 \delta \psi^{A}\Bigr]
\sqrt{{\rm det} \; g} \; d^{4}x \cr
&-\int_{\partial M}\Bigr[{_{e}n_{AA'}}
\Bigr(\delta {\widetilde \psi}^{A'}\Bigr)
\psi^{A}\Bigr]\sqrt{{\rm det} \; h} \; d^{3}x
+\int_{\partial M}\Bigr[{_{e}n_{AA'}}
{\widetilde \psi}^{A'}\Bigr(\delta \psi^{A}\Bigr)
\Bigr] \sqrt{{\rm det} \; h} \; d^{3}x \cr
&+\delta I_{B} \; \; \; \; ,
&(L4.3.11)\cr}
$$
where $I_{B}$ should be chosen in such a way that its
variation $\delta I_{B}$ combines with the sum of the two terms
on the second line of (L4.3.11) so as to specify what is fixed
on the boundary (see below). Indeed, setting
$\epsilon = \pm 1$ and using the boundary conditions (L4.3.5)
one finds
$$
{_{e}n_{AA'}}{\widetilde \psi}^{A'}
={\epsilon \over \sqrt{2}}\psi_{A}
-\epsilon \; {_{e}n_{AA'}}\Phi^{A'}
\; {\rm on} \; S^{3}
\; \; \; \; .
\eqno (L4.3.12)
$$
Thus, anticommutation rules for spinor fields [4] show that the
second line of equation (L4.3.11) reads
$$ \eqalignno{
\delta I_{\partial M} & \equiv
-\int_{\partial M}\Bigr[\Bigr(\delta {\widetilde \psi}^{A'}
\Bigr){_{e}n_{AA'}}\psi^{A}\Bigr]
\sqrt{{\rm det} \; h} \; d^{3}x
+\int_{\partial M}\Bigr[{_{e}n_{AA'}}{\widetilde \psi}^{A'}
\Bigr(\delta \psi^{A}\Bigr)\Bigr]
\sqrt{{\rm det} \; h} \; d^{3}x \cr
&=\epsilon \int_{\partial M}
{_{e}n_{AA'}}\Bigr[\Bigr(\delta \Phi^{A'}\Bigr)
\psi^{A}-\Phi^{A'}\Bigr(\delta \psi^{A}\Bigr)
\Bigr] \sqrt{{\rm det} \; h} \; d^{3}x
\; \; \; \; .
&(L4.3.13)\cr}
$$
Now it is clear that setting
$$
I_{B} \equiv \epsilon \, \int_{\partial M}
\; \Phi^{A'} {_{e}n_{AA'}}
\; \psi^A \sqrt{ {\rm det} \; h} \; d^3x
\; \; \; \; ,
\eqno (L4.3.14)
$$
enables one to specify $\Phi^{A'}$ on the boundary, since
$$
\delta \Bigr[ I_{\partial M} + I_{B} \Bigr] =
2 \epsilon \int_{\partial M} {_{e}n_{AA'}}
\Bigr( \delta \Phi^{A'} \Bigr) \psi^A \sqrt{{\rm det }\; h}\; d^3x
\; \; \; \; .
\eqno (L4.3.15)
$$
Hence the action integral (L4.3.10) appropriate for our boundary-value
problem is
$$ \eqalignno{
I_{E}&={i\over 2}\int_{M}\Bigr[{\widetilde \psi}^{A'}
\Bigr(\nabla_{AA'}\psi^{A}\Bigr)
-\Bigr(\nabla_{AA'}{\widetilde \psi}^{A'}\Bigr)
\psi^{A}\Bigr]\sqrt{{\rm det} \; g} \; d^{4}x \cr
&+{i\epsilon \over 2} \, \int_{\partial M}
\; \Phi^{A'} {_{e}n_{AA'}}
\; \psi^A \sqrt{ {\rm det} \; h} \; d^3x
\; \; \; \; .
&(L4.3.16)\cr}
$$
Note that, by virtue of (L4.3.5), equation (L4.3.13) may also be
cast in the form
$$
\delta I_{\partial M}
={1\over \sqrt{2}}\int_{\partial M}
\Bigr[{\widetilde \psi}^{A'}\Bigr(\delta \Phi_{A'}\Bigr)
-\Bigr(\delta {\widetilde \psi}^{A'}\Bigr)\Phi_{A'}
\Bigr] \sqrt{{\rm det} \; h} \; d^{3}x
\; \; \; \; ,
\eqno (L4.3.17)
$$
which implies that an equivalent form of $I_{B}$ is
$$
I_{B} \equiv {1\over \sqrt{2}}
\int_{\partial M}{\widetilde \psi}^{A'}
\; \Phi_{A'} \sqrt{{\rm det} \; h}
\; d^{3}x
\; \; \; \; .
\eqno (L4.3.18)
$$

The local boundary conditions studied at the classical level
in this Lecture, have been applied to one-loop quantum cosmology
[4]. Interestingly, our work seems to add evidence in favour
of quantum amplitudes having to respect the properties of the
classical boundary-value problem. In other words, if fermionic
fields are massless, their one-loop properties in the presence
of boundaries coincide in the case of spectral or local
boundary conditions [4], while we find that classical modes
for a regular solution of the Weyl equation obey the same
conditions on a three-sphere boundary with spectral or local
boundary conditions, provided the spinor field $\Phi^{A'}$
of (L4.3.5) is set to zero on $S^{3}$. We also hope that the
analysis presented in Eqs. (L4.3.10)-(L4.3.18) may clarify the
spin-${1\over 2}$
variational problem in the case of local boundary conditions
on a three-sphere.
\vskip 10cm
\centerline {\bf L4.4 Potentials for Massless Spin-${3\over 2}$ Fields}
\vskip 0.3cm
Recent work in the literature has studied the quantization
of gauge theories and supersymmetric field theories in the
presence of boundaries [4].
In particular, in the work described
in [4], two possible sets of
local boundary conditions were
studied. One of these, first proposed in anti-de Sitter
space-time, involves the normal to the boundary
and Dirichlet or Neumann conditions for spin $0$, the normal
and the field for massless spin-${1\over 2}$ fermions, the
normal and totally symmetric field strengths for spins
$1,{3\over 2}$ and $2$. Although more attention has been paid
to alternative local boundary conditions motivated by
supersymmetry, described in section L4.3,
the analysis of the former boundary
conditions remains of mathematical and physical interest by
virtue of its links with twistor theory [4]. The aim of
the following analysis is to derive further mathematical properties of
the corresponding boundary-value problem.

In section 5.7 of [4],
a flat Euclidean background bounded by
a three-sphere was studied. On the bounding $S^3$, the following
boundary conditions for a spin-$s$ field were required:
$$
2^{s} \; {_{e}}n^{AA'}...{_{e}}n^{LL'}
\; \phi_{A...L}= \pm {\widetilde \phi}^{A'...L'}
\; \; \; \; .
\eqno (L4.4.1)
$$
With our notation, ${_{e}}n^{AA'}$ is the Euclidean normal
to $S^3$ [4],
$\phi_{A...L}=\phi_{(A...L)}$
and ${\widetilde \phi}_{A'...L'}
={\widetilde \phi}_{(A'...L')}$ are totally symmetric
and independent (i.e. not related by any conjugation)
field strengths, which reduce to the massless
spin-${1\over 2}$ field for $s={1\over 2}$. Moreover,
the complex scalar field $\phi$ is such that its real
part obeys Dirichlet conditions on $S^3$ and its imaginary
part obeys Neumann conditions on $S^3$, or the other way
around, according to the value of the parameter
$\epsilon \equiv \pm 1$ occurring in (L4.4.1), as described
in [4].

We now focus on the totally
symmetric field strengths $\psi_{ABC}$ and
${\widetilde \psi}_{A'B'C'}$ for spin-${3\over 2}$ fields,
and we express them in terms of their potentials. The
corresponding theory in Minkowski space-time (and curved
space-time), first described by Penrose [5],
is adapted here to
the case of flat Euclidean four-space with flat connection $D$.
It turns out that ${\widetilde \psi}_{A'B'C'}$ can then be
obtained from two potentials defined as follows. The first
potential satisfies the properties
$$
\gamma_{A'B'}^{C}=\gamma_{(A'B')}^{C}
\; \; \; \; ,
\eqno (L4.4.2)
$$
$$
D^{AA'} \; \gamma_{A'B'}^{C}=0
\; \; \; \; ,
\eqno (L4.4.3)
$$
$$
{\widetilde \psi}_{A'B'C'}=D_{CC'} \; \gamma_{A'B'}^{C}
\; \; \; \; ,
\eqno (L4.4.4)
$$
with the gauge freedom of replacing it by
$$
{\widehat \gamma}_{A'B'}^{C} \equiv \gamma_{A'B'}^{C}
+D_{\; \; B'}^{C} \; {\widetilde \nu}_{A'}
\; \; \; \; ,
\eqno (L4.4.5)
$$
where ${\widetilde \nu}_{A'}$ satisfies the positive-helicity Weyl
equation
$$
D^{AA'} \; {\widetilde \nu}_{A'}=0
\; \; \; \; .
\eqno (L4.4.6)
$$
The second potential is defined by the conditions [5]
$$
\rho_{A'}^{BC}=\rho_{A'}^{(BC)}
\; \; \; \; ,
\eqno (L4.4.7)
$$
$$
D^{AA'} \; \rho_{A'}^{BC}=0
\; \; \; \; ,
\eqno (L4.4.8)
$$
$$
\gamma_{A'B'}^{C}=D_{BB'} \; \rho_{A'}^{BC}
\; \; \; \; ,
\eqno (L4.4.9)
$$
with the gauge freedom of being replaced by
$$
{\widehat \rho}_{A'}^{BC} \equiv \rho_{A'}^{BC}
+D_{\; \; A'}^{C} \; \chi^{B}
\; \; \; \; ,
\eqno (L4.4.10)
$$
where $\chi^{B}$ satisfies the negative-helicity
Weyl equation
$$
D_{BB'} \; \chi^{B}=0
\; \; \; \; .
\eqno (L4.4.11)
$$
Moreover, in flat Euclidean four-space the field strength
$\psi_{ABC}$ is expressed in terms of the potential
$\Gamma_{AB}^{C'}=\Gamma_{(AB)}^{C'}$, independent
of $\gamma_{A'B'}^{C}$, as
$$
\psi_{ABC}=D_{CC'} \; \Gamma_{AB}^{C'}
\; \; \; \; ,
\eqno (L4.4.12)
$$
with gauge freedom
$$
{\widehat \Gamma}_{AB}^{C'} \equiv \Gamma_{AB}^{C'}
+D_{\; \; B}^{C'} \; \nu_{A}
\; \; \; \; .
\eqno (L4.4.13)
$$
Thus, if we insert (L4.4.4) and (L4.4.12) into the boundary
conditions (L4.4.1) with $s={3\over 2}$, and require that
also the gauge-equivalent potentials (L4.4.5) and (L4.4.13)
should obey such boundary conditions on $S^3$, we
find that
$$
2^{3\over 2} \; {_{e}}n_{\; \; A'}^{A}
\; {_{e}}n_{\; \; B'}^{B}
\; {_{e}}n_{\; \; C'}^{C}
\; D_{CL'} \; D_{\; \; B}^{L'}
\; \nu_{A}=\epsilon \;
D_{LC'} \; D_{\; \; B'}^{L}
\; {\widetilde \nu}_{A'}
\; \; \; \; ,
\eqno (L4.4.14)
$$
on the three-sphere. Note that, from now on,
covariant derivatives appearing in boundary
conditions are first taken on the background and then
evaluated on $S^3$.
In the case of our flat background, (L4.4.14) is identically
satisfied since $D_{CL'} \; D_{\; \; \; B}^{L'} \; \nu_{A}$
and $D_{LC'} \; D_{\; \; B'}^{L} \; {\widetilde \nu}_{A'}$
vanish by virtue of spinor Ricci identities. In
a curved background, however, denoting by $\nabla$ the
corresponding curved connection, and defining
$\cstok{\ }_{AB} \equiv \nabla_{M'(A}
\nabla_{\; \; \; B)}^{M'} \; , \; \cstok{\ }_{A'B'} \equiv
\nabla_{X(A'} \; \nabla_{\; \; B')}^{X}$,
since the spinor Ricci identities we need are
$$
\cstok{\ }_{AB} \; \nu_{C}=\phi_{ABDC} \; \nu^{D}
-2\Lambda \; \nu_{(A} \; \epsilon_{B)C}
\; \; \; \; ,
\eqno (L4.4.15)
$$
$$
\cstok{\ }_{A'B'} \; {\widetilde \nu}_{C'}
={\widetilde \phi}_{A'B'D'C'} \;
{\widetilde \nu}^{D'} -2 {\widetilde \Lambda}
\; {\widetilde \nu}_{(A'} \; \epsilon_{B')C'}
\; \; \; \; ,
\eqno (L4.4.16)
$$
one finds that the corresponding boundary conditions
$$
2^{3\over 2} \; {_{e}}n_{\; \; A'}^{A}
\; {_{e}}n_{\; \; B'}^{B}
\; {_{e}}n_{\; \; C'}^{C}
\; \nabla_{CL'} \; \nabla_{\; \; \; B}^{L'}
\; \nu_{A}=\epsilon \; \nabla_{LC'}
\; \nabla_{\; \; B'}^{L}
\; {\widetilde \nu}_{A'}
\; \; \; \; ,
\eqno (L4.4.17)
$$
are identically satisfied if and only if one of the
following conditions holds: (i) $\nu_{A}=
{\widetilde \nu}_{A'}=0$; (ii) the Weyl spinors
$\phi_{ABCD},{\widetilde \phi}_{A'B'C'D'}$ and the
scalars $\Lambda,{\widetilde \Lambda}$
vanish everywhere. However,
since in a curved space-time with vanishing $\Lambda,
{\widetilde \Lambda}$, the potentials with the gauge
freedoms (L4.4.5) and (L4.4.13) only exist provided $D$
is replaced by $\nabla$ and the trace-free part
$\Phi_{ab}$ of the Ricci tensor vanishes as well [5],
the background four-geometry is actually
flat Euclidean four-space. Note that we require that
(L4.4.17) should be identically satisfied to avoid that,
after a gauge transformation, one obtains more boundary
conditions than the ones originally imposed. The curvature
of the background should not, itself, be subject to a
boundary condition.

The same result can be derived by using
the potential $\rho_{A'}^{BC}$ and its independent
counterpart $\Lambda_{A}^{B'C'}$. This spinor field
yields the $\Gamma_{AB}^{C'}$ potential by means of
$$
\Gamma_{AB}^{C'}=D_{BB'} \; \Lambda_{A}^{B'C'}
\; \; \; \; ,
\eqno (L4.4.18)
$$
and has the gauge freedom
$$
{\widehat \Lambda}_{A}^{B'C'} \equiv \Lambda_{A}^{B'C'}
+D_{\; \; A}^{C'} \; {\widetilde \chi}^{B'}
\; \; \; \; ,
\eqno (L4.4.19)
$$
where ${\widetilde \chi}^{B'}$ satisfies the positive-helicity
Weyl equation
$$
D_{BF'} \; {\widetilde \chi}^{F'}=0
\; \; \; \; .
\eqno (L4.4.20)
$$
Thus, if also the gauge-equivalent
potentials (L4.4.10) and (L4.4.19)
have to satisfy the boundary conditions (L4.4.1) on $S^3$, one
finds
$$
2^{3\over 2} \; {_{e}}n_{\; \; A'}^{A}
\; {_{e}}n_{\; \; B'}^{B}
\; {_{e}}n_{\; \; C'}^{C}
\; D_{CL'} \; D_{BF'} \;
D_{\; \; A}^{L'} \;
{\widetilde \chi}^{F'}
=\epsilon \; D_{LC'} \; D_{MB'} \;
D_{\; \; A'}^{L} \; \chi^{M}
\; \; \; \; ,
\eqno (L4.4.21)
$$
on the three-sphere. In a flat background, covariant derivatives
commute, hence (L4.4.21) is identically
satisfied by virtue of (L4.4.11)
and (L4.4.20). However, in the curved case the boundary conditions
(L4.4.21) are replaced by
$$
2^{3\over 2} \; {_{e}}n_{\; \; A'}^{A}
\; {_{e}}n_{\; \; B'}^{B}
\; {_{e}}n_{\; \; C'}^{C}
\; \nabla_{CL'} \; \nabla_{BF'}
\; \nabla_{\; \; A}^{L'}
\; {\widetilde \chi}^{F'}
=\epsilon \; \nabla_{LC'} \;
\nabla_{MB'} \; \nabla_{\; \; A'}^{L}
\; \chi^{M}
\; \; \; \; ,
\eqno (L4.4.22)
$$
on $S^3$, if the {\it local} expressions of $\psi_{ABC}$ and
${\widetilde \psi}_{A'B'C'}$ in terms of potentials still
hold [5].
By virtue of (L4.4.15)-(L4.4.16), where $\nu_{C}$ is replaced
by $\chi_{C}$ and ${\widetilde \nu}_{C'}$ is replaced by
${\widetilde \chi}_{C'}$, this means that the Weyl spinors
$\phi_{ABCD},{\widetilde \phi}_{A'B'C'D'}$ and the scalars
$\Lambda,{\widetilde \Lambda}$ should vanish, since one
should find
$$
\nabla^{AA'} \; {\widehat \rho}_{A'}^{BC}=0
\; \; \; \; , \; \; \; \;
\nabla^{AA'} \; {\widehat \Lambda}_{A}^{B'C'}=0
\; \; \; \; .
\eqno (L4.4.23)
$$
If we assume that
$\nabla_{BF'} \; {\widetilde \chi}^{F'}=0$ and
$\nabla_{MB'} \; \chi^{M}=0$, we have to show that (L4.4.22)
differs from (L4.4.21) by terms involving a part of the curvature
that is vanishing everywhere.
This is proved by using the basic rules
of two-spinor calculus and spinor Ricci identities [5].
Thus, bearing in mind that [5]
$$
\cstok{\ }^{AB} \; {\widetilde \chi}_{B'}
=\Phi_{\; \; \; \; L'B'}^{AB} \;
{\widetilde \chi}^{L'}
\; \; \; \; ,
\eqno (L4.4.24)
$$
$$
\cstok{\ }^{A'B'} \; \chi_{B}
={\widetilde \Phi}_{\; \; \; \; \; \; LB}^{A'B'}
\; \chi^{L}
\; \; \; \; ,
\eqno (L4.4.25)
$$
one finds
$$ \eqalignno{
\nabla^{BB'} \; \nabla^{CA'} \; \chi_{B}&=
\nabla^{(BB'} \; \nabla^{C)A'} \; \chi_{B}
+\nabla^{[BB'} \; \nabla^{C]A'} \; \chi_{B} \cr
&=-{1\over 2} \nabla_{B}^{\; \; B'} \;
\nabla^{CA'} \; \chi^{B}
+{1\over 2} {\widetilde \Phi}^{A'B'LC} \; \chi_{L}
\; \; \; \; .
&(L4.4.26)\cr}
$$
Thus, if ${\widetilde \Phi}^{A'B'LC}$ vanishes, also the left-hand side
of (L4.4.26) has to vanish since this leads to the equation
$
\nabla^{BB'} \; \nabla^{CA'} \; \chi_{B}
={1\over 2}
\nabla^{BB'} \; \nabla^{CA'} \; \chi_{B}
$. Hence (L4.4.26) is identically satisfied. Similarly, the
left-hand side of (L4.4.22) can be made to vanish identically
provided the additional condition $\Phi^{CDF'M'}=0$ holds.
The conditions
$$
\Phi^{CDF'M'}=0
\; \; \; \; , \; \; \; \;
{\widetilde \Phi}^{A'B'CL}=0
\; \; \; \; ,
\eqno (L4.4.27)
$$
when combined with the conditions
$$
\phi_{ABCD}={\widetilde \phi}_{A'B'C'D'}=0
\; \; \; \; , \; \; \; \;
\Lambda={\widetilde \Lambda}=0
\; \; \; \; ,
\eqno (L4.4.28)
$$
arising from (L4.4.23) for the local existence of
$\rho_{A'}^{BC}$ and $\Lambda_{A}^{B'C'}$ potentials,
imply that the whole Riemann curvature should vanish.
Hence, in the boundary-value problems we are interested
in, the only admissible background four-geometry
(of the Einstein type) is flat Euclidean four-space.

In conclusion, we have focused on the potentials for
spin-${3\over 2}$ field strengths in
flat or curved Riemannian four-space
bounded by a three-sphere. Remarkably, it turns out that
local boundary conditions involving
field strengths and normals can only be imposed in a
flat Euclidean background, for which the gauge freedom in the
choice of the potentials remains. Penrose [5]
found that $\rho$ potentials exist {\it locally} only in the
self-dual Ricci-flat case, whereas $\gamma$ potentials may be
introduced in the anti-self-dual case. Our result may be
interpreted as a further restriction provided by (quantum)
cosmology.

A naturally occurring question is whether the potentials studied
in this section can be used to perform one-loop calculations for
spin-${3\over 2}$ field strengths subject to (L4.4.1) on $S^3$.
The solution of this problem
might shed new light
on the quantization program for gauge theories in the presence
of boundaries [4-5].
\vskip 100cm
\centerline {\bf LECTURE 5.}
\vskip 1cm
\centerline {\bf L5.1 Self-Adjointness of Boundary-Value Problems
for the Dirac Operator}
\vskip 0.3cm
We now study in detail local boundary conditions described in
Lecture 4 in the case of the massless
spin-${1\over 2}$ field on a three-sphere of radius $a$, bounding a region
of Euclidean four-space centred on the origin. The field
$\Bigr(\psi^{A}, \; {\widetilde \psi}^{A'}\Bigr)$ may be expanded in terms of
harmonics on the family of three-spheres centred on the origin, as
(cf. (L4.3.1)-(L4.3.2))
$$
\psi^A={t^{-{\textstyle {3 \over 2}}}\over 2\pi}\sum_{npq}
\alpha_{n}^{pq}\Bigr[m_{np}(t)\rho^{nqA} +{\widetilde r}_{np}(t)
{\overline \sigma}^{nqA}\Bigr]
\; \; \; \; ,
\eqno (L5.1.1)
$$
$$
{\widetilde \psi}^{A'}={t^{-{\textstyle {3 \over 2}}}\over 2\pi}
\sum_{npq} \alpha_{n}^{pq}\left[{\widetilde m}_{np}(t) \bar \rho^{nqA'} +
r_{np}(t) \sigma^{nqA'}\right]
\; \; \; \; .
\eqno (L5.1.2)
$$
Here $t$ is the radius of a three-sphere, and the notation is the one
described in Lecture 4.
The {\it twiddle} symbol for ${\widetilde \psi}^{A'}$
and for the coefficients ${\widetilde m}_{np}$ and ${\widetilde r}_{np}$ does
not denote any conjugation operation. In fact, on analytic
continuation to a smooth and positive-definite metric, unprimed and primed
spinors transform under independent groups $SU(2)$ and
${\widetilde {SU(2)}}$.
Moreover, setting $\epsilon \equiv \pm 1$,
we know that supersymmetric local boundary conditions may be written
in the form
$$
\sqrt{2} \; {_{e}n_{A}^{\ A'}}\psi^A = \epsilon \; {\widetilde \psi}^{A'}
\; \; {\rm on} \; \; S^{3}
\; \; \; \; ,
\eqno (L5.1.3)
$$
and we also know that the harmonics ${\overline \rho}^{nqA'}$ and
${\overline \sigma}^{nqA}$ may be re-expressed in terms of the harmonics
$n_{A}^{\; \; A'}\rho^{npA}$ and $n_{\; \; A'}^{A}\sigma^{npA'}$ using
the relations [4]
$$
\bar \rho^{nqA'}=2n_{A}^{\ A'}\sum_{d}\rho^{ndA} (A_{n}^{-1} H_{n})^{dq}
\; \; \; \; ,
\eqno (L5.1.4)
$$
$$
{\overline \sigma}^{nqA}=2n_{\ A'}^{A}\sum_{d}\sigma^{ndA'}(A_{n}^{-1} H_{n})
^{dq}
\; \; \; \; .
\eqno (L5.1.5)
$$
Boundary conditions of this
kind have also been studied in Luckock 1991,
where (though using a different
formalism) the more general possibility of an $\epsilon$ which is a
complex-valued function of position on the boundary has been considered.
We shall see later that self-adjointness of the boundary-value problem
implies reality of $\epsilon$.

Let us now recall that $\alpha_{n}=
\pmatrix {1&1\cr 1&-1\cr}$, $A_{n}=\sqrt{2} \pmatrix {0&1\cr -1&0\cr}$,
$A_{n}^{-1}={1\over \sqrt{2}} \pmatrix {0&-1\cr 1&0\cr} = A_{n}^{-1}H_{n}$.
For simplicity, consider first that part of (L5.1.3)
which involves the $\rho$ harmonics.
The relations (L5.1.1)-(L5.1.4) yield therefore for each $n$
$$
-i\sum_{pq} \pmatrix {1&1\cr 1&-1\cr}^{pq}m_{np}(a)\rho^{nqA}=
\epsilon \sum_{pq} \pmatrix {1&1\cr 1&-1\cr}^{pq}\widetilde m_{np}(a)
\sum_{d}\rho^{ndA} \pmatrix {0&-1\cr 1&0\cr}^{dq} .
\eqno (L5.1.6)
$$
In the typical case of the indices $p,q=1,2$, this implies
$$ \eqalignno{
-im_{n1}(a)\Bigr(\rho^{n1A}+\rho^{n2A}\Bigr)-im_{n2}(a)
\Bigr(\rho^{n1A}-\rho^{n2A}\Bigr)&=
\epsilon \widetilde m_{n1}(a)\Bigr(\rho^{n2A}-\rho^{n1A}\Bigr)\cr
&+\epsilon \widetilde m_{n2}(a)(\rho^{n2A}+\rho^{n1A})
&({\rm L}5.1.7)\cr}
$$
Similar equations hold for adjacent indices $p,q=2k+1,2k+2$
($k=0,1,...,{n\over 2}(n+3)$).
Since the harmonics $\rho^{n1A}$ and $\rho^{n2A}$ on the bounding
three-sphere are linearly independent, one has the following system:
$$
-i \Bigr[m_{n1}(a)+m_{n2}(a)\Bigr]=\epsilon \Bigr[\widetilde m_{n2}(a)
-{\widetilde m}_{n1}(a)\Bigr]
\; \; \; \; ,
\eqno (L5.1.8)
$$
$$
-i \Bigr[m_{n1}(a)-m_{n2}(a)\Bigr]=\epsilon \Bigr[\widetilde m_{n2}(a)
+{\widetilde m}_{n1}(a)\Bigr]
\; \; \; \; ,
\eqno (L5.1.9)
$$
whose solution is
$$
-im_{n1}(a)=\epsilon \; \widetilde m_{n2}(a)
\; \; \; \; ,
\eqno (L5.1.10)
$$
$$
im_{n2}(a)=\epsilon \; \widetilde m_{n1}(a)
\; \; \; \; .
\eqno (L5.1.11)
$$
In the same way, the part of (L5.1.3) involving the $\sigma$ harmonics
leads to
$$
\epsilon
\sum_{pq} \pmatrix {1&1\cr 1&-1\cr}^{pq}r_{np}(a)\sigma^{nqA'}=
i\sum_{pq}\pmatrix {1&1\cr 1&-1\cr}^{pq}\widetilde r_{np}(a)
\sum_{d} \sigma^{ndA'}\pmatrix {0&-1\cr 1&0\cr}^{dq} ,
\eqno (L5.1.12)
$$
which implies, for example
$$ \eqalignno{
\epsilon
\Bigr[r_{n1}(a)+r_{n2}(a)\Bigr]\sigma^{n1A'}+
\epsilon
\Bigr[r_{n1}(a)-r_{n2}(a)\Bigr]\sigma^{n2A'}&=
-i\Bigr[\widetilde r_{n1}(a)-\widetilde r_{n2}(a)\Bigr]\sigma^{n1A'}\cr
&+i[\widetilde r_{n1}(a)+\widetilde r_{n2}(a)]\sigma^{n2A'}
&({\rm L}5.1.13)\cr}
$$
so that we finally get
$$
-i\widetilde r_{n1}(a)=\epsilon \; r_{n2}(a)
\; \; \; \; ,
\eqno (L5.1.14)
$$
$$
i\widetilde r_{n2}(a)=\epsilon \; r_{n1}(a)
\; \; \; \; ,
\eqno (L5.1.15)
$$
and similar equations for other adjacent indices $p,q$. Thus, defining
$$
x \equiv m_{n1}\; \; \; \; , \; \; \; \; X \equiv m_{n2}
\; \; \; \; ,
\eqno (L5.1.16)
$$
$$
\widetilde x \equiv \widetilde m_{n1}\; \; \; \; , \; \; \; \;
\widetilde X \equiv \widetilde m_{n2}
\; \; \; \; ,
\eqno (L5.1.17)
$$
$$
y \equiv r_{n1}\; \; \; \; , \; \; \; \; Y \equiv r_{n2}
\; \; \; \; ,
\eqno (L5.1.18)
$$
$$
\widetilde y \equiv \widetilde r_{n1}\; \; \; \; , \; \; \; \;
\widetilde Y \equiv \widetilde r_{n2}
\; \; \; \; ,
\eqno (L5.1.19)
$$
we may cast (L5.1.10)-(L5.1.11) and
(L5.1.14)-(L5.1.15) in the form
$$
-ix(a)=\epsilon \; \widetilde X(a) \; \; \; \; , \; \; \; \;
iX(a)=\epsilon \; \widetilde x(a)
\; \; \; \; ,
\eqno (L5.1.20)
$$
$$
-i\widetilde y(a)=\epsilon \; Y(a) \; \; \; \; , \; \; \; \;
i\widetilde Y(a)=\epsilon \; y(a)
\; \; \; \; .
\eqno (L5.1.21)
$$
Again, similar equations hold relating $m_{np},{\widetilde m}_{np},
r_{np}$ and ${\widetilde r}_{np}$ at the boundary for adjacent indices
$p=2k+1,2k+2$. The task now
remains to work out the eigenvalue condition for this problem in
the massless case. Indeed, we know that the Euclidean action $I_{E}$ for a
massless spin-${1\over 2}$ field is a sum of terms of the kind [4]
$$
I_{n,E}(x,\widetilde x,y,\widetilde y)=
\int_{0}^{a} d\tau \left[{1\over 2}\Bigr(\widetilde x \dot x
+x \dot {\widetilde x}
+\widetilde y \dot y +y \dot {\widetilde y}\Bigr)
-{1\over \tau}\left(n+{3\over 2}\right)(\widetilde x x +\widetilde y y)
\right] \; .
\eqno (L5.1.22)
$$
Thus, writing $\kappa_{n} \equiv
n+{3\over 2}$ and introducing, $\forall n \geq 0$, the operators
$$
L_{n}\equiv {d\over d\tau}-{\kappa_{n} \over \tau} \; \; \; \; ,
\; \; \; \; M_{n}\equiv {d\over d\tau}+{\kappa_{n} \over \tau}
\; \; \; \; ,
\eqno (L5.1.23)
$$
the eigenvalue equations are found to be
$$
L_{n}x=E {\widetilde x} \; \; \; \; , \; \; \; \;
M_{n}{\widetilde x}=-Ex
\; \; \; \; ,
\eqno (L5.1.24)
$$
$$
L_{n}y=E{\widetilde y} \; \; \; \; , \; \; \; \;
M_{n}{\widetilde y}=-Ey
\; \; \; \; ,
\eqno (L5.1.25)
$$
$$
L_{n}X=E{\widetilde X} \; \; \; \; , \; \; \; \;
M_{n}{\widetilde X}=-EX
\; \; \; \; ,
\eqno (L5.1.26)
$$
$$
L_{n}Y=E{\widetilde Y} \; \; \; \; , \; \; \; \;
M_{n}{\widetilde Y}=-EY
\; \; \; \; .
\eqno (L5.1.27)
$$
We now define $\forall n \geq 0$ the differential operators
$$
P_{n}\equiv {d^{2}\over d\tau^{2}}+
\left[E^{2}-{((n+2)^{2}-{1\over 4})\over \tau^{2}}\right]
\; \; \; \; ,
\eqno (L5.1.28)
$$
$$
Q_{n}\equiv {d^{2}\over d\tau^{2}}+
\left[E^{2}-{((n+1)^{2}-{1\over 4})\over \tau^{2}}\right]
\; \; \; \; .
\eqno (L5.1.29)
$$
Equations (L5.1.24)-(L5.1.27) lead straightforwardly to the following
second-order equations, $\forall n \geq 0$ :
$$
P_{n}{\widetilde x}=P_{n}{\widetilde X}=P_{n}{\widetilde y}
=P_{n}{\widetilde Y}=0
\; \; \; \; ,
\eqno (L5.1.30)
$$
$$
Q_{n}y=Q_{n}Y=Q_{n}x=Q_{n}X=0
\; \; \; \; .
\eqno (L5.1.31)
$$
The solutions of (L5.1.30)-(L5.1.31)
which are regular at the origin are
$$
\widetilde x=C_{1}\sqrt{\tau}J_{n+2}(E\tau) \; \; \; \; , \; \; \; \;
\widetilde X=C_{2}\sqrt{\tau}J_{n+2}(E\tau)
\; \; \; \; ,
\eqno (L5.1.32)
$$
$$
x=C_{3}\sqrt{\tau}J_{n+1}(E\tau) \; \; \; \; , \; \; \; \;
X=C_{4}\sqrt{\tau}J_{n+1}(E\tau)
\; \; \; \; ,
\eqno (L5.1.33)
$$
$$
\widetilde y=C_{5}\sqrt{\tau}J_{n+2}(E\tau) \; \; \; \; ,
\; \; \; \;
\widetilde Y=C_{6}\sqrt{\tau}J_{n+2}(E\tau)
\; \; \; \; ,
\eqno (L5.1.34)
$$
$$
y=C_{7}\sqrt{\tau}J_{n+1}(E\tau) \; \; \; \; , \; \; \; \;
Y=C_{8}\sqrt{\tau}J_{n+1}(E\tau)
\; \; \; \; .
\eqno (L5.1.35)
$$
To find the condition obeyed by $E$,
we must now insert (L5.1.32)-(L5.1.35)
into the boundary conditions (L5.1.20)-(L5.1.21),
taking into account also the
first-order system given by (L5.1.24)-(L5.1.27).
This gives the following eight equations:
$$
-iC_{3}J_{n+1}(Ea)=\epsilon \; C_{2}J_{n+2}(Ea)
\; \; \; \; ,
\eqno (L5.1.36)
$$
$$
iC_{4}J_{n+1}(Ea)=\epsilon \; C_{1}J_{n+2}(Ea)
\; \; \; \; ,
\eqno (L5.1.37)
$$
$$
-iC_{5}J_{n+2}(Ea)=\epsilon \; C_{8}J_{n+1}(Ea)
\; \; \; \; ,
\eqno (L5.1.38)
$$
$$
iC_{6}J_{n+2}(Ea)=\epsilon \; C_{7}J_{n+1}(Ea)
\; \; \; \; ,
\eqno (L5.1.39)
$$
$$
C_{1}=-{{EC_{3}J_{n+1}(Ea)}\over \left[
{E\dot J_{n+2}(Ea)+(n+2){\textstyle {J_{n+2}(Ea)\over a}}}\right]}
\; \; \; \; ,
\eqno (L5.1.40)
$$
$$
C_{2}=-{{EC_{4}J_{n+1}(Ea)}\over \left[
{E\dot J_{n+2}(Ea)+(n+2){\textstyle {J_{n+2}(Ea)\over a}}}\right]}
\; \; \; \; ,
\eqno (L5.1.41)
$$
$$
C_{7}={{EC_{5}J_{n+2}(Ea)}\over \left[
{E\dot J_{n+1}(Ea)-(n+1){\textstyle {J_{n+1}(Ea)\over a}}}\right]}
\; \; \; \; ,
\eqno (L5.1.42)
$$
$$
C_{8}={{EC_{6}J_{n+2}(Ea)}\over \left[
{E\dot J_{n+1}(Ea)-(n+1){\textstyle {J_{n+1}(Ea)\over a}}}\right]}
\; \; \; \; .
\eqno (L5.1.43)
$$
Note that these give separate relations among the constants $C_{1},C_{2},
C_{3},C_{4}$ and among $C_{5},C_{6},C_{7},C_{8}$. For example,
using (L5.1.36)-(L5.1.37),
(L5.1.40)-(L5.1.41) to eliminate $C_{1},C_{2},C_{3},C_{4}$,
and the useful identities
$$
Ea\dot J_{n+1}(Ea)-(n+1)J_{n+1}(Ea)=-EaJ_{n+2}(Ea)
\; \; \; \; ,
\eqno (L5.1.44)
$$
$$
Ea\dot J_{n+2}(Ea)+(n+2)J_{n+2}(Ea)=EaJ_{n+1}(Ea)
\; \; \; \; ,
\eqno (L5.1.45)
$$
one finds
$$
-i\epsilon {J_{n+1}(Ea)\over J_{n+2}(Ea)}={\epsilon}^{2}{C_{2}\over C_{3}}
={\epsilon}^{2}{C_{4}\over C_{1}}
=-i{\epsilon}^{3} {J_{n+2}(Ea)\over J_{n+1}(Ea)}
\; \; \; \; ,
\eqno (L5.1.46)
$$
which implies (since $\epsilon \equiv \pm 1$)
$$
{\Bigr[J_{n+1}(Ea)\Bigr]}^{2}-{\Bigr[J_{n+2}(Ea)\Bigr]}^{2}=0
\; \; \; \; , \; \; \; \;
\forall n \geq 0
\; \; \; \; .
\eqno (L5.1.47)
$$
The desired set of eigenvalue conditions can also be written in the form
$$
J_{n+1}(Ea)=\pm J_{n+2}(Ea) \; \; \; \; , \; \; \; \;
\forall n \geq 0
\; \; \; \; .
\eqno (L5.1.48)
$$
Exactly the same set of eigenvalue conditions arises from
eliminating $C_{5},C_{6},C_{7},C_{8}$. The limiting behaviour of the
eigenvalues can be found under certain approximations. For example,
if $n$ is fixed and $\mid z \mid \rightarrow \infty$, one has the standard
asymptotic expansion
$$
J_{n}(z)\sim \sqrt{{2\over {\pi z}}}
\cos \left(z-{n\pi \over 2}-{\pi \over 4}\right)
+{\rm O} \Bigr(z^{-{3\over 2}}\Bigr)
\; \; \; \; .
\eqno (L5.1.49)
$$
Thus, writing (L5.1.48) in the form $J_{n+1}(E)=
{\hat \kappa} J_{n+2}(E)$, where
${\hat \kappa} \equiv \pm 1$ and we
set $a=1$ for simplicity,
two asymptotic sets of eigenvalues result [4]
$$
E^{+} \sim \pi \left({n\over 2}+L\right)
\; \; {\rm if} \; \;
{\hat \kappa}=1
\; \; \; \; ,
\eqno (L5.1.50)
$$
$$
E^{-} \sim \pi \left ({n\over 2}+M+{1\over 2}\right)
\; \; {\rm if} \; \;
{\hat \kappa}=-1
\; \; \; \; ,
\eqno (L5.1.51)
$$
where $L$ and $M$ are large integers (both positive and negative).
One can also obtain an
estimate of the smallest eigenvalues, for a given large $n$.
These eigenvalues have the asymptotic form [4]
$$
\mid E \mid \sim \Bigr[n+{\rm o}(n)\Bigr]
\; \; \; \; .
\eqno (L5.1.52)
$$

In general it is very difficult to solve numerically (L5.1.48), because the
recurrence relations which enable one to compute Bessel functions starting
from $J_{0}$ and $J_{1}$ are a source of large errors when the argument
is comparable with the order. Alternatively, one can easily work out a
fourth-order differential equation whose solution is of the form
$\sqrt{x}\Bigr(J_{n+1}(x)\mp J_{n+2}(x)\Bigr)$. The coefficients of this
equation do not depend on the Bessel functions, but unfortunately they
involve the eigenvalues $E$. This is why we have not been able to compute
numerically the eigenvalues. However, we should say that a more successful
numerical study has been carried out in Berry and Mondragon 1987. In that
paper, the authors study eigenvalues of the Dirac operator with local
boundary conditions, in the case of neutrino billiards. This corresponds
to massless spin-${1\over 2}$ particles moving under the action of a
potential describing a hard wall bounding a finite domain. The
authors end up with an eigenvalue condition of the kind
$J_{l}(k_{nl})=J_{l+1}(k_{nl})$, and compute the lowest 2600 positive
eigenvalues $k_{nl}$.

We now study the fermionic path integral
$$
K \equiv \int e^{-I_{E}}\; D\psi^{A} \; D{\widetilde \psi}^{A'}
\; \; \; \; ,
\eqno (L5.1.53)
$$
taken over the class of massless spin-${1\over 2}$ fields
$\Bigr(\psi^{A}, \; {\widetilde \psi}^{A'}\Bigr)$ which obey (L5.1.3) on the
bounding $S^3$. Here, with our conventions (cf. (L4.3.16))
$$
I_{E} \equiv {i\over 2} \int d^{4}x\sqrt{g}\left[\widetilde
\psi^{A'}\left(\nabla_{AA'}\psi^{A}\right)-\left(\nabla_{AA'}\widetilde
\psi^{A'}\right)\psi^{A}\right]
\; \; \; \;
\eqno (L5.1.54)
$$
is the Euclidean action for a massless spin-${1\over 2}$ field
$\Bigr(\psi^{A}, \; {\widetilde \psi}^{A'}\Bigr)$. The fermionic fields are
taken to be anti-commuting, and Berezin integration is being used
$$
\int dy =0 \; \; \; \; , \; \; \; \; \int y \; dy =1
\; \; \; \; .
\eqno (L5.1.55)
$$
Let us now assume provisionally that
$\psi^A$ and $\widetilde \psi^{A'}$ can be expanded in a complete set of
eigenfunctions $\Bigr \{\psi_{n}^{A}, \; {\widetilde \psi}_{n}^{A'}\Bigr \}$
obeying the eigenvalue
equations which arise from variation of the action (L5.1.54)
$$
\nabla_{AA'}\psi_{n}^A=\lambda_{n}\widetilde \psi_{nA'}
\; \; \; \; ,
\eqno (L5.1.56)
$$
$$
\nabla_{AA'}\widetilde \psi_{n}^{A'}=\lambda_{n}\psi_{nA}
\; \; \; \; ,
\eqno (L5.1.57)
$$
and the boundary condition $\sqrt{2} \; {_{e}n_{AA'}}\psi_{n}^{A}
=\epsilon \; {\widetilde \psi}_{nA'}$
on $S^3$ (cf. (L5.1.3)). As with a bosonic one-loop path integral, one would
like to be able to express the fermionic path integral
$K$ in terms of a suitable product of eigenvalues. One then needs the
cross-terms in $I_{E}$ to vanish. Indeed, the typical cross-term $\Sigma$
appearing in $I_{E}$ is
$$ \eqalignno{
\Sigma&={i\over 2}\int d^{4}x \sqrt{g}\Bigr[\lambda_{n}
{\widetilde \psi}_{m}^{A'}
\widetilde \psi_{nA'}+\lambda_{m}\psi_{n}^{A}\psi_{mA}
+\lambda_{m}\widetilde \psi_{m}^{A'}\widetilde \psi_{nA'}
+\lambda_{n}\psi_{n}^{A}\psi_{mA}\Bigr]\cr
&={i\over 2}\int d^{4}x \sqrt{g}\Bigr[(\lambda_{n}+\lambda_{m})
\widetilde \psi_{m}^{A'}\widetilde \psi_{nA'}
+(\lambda_{n}+\lambda_{m})\psi_{n}^{A}\psi_{mA}\Bigr]
\; \; \; \; .
&(L5.1.58)\cr}
$$
In deriving (L5.1.58), where $n \not = m$, we have raised and lowered spinor
indices, and used the anticommutation relations obeyed by our spinor fields.
However, commutation can be assumed when the path integral is not involved,
as we shall do later.
The use of the eigenvalue equations (L5.1.56-57) shows now that the square
bracket in (L5.1.58) may be cast in the form $a\nabla_{AA'}\left(\psi_{m}^{A}
\widetilde \psi_{n}^{A'}\right)+b\nabla_{AA'}\left(\psi_{n}^{A}\widetilde
\psi_{m}^{A'}\right)$, with $a=-b={(\lambda_{n}+\lambda_{m})
\over (\lambda_{n}-\lambda_{m})}$, provided $\lambda_{n} \not = \lambda_{m}$.
Thus $\Sigma$ becomes
$$
\Sigma={i\over 2}
{(\lambda_{n}+\lambda_{m})\over (\lambda_{n}-\lambda_{m})}
\left[\int_{\partial M}d^{3}x \; \sqrt{h}\; {_{e}n_{AA'}}\psi_{m}^{A}
\widetilde \psi_{n}^{A'}
-\int_{\partial M}d^{3}x \; \sqrt{h}\; {_{e}n_{AA'}}\psi_{n}^{A}
\widetilde \psi_{m}^{A'}\right]=0 \; ,
\eqno (L5.1.59)
$$
by virtue of the local boundary conditions (L5.1.3). In the
degenerate case $\lambda_{n}=\lambda_{m}$, a linear transformation within
the degenerate eigenspace can be found such that the cross-terms again
vanish. It is indeed
well-known that every Hermitian matrix can be cast in diagonal form with
real eigenvalues. Thus the
asymptotic calculations (L5.1.50)-(L5.1.52),
and the property that $I_{E}$ can be written as a diagonal expression in
terms of a sum over eigenfunctions suggest that the Dirac action used here,
subject to local boundary conditions, can be expressed in terms of a
self-adjoint differential operator acting on fields
$\Bigr(\psi^{A}, \; {\widetilde \psi}^{A'}\Bigr)$. We shall now prove
that this is indeed the case.

So far we have seen that the framework for the formulation of
local boundary conditions involving normals and field strengths
or fields is the Euclidean regime, where one deals with Riemannian metrics.
Thus, we will pay special
attention to the conjugation of $SU(2)$ spinors in Euclidean four-space.
In fact such a conjugation will play a key role in proving self-adjointness.
For this purpose, it can be useful to recall at first
some basic results about $SU(2)$ spinors on an abstract Riemannian
three-manifold $\left(\Sigma,h\right)$. In that case, one considers
a bundle over the three-manifold, each fibre of which is isomorphic
to a two-dimensional complex vector space $W$. It is then possible to define
a nowhere vanishing antisymmetric $\epsilon_{AB}$ (the usual one of Lecture 4)
so as to raise and lower internal indices, and a positive-definite
Hermitian inner product on each fibre: $(\psi , \phi)={\overline \psi}^{A'}
G_{A'A}\phi^{A}$. The requirements of Hermiticity and positivity imply
respectively that ${\overline G}_{A'A}=G_{A'A}$, ${\overline \psi}^{A'}
G_{A'A}\psi^{A}>0$,$\forall$ $\psi^{A} \not = 0$. This $G_{A'A}$ converts
primed indices to unprimed ones, and it is given by $i\sqrt{2} \; n_{AA'}$.
Given the space H of all objects
$\alpha_{\; \; B}^{A}$ such that $\alpha_{\; \; A}^{A}=0$ and
$\left({\alpha}^{\dagger}\right)_{\; \; B}^{A}=-\alpha_{\; \; B}^{A}$,
one finds there always exists a $SU(2)$ soldering form $\sigma_{\; \; A}
^{a\; \; \; B}$ (i.e. a global isomorphism) between $H$ and the tangent
space on $\left(\Sigma,h\right)$ such that
$h^{ab}=-\sigma_{\; \; A}^{a\; \; \; B}\sigma_{\; \; B}^{b\; \; \; A}$.
Therefore one also finds $\sigma_{\; \; A}^{a\; \; \; A}=0$ and
${\left(\sigma_{\; \; A}^{a\; \; \; B}\right)}^{\dagger}=
-\sigma_{\; \; A}^{a\; \; \; B}$. One then defines $\psi^{A}$ an
$SU(2)$ spinor on $\left(\Sigma,h\right)$. A basic remark is that
$SU(2)$ transformations are those $SL(2,C)$ transformations which preserve
$n^{AA'}=n^{a}\sigma_{a}^{\; \; AA'}$, where $n^{a}=(1, 0, 0, 0)$
is the normal to
$\Sigma$. The Euclidean conjugation used here (not to be confused
with complex conjugation in Minkowski space-time) is such that
(see now section L4.1)
$$
{\left(\psi_{A} +\lambda \phi_{A}\right)}^{\dagger}=
{\psi_{A}}^{\dagger}
+{\lambda}^{*} {\phi_{A}}^{\dagger} \; \; \; \; , \; \; \; \;
{\left({\psi_{A}}^{\dagger}\right)}^{\dagger}=-\psi_{A}
\; \; \; \; ,
\eqno (L5.1.60)
$$
$$
{\epsilon_{AB}}^{\dagger}=\epsilon_{AB} \; \; \; \; , \; \; \; \;
{\left(\psi_{A}\phi_{B}\right)}^{\dagger}=
{\psi_{A}}^{\dagger}{\phi_{B}}^{\dagger}
\; \; \; \; ,
\eqno (L5.1.61)
$$
$$
{\left(\psi_{A}\right)}^{\dagger}\psi^{A}>0 \; \; \; \; , \; \; \; \;
\forall \; \psi_{A} \not =0
\; \; \; \; .
\eqno (L5.1.62)
$$
In (L5.1.60) and in the following pages, the symbol $*$ denotes complex
conjugation of scalars.
How to generalize this picture to the Euclidean four-space ? For this purpose,
let us now focus our attention on states that are pairs of
spinor fields, defining
$$
w\equiv \left(\psi^{A}, \; \widetilde \psi^{A'}\right) \; \; \; \; ,
\; \; \; \;
z\equiv \left(\phi^{A}, \; \widetilde \phi^{A'}\right)
\; \; \; \; ,
\eqno (L5.1.63)
$$
on the ball of radius $a$ in Euclidean four-space, subject always to the
boundary conditions (L5.1.3).
$w$ and $z$ are subject also to suitable
differentiability conditions, to be specified later. Let us also define
the operator $C$
$$
C\; : \left(\psi^{A}, \; \widetilde \psi^{A'}\right)\rightarrow
\left(\nabla_{\; \; B'}^{A}\widetilde \psi^{B'},
\nabla_{B}^{\; \; A'}\psi^{B}\right)
\; \; \; \; ,
\eqno (L5.1.64)
$$
and the {\it dagger} operation
$$
{\left(\psi^{A}\right)}^{\dagger}\equiv \epsilon^{AB}\delta_{BA'}
{\overline \psi}^{A'}
\; \; \; \; , \; \; \; \; \; \; \; \;
{\left({\widetilde \psi}^{A'}\right)}^{\dagger} \equiv
\epsilon^{A'B'}\delta_{B'A}{\overline {\widetilde \psi}}^{A}
\; \; \; \; .
\eqno (L5.1.65)
$$
The consideration of $C$ is suggested of course by the action (L5.1.54) and
by the eigenvalue equations (L5.1.56)-(L5.1.57).
In (L5.1.65), $\delta_{BA'}$ is an identity matrix playing the same role of
$G_{AA'}$ for $SU(2)$ spinors on $\left(\Sigma,h\right)$, so that
$\delta_{BA'}$ is preserved by $SU(2)$ transformations.
Moreover, the {\it bar} symbol
$\overline {\psi^{A}}={\overline \psi}^{A'}$ denotes the usual complex
conjugation of $SL(2,C)$ spinors. Hence one finds
$$
{\left({\left(\psi^{A}\right)}^{\dagger}\right)}^{\dagger}=
\epsilon^{AC}\delta_{CB'}\overline {\left(\psi^{B \dagger}\right)'}
=\epsilon^{AC}\delta_{CB'}\epsilon^{B'D'}\delta_{D'F}\psi^{F}
=-\psi^{A}
\; \; \; \; ,
\eqno (L5.1.66)
$$
in view of the definition of $\epsilon^{AB}$. Thus, the
{\it dagger} operation defined in (L5.1.65) is anti-involutory, because,
when applied twice to $\psi^{A}$, it yields $-\psi^{A}$.

As anticipated
after (L5.1.58), from now on we study commuting spinors, for simplicity
of exposition of the self-adjointness. It is easy to check that the
{\it dagger}, also called in the literature Euclidean conjugation
(Lecture 4), satisfies all properties (L5.1.60)-(L5.1.62).
We can now define the scalar product
$$
(w,z)\equiv \int_{M}\left[\psi_{A}^{\dagger}
\phi^{A}+{\widetilde \psi_{A'}}^{\dagger}
{\widetilde \phi}^{A'}\right]\sqrt{g} \; d^{4}x
\; \; \; \; .
\eqno (L5.1.67)
$$
This is indeed a scalar product, because it satisfies all following
properties for all vectors $u$, $v$, $w$ and $\forall \lambda \in \; C$ :
$$
(u,u)>0 \; \; \; \; , \; \; \; \; \forall u \not =0
\; \; \; \; ,
\eqno (L5.1.68)
$$
$$
(u,v+w)=(u,v)+(u,w)
\; \; \; \; ,
\eqno (L5.1.69)
$$
$$
(u,\lambda v)=\lambda(u,v) \; \; \; \; , \; \; \; \;
(\lambda u,v)={\lambda}^{*}(u,v)
\; \; \; \; ,
\eqno (L5.1.70)
$$
$$
(v,u)={(u,v)}^{*}
\; \; \; \; .
\eqno (L5.1.71)
$$
We are now aiming to check that $C$ or $iC$ is a symmetric operator,
i.e. that
$\left(Cz,w\right)=\left(z,Cw\right)$ or $\left(iCz,w\right)=
\left(z,iCw\right), \forall z,w$. This will be used in the course of
proving further that the symmetric operator has self-adjoint extensions.
In order to prove this result it is clear, in view
of (L5.1.67), we need to know the properties of the spinor covariant
derivative acting on $SU(2)$ spinors. In the case of $SL(2,C)$ spinors
it is known this derivative
is a linear, torsion-free map $\nabla_{AA'}$ which satisfies the Leibniz
rule, annihilates $\epsilon_{AB}$ and is real (i.e. $\psi=\nabla_{AA'}
\theta \Rightarrow {\overline \psi}=\nabla_{AA'}{\overline \theta}$).
Moreover, we know that (cf. (L1.4.4))
$$
\nabla^{AA'}=e_{\; \; \; \; \; \; \mu}^{AA'}\nabla^{\mu}=
e_{\; \; \mu}^{a} \; \sigma_{a}^{\; \; AA'}\nabla^{\mu}
\; \; \; \; .
\eqno (L5.1.72)
$$
In Euclidean four-space, we use both (L5.1.72) and the relation
$$
\sigma_{\mu AC'} \; \sigma_{\nu B}^{\; \; \; \; \; C'}
+\sigma_{\nu BC'} \; \sigma_{\mu A}^{\; \; \; \; \; C'}
= \delta_{\mu \nu}\epsilon_{AB}
\; \; \; \; ,
\eqno (L5.1.73)
$$
where $\delta_{\mu\nu}$ has signature $(+,+,+,+)$. This implies that
$\sigma_{0}=-{i\over \sqrt{2}}I$, $\sigma_{i}={\Sigma_{i}\over \sqrt{2}}$,
$\forall i=1, 2, 3$, where $\Sigma_{i}$ are the Pauli matrices. Now, in
view of (L5.1.64) and (L5.1.67) one finds
$$
(Cz,w)=
\int_{M}{\left(\nabla_{AB'}\phi^{A}\right)}^{\dagger}
{\widetilde \psi}^{B'}
\sqrt{g}d^{4}x
+\int_{M}{\left(\nabla_{BA'}{\widetilde \phi}^{A'}\right)}^{\dagger}
\psi^{B}\sqrt{g}d^{4}x
\; \; ,
\eqno (L5.1.74)
$$
whereas, using the Leibniz rule in computing $\nabla_{\; \; B'}^{A}
\left(\phi_{A}^{\dagger}{\widetilde \psi}^{B'}\right)$ and
$\nabla_{B}^{\; \; A'}
\left(\left({\widetilde \phi}_{A'}\right)^{\dagger}\psi^{B}
\right)$, and integrating by parts, one finds
$$ \eqalignno{
(z,Cw)&=
\int_{M}\left(\nabla_{AB'}\phi^{A \dagger}\right)
{\widetilde \psi}^{B'}\sqrt{g}d^{4}x
+\int_{M}\left(\nabla_{BA'}
\left({\widetilde \phi}^{A'}\right)^{\dagger}\right)
\psi^{B}\sqrt{g}d^{4}x \cr
&-\int_{\partial M}(_{e}n_{AB'})\phi^{A \dagger}{\widetilde \psi}^{B'}
\sqrt{h}d^{3}x
-\int_{\partial M}(_{e}n_{BA'}){\left({\widetilde \phi}^{A'}\right)}
^{\dagger}\psi^{B}\sqrt{h}d^{3}x.
&(L5.1.75)\cr}
$$
Now we use (L5.1.65), the identity
$$
{\left(_{e}n^{AA'}\phi_{A}\right)}^{\dagger}=
\epsilon^{A'B'}\delta_{B'C}\;{\overline {_{e}n^{DC'}}}\;
{\overline {\phi_{D}}}=-\epsilon^{A'B'}\delta_{B'C}\left(_{e}n^{CD'}\right)
{\overline \phi}_{D'}
\; \; \; \; ,
\eqno (L5.1.76)
$$
section L4.1 and the boundary
conditions on $S^{3}$: $\sqrt{2}\; {_{e}n^{CB'}}\psi_{C}
={\widetilde \psi}^{B'}$, $\sqrt{2}\; {_{e}n^{AA'}}\phi_{A}
={\widetilde \phi}^{A'}$. In so doing, the sum of the
boundary terms in (L5.1.75) is found to vanish. This implies in turn that
equality of the volume integrands is sufficient to show that
$(Cz,w)$ and $(z,Cw)$ are equal. For
example, one finds in flat space, using also (L5.1.65) :
${\left(\nabla_{BA'}{\widetilde \phi}^{A'}\right)}^{\dagger}=\delta_{BF'}
{\overline \sigma}_{\; \; C}^{F'\; \; a}\partial_{a}
\left({\overline {\widetilde \phi}}^{C}\right)$, whereas :
$\left(\nabla_{BA'}\left({\widetilde \phi}^{A'}\right)^{\dagger}\right)=
-\delta_{CF'}\sigma_{B}^{\; \; F'a}\partial_{a}
\left({\overline {\widetilde \phi}}^{C}\right)$. In other words, we are led
to study the condition
$$
\delta_{BF'} \; {\overline \sigma}_{\; \; \; C}^{F' \; \; \; a}=
\pm \delta_{BF'} \; \sigma_{C}^{\; \; F'a}
\; \; \; \; ,
\eqno (L5.1.77)
$$
$\forall$ $a=0, 1, 2, 3$. Now, using the relations
$$
\sqrt{2}\; \sigma_{AA'}^{\; \; \; \; \; 0}=
\pmatrix {-i&0\cr 0&-i\cr} \; \; \; \; , \; \; \; \;
\sqrt{2}\; \sigma_{AA'}^{\; \; \; \; \; 1}=
\pmatrix {0&1\cr 1&0\cr}
\; \; \; \; ,
\eqno (L5.1.78)
$$
$$
\sqrt{2}\; \sigma_{AA'}^{\; \; \; \; \; 2}=
\pmatrix {0&-i\cr i&0\cr} \; \; \; \; , \; \; \; \;
\sqrt{2}\; \sigma_{AA'}^{\; \; \; \; \; 3}=
\pmatrix {1&0\cr 0&-1\cr}
\; \; \; \; ,
\eqno (L5.1.79)
$$
$$
\sigma_{\; \; A'}^{A\; \; \; a}=\epsilon^{AB}
\sigma_{BA'}^{\; \; \; \; \; \; a}
\; \; \; \; , \; \; \; \;
\sigma_{A}^{\; \; A'a}=
-\sigma_{AB'}^{\; \; \; \; \; \; a} \; \epsilon^{B'A'}
\; \; \; \; ,
\eqno (L5.1.80)
$$
one finds that the complex conjugate of $\sigma_{\; \; A'}^{A\; \; \; a}$ is
always equal to $\sigma_{A}^{\; \; A'a}$, which is not in agreement with
the choice of the $(-)$ sign on the right-hand side of (L5.1.77). This
implies in turn that the symmetric operator we are looking for is $iC$,
where $C$ has been defined in (L5.1.64). The generalization to a curved
four-dimensional Riemannian space is
obtained via the relation $e_{\; \; \; \; \; \mu}^{AA'}=e_{\; \; \mu}^{a}
\; \sigma_{a}^{\; \; AA'}$ (see problem (P.14)).

Now, it is known that every symmetric
operator has a closure, and the operator and its closure have the same closed
extensions. Moreover, a closed symmetric operator on a
Hilbert space is self-adjoint if and only if its spectrum is a subset of the
real axis. To prove
self-adjointness for our boundary-value problem,
we may recall an important result due to von
Neumann [4]. This theorem states that, given a
symmetric operator $A$ with domain $D(A)$, if a map $F : D(A)\rightarrow
D(A)$ exists such that
$$
F(\alpha w + \beta z)=\alpha^{*}F(w)+\beta^{*}F(z)
\; \; \; \; ,
\eqno (L5.1.81)
$$
$$
(w,w)=(Fw,Fw)
\; \; \; \; ,
\eqno (L5.1.82)
$$
$$
F^{2}= \pm I
\; \; \; \; ,
\eqno (L5.1.83)
$$
$$
FA=AF
\; \; \; \; ,
\eqno (L5.1.84)
$$
then $A$ has self-adjoint extensions. In our case, denoting by $D$ the
operator (cf. (L5.1.65))
$$
D : \left(\psi^{A}, \; {\widetilde \psi}^{A'}\right)\rightarrow
\left( {\Bigr(\psi^{A}\Bigr)}^{\dagger},
{\left({\widetilde \psi}^{A'}\right)}^{\dagger}\right)
\; \; \; \; ,
\eqno (L5.1.85)
$$
let us focus our attention on the operators $F=iD$ and $A=iC$. The
operator $F$ maps indeed $D(A)$ to $D(A)$. In fact, bearing in mind the
definitions
$$
G\equiv \left \{ \varphi =\Bigr(\phi^{A}, \; {\widetilde \phi}^{A'}\Bigr) :
\varphi \; {\rm is} \; {\rm at} \; {\rm least} \;
C^{1} \right \}
\; \; \; \; ,
\eqno (L5.1.86)
$$
$$
D(A)\equiv \left \{ \varphi \in G : \sqrt{2} \; {_{e}n^{AA'}\phi_{A}}
=\epsilon \; {\widetilde \phi}^{A'} \;
{\rm on} \; S^{3} \right \}
\; \; \; \; ,
\eqno (L5.1.87)
$$
one finds that $F$ maps $\Bigr(\phi^{A}, \; {\widetilde \phi}^{A'}\Bigr)$ to
$\Bigr(\beta^{A}, \; {\widetilde \beta}^{A'}\Bigr)=
\left(i {\Bigr(\phi^{A}\Bigr)}^{\dagger},
i{\Bigr({\widetilde \phi}^{A'}\Bigr)}^{\dagger}\right)$ such that
$$
\sqrt{2} \; {_{e}n^{AA'}\beta_{A}}=\gamma \; {\widetilde \beta}^{A'}
\; {\rm on}
\; S^{3}
\; \; \; \; ,
\eqno (L5.1.88)
$$
where $\gamma =\epsilon^{*}$. The boundary condition (L5.1.88) is clearly
of the type which occurs in (L5.1.87) provided $\epsilon$ is real,
and the differentiability of
$\Bigr(\beta^{A}, \; {\widetilde \beta}^{A'}\Bigr)$ is not affected by the
action of $F$ (cf. (L5.1.85)). In deriving (L5.1.88), we have used the
result for ${\Bigr({_{e}n^{AA'}\phi_{A}}\Bigr)}^{\dagger}$
obtained in (L5.1.76).
It is worth emphasizing that the requirement of self-adjointness
enforces the choice of a real function $\epsilon$, which is a constant
in our case. Moreover, in view of
(L5.1.66), one immediately sees that (L5.1.81) and (L5.1.83) hold when
$F=iD$, provided we write (L5.1.83) as $F^{2}=-I$. This is indeed a crucial
point which deserves special attention. Condition (L5.1.83) is written in
Reed and Simon 1975 as $F^{2}=I$, and examples are later given (see page
144 therein) where $F$ is complex conjugation. But we are formulating
our problem in the Euclidean regime, where we have seen that the only
possible conjugation is the {\it dagger} operation, which is
anti-involutory on spinors with an
odd number of indices. Thus, we are here
generalizing von Neumann's theorem in the following way. If $F$ is a map
$D(A)\rightarrow D(A)$ which satisfies
(L5.1.81)-(L5.1.84), then the same is clearly
true of ${\widetilde F}=-iD=-F$. Hence
$$
-F \; D(A) \subseteq D(A)
\; \; \; \; ,
\eqno (L5.1.89)
$$
$$
F \; D(A) \subseteq D(A)
\; \; \; \; .
\eqno (L5.1.90)
$$
Acting with $F$ on both sides of (L5.1.89), one finds
$$
D(A) \subseteq F \; D(A)
\; \; \; \; ,
\eqno (L5.1.91)
$$
using the property $F^{2}=-I$. But then the relations (L5.1.90-91)
imply that $F \; D(A)=D(A)$, so that $F$ takes $D(A)$ onto $D(A)$ also in
the case of the anti-involutory Euclidean conjugation that we called
{\it dagger}. Comparison with the proof presented at the beginning of
page 144 in Reed and Simon 1975 shows that this is all what we need so as
to generalize von Neumann's theorem to the Dirac operator acting on
$SU(2)$ spinors in Euclidean four-space (one later uses properties
(L5.1.84), (L5.1.81) and (L5.1.82) as well in completing the proof).

It remains to verify conditions (L5.1.82) and (L5.1.84). First, note that
$$ \eqalignno{
\left(Fw,Fw\right)&=
\left(iDw,iDw\right) \cr
&=\int_{M}{\left(i\psi_{A}^{\dagger}\right)}^{\dagger}
i {\Bigr(\psi^{A}\Bigr)}^{\dagger} \sqrt{g} \; d^{4}x
+\int_{M}{\left(i{\widetilde \psi}_{A'}^{\dagger}\right)}^{\dagger}
i{\left({\widetilde \psi}^{A'}\right)^{\dagger}}
\sqrt{g} \; d^{4}x\cr
&=\left(w,w\right)
\; \; \; \; ,
&(L5.1.92)\cr}
$$
where we have used (L5.1.66)-(L5.1.67) and
the commutation property of our spinors. Second, one finds
$$
FAw=
\left(iD\right)\left(iC\right)w
=i{\left[i\left(\nabla_{\; \; B'}^{A}{\widetilde \psi}^{B'},
\nabla_{B}^{\; \; A'}\psi^{B}\right)\right]}^{\dagger}
={\left(\nabla_{\; \; B'}^{A}{\widetilde \psi}^{B'},
\nabla_{B}^{\; \; A'}\psi^{B}\right)}^{\dagger} \; ,
\eqno (L5.1.93)
$$
$$
AFw=
\left(iC\right)\left(iD\right)w
=iCi\left(\psi^{A \dagger},
{\left({\widetilde \psi}^{A'}\right)}^{\dagger}\right)
=-\left(\nabla_{\; \; B'}^{A}
{\left({\widetilde \psi}^{B'}\right)}^{\dagger},
\nabla_{B}^{\; \; A'}\psi^{B \dagger}\right) \; ,
\eqno (L5.1.94)
$$
which in turn implies that also (L5.1.84) holds in view of
what we found just before (L5.1.77) and after (L5.1.80).
To sum up, we have proved that the operator
$iC$ arising in our boundary-value problem is symmetric and has self-adjoint
extensions (see now problem (P.15)).
Hence the eigenvalues of $iC$ are real, and the eigenvalues $\lambda_{n}$
of $C$ are purely imaginary. This is in agreement with (L5.1.48), because
there $E=i\lambda_{n}=-Im(\lambda_{n})$. Further, they occur in equal and
opposite pairs in our example of flat Euclidean four-space bounded by a
three-sphere. Hence the fermionic one-loop path integral
$K$ is formally given by the dimensionless product
$$
K=
{\prod_{n}\left({\mid \lambda_{n} \mid \over {\widetilde \mu}}\right)}
\; \; \; \; ,
\eqno (L5.1.95)
$$
where ${\widetilde \mu}$ is a normalization constant with dimensions of mass,
and the right-hand side of (L5.1.95) should be interpreted as
explained in section 4.3 of [4].
This one-loop $K$ can be regularized by using the zeta-function
$$
\zeta_{M}(s)\equiv \sum_{n,k}d_{k}(n)
{\Bigr({\mid \lambda_{n,k} \mid}^{2}\Bigr)}^{-s}
\; \; \; \; .
\eqno (L5.1.96)
$$
With this more
accurate notation, we write $d_{k}(n)$ for the degeneracy of the eigenvalues
${\mid \lambda_{n,k} \mid}$.
Two indices are used because, for each
value of the integer $n$, there are infinitely many eigenvalues labeled
by $k$.
\vskip 10cm
\centerline {\bf L5.2 $\zeta$-Function Calculations for the Dirac Operator}
\vskip 0.3cm
In section L5.1 we derived the eigenvalue condition (L5.1.48) which, setting
$a=1$ for simplicity, can be written in the non-linear form (L5.1.47)
$$
F(E) \equiv {\Bigr[J_{n+1}(E)\Bigr]}^{2}-{\Bigr[J_{n+2}(E)\Bigr]}^{2}=0
\; \; \; \; , \; \; \; \;
\forall n \geq 0
\; \; \; \; .
\eqno (L5.2.1)
$$
The function $F$ is the product of the entire functions (functions analytic
in the whole complex plane)
$F_{1} \equiv J_{n+1}-J_{n+2}$ and $F_{2} \equiv J_{n+1}+J_{n+2}$,
which can be written in the form
$$
F_{1}(z) \equiv J_{n+1}(z)-J_{n+2}(z)=\gamma_{1}z^{(n+1)}e^{g_{1}(z)}
\prod_{i=1}^{\infty}\left(1-{z\over \mu_{i}}\right)e^{z\over \mu_{i}}
\; \; \; \; ,
\eqno (L5.2.2)
$$
$$
F_{2}(z) \equiv J_{n+1}(z)+J_{n+2}(z)=\gamma_{2}z^{(n+1)}e^{g_{2}(z)}
\prod_{i=1}^{\infty}\left(1-{z\over \nu_{i}}\right)e^{z\over \nu_{i}}
\; \; \; \; .
\eqno (L5.2.3)
$$
In (L5.2.2-3), $\gamma_{1}$ and $\gamma_{2}$ are constants,
$g_{1}$ and $g_{2}$ are entire functions, the
$\mu_{i}$ are the (real) zeros of $F_{1}$ and the $\nu_{i}$ are the (real)
zeros of $F_{2}$. In fact,
$F_{1}$ and $F_{2}$ are entire functions whose
canonical product has genus 1. Namely, in light of the asymptotic behaviour
of the eigenvalues (cf. (L5.1.50-51)), we know that
$\sum_{i=1}^{\infty}{1\over \mid \mu_{i} \mid}=\infty$ and
$\sum_{i=1}^{\infty}{1\over \mid \nu_{i} \mid}=\infty$, whereas
$\sum_{i=1}^{\infty}{1\over {\mid \mu_{i} \mid}^{2}}$ and
$\sum_{i=1}^{\infty}{1\over {\mid \nu_{i} \mid}^{2}}$ are convergent. This is
why $e^{z\over \mu_{i}}$ and $e^{z\over \nu_{i}}$ must appear in
(L5.2.2-3), which are called the canonical-product representations of
$F_{1}$ and $F_{2}$. The genus of the canonical product for $F_{1}$ is the
minimum integer $h$ such that $\sum_{i=1}^{\infty}{1\over
{\mid \mu_{i} \mid}^{h+1}}$ converges, and similarly for $F_{2}$,
replacing $\mu_{i}$ with $\nu_{i}$. If the genus is equal to $1$, this
ensures that no higher powers of ${z\over \mu_{i}}$ and
${z\over \nu_{i}}$ are needed in the argument of the exponential.
However, there is a very simple
relation between $\mu_{i}$ and $\nu_{i}$. In fact, if $J_{n+1}$ is an
even function (i.e. if $n$ is odd),
then $J_{n+2}$ is an odd function, and viceversa. Thus
$$
J_{n+1}(-z)-J_{n+2}(-z)=J_{n+1}(z)+J_{n+2}(z) \; \;
{\rm if} \; \; n \; \;
{\rm is} \; \; {\rm odd}
\; \; ,
\eqno (L5.2.4)
$$
$$
J_{n+1}(-z)-J_{n+2}(-z)=-J_{n+1}(z)-J_{n+2}(z) \; \;
{\rm if} \; \; n \; \;
{\rm is} \; \; {\rm even}
\; \; .
\eqno (L5.2.5)
$$
The relations (L5.2.4-5) imply that the zeros of $F_{1}$ are minus
the zeros of $F_{2}$ : $\mu_{i}=-\nu_{i}$, $\forall i$. This is of course
just the symmetry property of the eigenvalues pointed out in section L5.1,
following (L5.1.94). This implies in turn that (cf. (L5.2.1))
$$
F(z)={\widetilde \gamma} z^{2(n+1)}e^{(g_{1}+g_{2})(z)}
\prod_{i=1}^{\infty}\left(1-{z^{2}\over \mu_{i}^{2}}\right)
\; \; \; \; ,
\eqno (L5.2.6a)
$$
where ${\widetilde \gamma} \equiv
\gamma_{1}\gamma_{2}$, and $\mu_{i}^{2}$
are the positive zeros of $F(z)$. It turns out that the function
$(g_{1}+g_{2})$ is actually a constant, so that we can write
$$
F(z)=F(-z)=
\gamma z^{2(n+1)}\prod_{i=1}^{\infty}\left(1-{z^{2}\over \mu_{i}^{2}}
\right)
\; \; \; \; ,
\eqno (L5.2.6b)
$$
In fact, the following theorem holds [4].
\vskip 0.3cm
\noindent
{\bf Theorem L5.2.1} Let $f$ be an entire function. If $\forall \epsilon >0$
$\exists \; A_{\epsilon}$ such that
$$
\log \max \Bigr \{1, \mid f(z) \mid \Bigr \} \leq A_{\epsilon}
{\mid z \mid}^{1+ \epsilon}
\; \; \; \; ,
\eqno (L5.2.7)
$$
then $f$ can be expressed in terms of its zeros as
$$
f(z)=e^{A+Bz}\prod_{i=1}^{\infty}\left(1-{z\over \widetilde \nu_{i}}\right)
e^{z\over \widetilde \nu_{i}}
\; \; \; \; .
\eqno (L5.2.8)
$$
If we now apply theorem L5.2.1 to the functions $F_{1}(z)z^{-(n+1)}$ and
$F_{2}(z)z^{-(n+1)}$ (cf. (L5.2.2)-(L5.2.3)),
we discover that the well-known formula
$$
J_{n}(z)={i^{-n}\over \pi}\int_{0}^{\pi}
e^{iz \cos \theta}\cos (n\theta) \; d \theta
\; \; \; \; ,
\eqno (L5.2.9)
$$
leads to the fulfillment of (L5.2.7) for $F_{1}(z)z^{-(n+1)}$ and
$F_{2}(z)z^{-(n+1)}$. Hence these functions satisfy (L5.2.8) with
constants $A_{1}$ and $B_{1}$ for $F_{1}(z)z^{-(n+1)}$, and constants
$A_{2}$ and $B_{2}=-B_{1}$ for $F_{2}(z)z^{-(n+1)}$. The fact that
$B_{2}=-B_{1}$ is well-understood
if we look again at (L5.2.4)-(L5.2.5).
I am most indebted to Dr. R. Pinch for providing this argument.

The application of the method of section
7.3 of [4] (see now the appendix)
makes it necessary to re-express the function $F$ in
(L5.2.1) in terms of Bessel functions and their first derivatives of the same
order. Thus, defining $m\equiv n+2$, and using the identity
$$
J_{l}'(x)=J_{l-1}(x)-{l\over x}J_{l}(x)
\; \; \; \; ,
\eqno (L5.2.10)
$$
one finds
$$ \eqalignno{
J_{m-1}^{2}(x)-J_{m}^{2}(x)&=\left(J_{m}'+{m\over x}J_{m}-J_{m}\right)
\left(J_{m}'+{m\over x}J_{m}+J_{m}\right)\cr
&={J_{m}'}^{2}+\left({m^{2}\over x^{2}}-1\right)J_{m}^{2}+
2{m\over x}J_{m}J_{m}'
\; \; \; \; .
& (L5.2.11)\cr}
$$
This is why, making the analytic continuation $x\rightarrow ix$, and then
defining $\alpha_{m}\equiv
\sqrt{m^{2}+x^{2}}$ and using the uniform asymptotic
expansions of Bessel functions and their first derivatives we obtain
$$
J_{m-1}^{2}(ix)-J_{m}^{2}(ix) \sim {(ix)^{2(m-1)}\over 2\pi}\alpha_{m}
e^{2\alpha_{m}}e^{-2m \log(m+\alpha_{m})}
\left[\Sigma_{1}^{2}+\Sigma_{2}^{2}
+2{m\over \alpha_{m}}\Sigma_{1}\Sigma_{2}
\right] \; .
\eqno (L5.2.12)
$$
Thus we only need to multiply the left-hand side of (L5.2.12)
by $(ix)^{-2(m-1)}$ so as to get a function of the kind
${\alpha_{m} \over 2\pi}e^{2\alpha_{m}}
e^{-2m \log(m+\alpha_{m})}{\widetilde \Sigma}$,
similarly to section 7.3 of [4]. We now define
$$
t \equiv {m \over \alpha_{m}} \; \; \; \; , \; \; \; \;
{\widetilde \Sigma}\equiv \Sigma_{1}^{2}+\Sigma_{2}^{2}+2t\Sigma_{1}\Sigma_{2}
\; \; \; \; ,
\eqno (L5.2.13)
$$
and study the asymptotic expansion of $\log({\widetilde \Sigma})$
in the relation
$$
\log \left[(ix)^{-2(m-1)} \left(J_{m-1}^{2}-J_{m}^{2}\right)(ix)\right]
\sim  -\log(2\pi)+\log(\alpha_{m})+2\alpha_{m}
-2m \log(m+\alpha_{m})+
\log({\widetilde \Sigma}).
\eqno (L5.2.14)
$$
This is obtained bearing in mind that the functions $\Sigma_{1}$
and $\Sigma_{2}$ in (L5.2.12)-(L5.2.13)
have asymptotic series given by
$$
\Sigma_{1} \sim \sum_{k=0}^{\infty}{u_{k}({m\over \alpha_{m}})\over
m^{k}} \sim \left[1+{a_{1}(t)\over \alpha_{m}}
+{a_{2}(t)\over \alpha_{m}^{2}}
+{a_{3}(t)\over \alpha_{m}^{3}} +...\right]
\; \; \; \; ,
\eqno (L5.2.15)
$$
$$
\Sigma_{2} \sim \sum_{k=0}^{\infty}{v_{k}({m\over \alpha_{m}})\over
m^{k}} \sim \left[1+{b_{1}(t)\over \alpha_{m}}
+{b_{2}(t)\over \alpha_{m}^{2}}
+{b_{3}(t)\over \alpha_{m}^{3}} +...\right]
\; \; \; \; ,
\eqno (L5.2.16)
$$
so that
$$
a_{i}(t)={u_{i}({m\over \alpha_{m}})\over
({m\over \alpha_{m}})^{i}}={u_{i}(t)\over
t^{i}} \; \; \; \; , \; \; \; \; \;
b_{i}(t)={v_{i}({m\over \alpha_{m}})\over
({m\over \alpha_{m}})^{i}}={v_{i}(t)\over
t^{i}} \; \; \; \; , \; \; \; \; \forall i \geq 0
\; \; \; \; ,
\eqno (L5.2.17)
$$
where $u_{i}(t)$ and $v_{i}(t)$ are the polynomials already defined in
(4.4.17)-(4.4.22) of [4].
The relations (L5.2.13)-(L5.2.16) lead to the asymptotic expansion
$$
{\widetilde \Sigma} \sim \left[c_{0}+{c_{1}\over \alpha_{m}}
+{c_{2}\over \alpha_{m}^{2}}
+{c_{3}\over \alpha_{m}^{3}}+... \right]
\; \; \; \; ,
\eqno (L5.2.18)
$$
where
$$
c_{0}=2(1+t) \; \; \; \; , \; \; \; \; c_{1}=2(1+t)(a_{1}+b_{1})
\; \; \; \; ,
\eqno (L5.2.19)
$$
$$
c_{2}=a_{1}^{2}+b_{1}^{2}+2(1+t)(a_{2}+b_{2})+2ta_{1}b_{1}
\; \; \; \; ,
\eqno (L5.2.20)
$$
$$
c_{3}=2(1+t)(a_{3}+b_{3})+2(a_{1}a_{2}+b_{1}b_{2})+2t(a_{1}b_{2}+a_{2}b_{1})
\; \; \; \; .
\eqno (L5.2.21)
$$
Higher-order terms have not been computed in (L5.2.18) because they do not
affect the result for $\zeta(0)$, as we will prove in detail in section L5.6.
Since $t$ is variable, it is necessary to define
$$
\Sigma \equiv {{\widetilde \Sigma}\over c_{0}} \sim \left[
1+{({c_{1}\over c_{0}})\over \alpha_{m}}
+{({c_{2}\over c_{0}})\over \alpha_{m}^{2}}
+{({c_{3}\over c_{0}})\over \alpha_{m}^{3}}+... \right]
\; \; \; \; .
\eqno (L5.2.22)
$$
We can now expand $\log(\Sigma)$ according to the usual algorithm
valid as $\omega \rightarrow 0$
$$
\log(1+\omega)= \omega-{{\omega}^{2}\over 2}+{{\omega}^{3}\over 3}
-{{\omega}^{4}\over 4}
+{{\omega}^{5}\over 5}+...
\; \; \; \; .
\eqno (L5.2.23)
$$
Hence we find
$$
\log({\widetilde \Sigma})=\Bigr[\log(c_{0})+\log(\Sigma)\Bigr]\sim
\left[\log(c_{0})+{A_{1}\over \alpha_{m}}
+{A_{2}\over \alpha_{m}^{2}}+
{A_{3}\over \alpha_{m}^{3}}+... \right]
\; \; \; \; ,
\eqno (L5.2.24)
$$
where
$$
A_{1}=\left({c_{1}\over c_{0}}\right) \; \; \; \; \; \; , \; \; \; \; \; \;
A_{2}=\left({c_{2}\over c_{0}}\right)-
{{\left({c_{1}\over c_{0}}\right)}^{2}\over 2}
\; \; \; \; \; \; ,
\eqno (L5.2.25)
$$
$$
A_{3}=\left({c_{3}\over c_{0}}\right)-\left({c_{1}\over c_{0}}\right)
\left({c_{2}\over c_{0}}\right)+
{{\left({c_{1}\over c_{0}}\right)}^{3}\over 3}
\; \; \; \; .
\eqno (L5.2.26)
$$
Using (4.4.17)-(4.4.22) of [4] and (L5.2.17)-(L5.2.26),
one finds after a lengthy
calculation the following fundamental result:
$$
A_{1}=\sum_{r=0}^{2}k_{1r}t^{r} \; \; \; \; , \; \; \; \;
A_{2}=\sum_{r=0}^{4}k_{2r}t^{r}
\; \; \; \; , \; \; \; \;
A_{3}=\sum_{r=0}^{6}k_{3r}t^{r}
\; \; \; \; ,
\eqno (L5.2.27)
$$
where
$$
k_{10}=-{1\over 4} \; \; \; \; , \; \; \; \; k_{11}=0
\; \; \; \; , \; \; \; \;
k_{12}={1\over 12}
\; \; \; \; ,
\eqno (L5.2.28)
$$
$$
k_{20}=0 \; \; \; \; , \; \; \; \; k_{21}=-{1\over 8}
\; \; \; \; , \; \; \; \;
k_{22}=k_{23}={1\over 8} \; \; \; \; , \; \; \; \;
k_{24}=-{1\over 8}
\; \; \; \; ,
\eqno (L5.2.29)
$$
$$
k_{30}={5\over 192} \; \; \; \; , \; \; \; \;
k_{31}=-{1\over 8} \; \; \; \; , \; \; \; \;
k_{32}={9\over 320} \; \; \; \; , \; \; \; \;
k_{33}={1\over 2}
\; \; \; \; ,
\eqno (L5.2.30)
$$
$$
k_{34}=-{23\over 64} \; \; \; \; , \; \; \; \;
k_{35}=-{3\over 8} \; \; \; \; , \; \; \; \;
k_{36}={179\over 576}
\; \; \; \; .
\eqno (L5.2.31)
$$
The relations (L5.2.13-14), (L5.2.19) and (L5.2.27-31) finally lead to the
formula
$$
\log \left[(ix)^{-2(m-1)}\left(J_{m-1}^{2}-J_{m}^{2}\right)(ix)\right]
\sim \sum_{i=1}^{5}S_{i}(m,\alpha_{m}(x))+ \;
{\rm higher-order} \; {\rm terms} \; ,
\eqno (L5.2.32)
$$
where
$$
S_{1}(m,\alpha_{m}(x))\equiv -\log(\pi)+2\alpha_{m}
\; \; \; \; ,
\eqno (L5.2.33)
$$
$$
S_{2}(m,\alpha_{m}(x))\equiv
-(2m-1)\log(m+\alpha_{m})
\; \; \; \; ,
\eqno (L5.2.34)
$$
$$
S_{3}(m,\alpha_{m}(x))\equiv
\sum_{r=0}^{2}k_{1r}m^{r}\alpha_{m}^{-r-1}
\; \; \; \; ,
\eqno (L5.2.35)
$$
$$
S_{4}(m,\alpha_{m}(x))\equiv
\sum_{r=0}^{4}k_{2r}m^{r}\alpha_{m}^{-r-2}
\; \; \; \; ,
\eqno (L5.2.36)
$$
$$
S_{5}(m,\alpha_{m}(x))\equiv
\sum_{r=0}^{6}k_{3r}m^{r}\alpha_{m}^{-r-3}
\; \; \; \; .
\eqno (L5.2.37)
$$
This is why, defining
$$
W_{\infty}\equiv \sum_{m=0}^{\infty}\left(m^{2}-m\right)
{\left({1\over 2x}{d\over dx}\right)}^{3}\left[\sum_{i=1}^{5}
S_{i}(m,\alpha_{m}(x))\right]=\sum_{i=1}^{5}W_{\infty}^{(i)}
\; \; \; \; ,
\eqno (L5.2.38)
$$
we find
$$ \eqalignno{
\Gamma(3)\zeta(3,x^{2})&=\sum_{m=2}^{\infty}\left(m^{2}-m\right)
{\left({1\over 2x}{d\over dx}\right)}^{3}\log \left[(ix)^{-2(m-1)}
\left(J_{m-1}^{2}-J_{m}^{2}\right)(ix)\right]\cr
& \sim W_{\infty}+\sum_{n=5}^{\infty}{\hat q}_{n}x^{-2-n}
\; \; \; \; .
&(L5.2.39)\cr}
$$

In deriving the fundamental formulae
(L5.2.29)-(L5.2.31), the relevant
intermediate steps are the following
(see again (L5.2.17)-(L5.2.26)):
$$
{c_{2}\over c_{0}}=a_{2}+b_{2}+a_{1}b_{1}+
{(a_{1}-b_{1})^{2}\over 2(1+t)}
\; \; \; \; ,
\eqno (L5.2.40)
$$
$$
(a_{1}-b_{1})^{2}={1\over 4}(1-t)^{2}(1+t)^{2}
\; \; \; \; ,
\eqno (L5.2.41)
$$
$$
{c_{2}\over c_{0}}={1\over 32}-{t\over 8}+{5\over 48}t^{2}+{t^{3}\over 8}
-{35\over 288}t^{4}
\; \; \; \; ,
\eqno (L5.2.42)
$$
$$
-\left({c_{1}\over c_{0}}\right)
\left({c_{2}\over c_{0}}\right)
+{{\left({c_{1}\over c_{0}}\right)}^{3}\over 3}={1\over 384}-{t\over 32}
+{11\over 384}t^{2}+{t^{3}\over 24}-{47\over 1152}t^{4}-{t^{5}\over 96}
+{107\over 10368}t^{6} \; ,
\eqno (L5.2.43)
$$
$$
{c_{3}\over c_{0}}=a_{3}+b_{3}+a_{1}b_{2}+a_{2}b_{1}+
{\Bigr[(a_{2}-b_{2})(a_{1}-b_{1})\Bigr]\over (1+t)}
\; \; \; \; ,
\eqno (L5.2.44)
$$
$$
a_{3}+b_{3}+a_{1}b_{2}+a_{2}b_{1}=-{9\over 128}+{293\over 640}t^{2}
-{787\over 1152}t^{4}+{3115\over 10368}t^{6}
\; \; \; \; ,
\eqno (L5.2.45)
$$
$$
{\Bigr[(a_{2}-b_{2})(a_{1}-b_{1})\Bigr]\over (1+t)}
={3\over 32}-{3\over 32}t-{11\over 24}t^{2}
+{11\over 24}t^{3}+{35\over 96}t^{4}-{35\over 96}t^{5}
\; \; \; \; .
\eqno (L5.2.46)
$$
\vskip 0.3cm
\centerline {\bf L5.3 Contribution of $W_{\infty}^{(1)}$
and $W_{\infty}^{(2)}$}
\vskip 0.3cm
The term $W_{\infty}^{(1)}$ does not contribute to $\zeta(0)$. In fact,
using (A.5)-(A.6), jointly with
(L5.2.33), (L5.2.38), one finds
$$ \eqalignno{
W_{\infty}^{(1)}&={3\over 4}\sum_{m=0}^{\infty}\left(m^{2}-m\right)
\alpha_{m}^{-5}
\cr &\sim {x^{-2}\over 4}
-{3\over 4}{x^{-3}\over \Gamma({1\over 2})}
\sum_{r=0}^{\infty}{2^{r}\over r!}{\widetilde B}_{r}x^{-r}
{\Gamma\left({r\over 2}+{1\over 2}\right)
\Gamma\left({r\over 2}+{3\over 2}\right)\over
2\Gamma \left({5\over 2}\right)}
\cos \left({r\pi \over 2}\right)
\; \; \; \; ,
&(L5.3.1)\cr}
$$
which implies that $x^{-6}$ does not appear.

Now the series for $W_{\infty}^{(2)}$ is convergent, as may be checked using
(L5.2.34), (L5.2.38). However, when the sum over $m$ is rewritten
using the splitting (A.7), the individual pieces become divergent.
These {\it fictitious} divergences may be regularized dividing
by $\alpha_{m}^{2s}$, summing using (A.5)-(A.6), and then taking the
limit $s \rightarrow 0$. With this understanding, and
using the sums $\rho_{i}$ defined in
(F.15)-(F.18) of [4] one finds
$$ \eqalignno{
W_{\infty}^{(2)}&=-2x^{-6}\rho_{1}+2x^{-6}\rho_{2}+x^{-4}\rho_{3}
+{3\over 4}x^{-2}\rho_{4}+3x^{-6}\rho_{5}-3x^{-6}\rho_{6}\cr
&-{3\over 2}x^{-4}\rho_{7}-{9\over 8}x^{-2}\rho_{8}-x^{-6}\rho_{9}
+x^{-6}\rho_{10}+{x^{-4}\over 2}\rho_{11}+{3\over 8}x^{-2}\rho_{12}
\; \; .
&(L5.3.2)\cr}
$$
When odd powers of $m$ greater than $1$ occur, we can still apply (A.6)
after re-expressing $m^{2}$ as $\alpha_{m}^{2}-x^{2}$. Thus,
applying again the contour formulae (A.5)-(A.6), only $\rho_{1}$
and $\rho_{9}$ are found to contribute to $\zeta^{(2)}(0)$, leading to
$$
\zeta^{(2)}(0)=-{1\over 120}+{1\over 24}={1\over 30}
\; \; \; \; .
\eqno (L5.3.3)
$$
\vskip 0.3cm
\centerline {\bf L5.4 Effect of $W_{\infty}^{(3)}, W_{\infty}^{(4)}$ and
$W_{\infty}^{(5)}$}
\vskip 0.3cm
The term $W_{\infty}^{(3)}$ does not contribute to $\zeta(0)$. In fact,
using the relations (L5.2.35) and (L5.2.38) one finds
$$
W_{\infty}^{(3)}=-{1\over 8}\sum_{r=0}^{2}k_{1r}(r+1)(r+3)(r+5)
\left[\sum_{m=0}^{\infty}
\left(m^{2}-m\right)m^{r}\alpha_{m}^{-r-7}\right]
\; \; \; \; .
\eqno (L5.4.1)
$$
In light of (A.5)-(A.6), $x^{-6}$ does not appear in the asymptotic
expansion of (L5.4.1) at large $x$.

For the term $W_{\infty}^{(4)}$ a remarkable cancellation occurs. In fact,
using (L5.2.36) and (L5.2.38) one finds
$$
W_{\infty}^{(4)}=-{1\over 8}\sum_{r=0}^{4}k_{2r}(r+2)(r+4)(r+6)
\left[\sum_{m=0}^{\infty}
\left(m^{2}-m\right)m^{r}\alpha_{m}^{-r-8}\right]
\; \; \; \; .
\eqno (L5.4.2)
$$
The application of (A.5)-(A.6) and (L5.2.29) leads to
$$
\zeta^{(4)}(0)={1\over 2}\sum_{r=0}^{4}k_{2r}=0
\; \; \; \; .
\eqno (L5.4.3)
$$

Finally, using (L5.2.37)-(L5.2.38) one finds
$$
W_{\infty}^{(5)}=-{1\over 8}\sum_{r=0}^{6}k_{3r}(r+3)(r+5)(r+7)
\left[\sum_{m=0}^{\infty}
\left(m^{2}-m\right)m^{r}\alpha_{m}^{-r-9}\right]
\; \; \; \; .
\eqno (L5.4.4)
$$
Again, the contour formulae (A.5)-(A.6) lead to
$$
\zeta^{(5)}(0)=-{1\over 2}\sum_{r=0}^{6}k_{3r}=-{1\over 360}
\; \; \; \; ,
\eqno (L5.4.5)
$$
in light of (L5.2.30)-(L5.2.31).
\vskip 0.3cm
\centerline {\bf L5.5 Vanishing Effect of Higher-Order Terms}
\vskip 0.3cm
We now prove the statement made after (L5.2.21), i.e. that
we do not need to compute the explicit form of $c_{k}$ in (L5.2.18),
$\forall k>3$. In fact, the formulae
(L5.2.25)-(L5.2.26) can be completed by
$$
A_{4}=\left({c_{4}\over c_{0}}\right)
-{{\left({c_{2}\over c_{0}}\right)}^{2}\over 2}
-\left({c_{1}\over c_{0}}\right)\left({c_{3}\over c_{0}}\right)
+{\left({c_{1}\over c_{0}}\right)}^{2}
\left({c_{2}\over c_{0}}\right)
-{{\left({c_{1}\over c_{0}}\right)}^{4}\over 4} \; \; ,
\eqno (L5.5.1)
$$
plus infinitely many others, and the general term has the structure
$$
A_{n}=\sum_{p=1}^{l}h_{np}(1+t)^{-p}+\sum_{r=0}^{2n}k_{nr}t^{r} \; \; \; \; ,
\; \; \; \; \forall n \geq 1
\; \; \; \; ,
\eqno (L5.5.2)
$$
where $l<n$, the $h_{np}$ are constants, and
$r$ assumes both odd and even values. The integer $n$ appearing in (L5.5.2)
should not be confused with the one occurring in (L5.2.1) and in the
definition of $m$.
We have indeed proved that $h_{11}=h_{21}=h_{31}=0$, but the calculation
of $h_{np}$ for all values of $n$ is not obviously feasible.
However, we will show that the exact
value of $h_{np}$ does not affect the $\zeta(0)$ value.
Thus, $\forall n>3$, we must study
$$
H_{\infty}^{(n)}\equiv \sum_{m=0}^{\infty}\left(m^{2}-m\right)
{\left({1\over 2x}{d\over dx}\right)}^{3}
\left[{A_{n}\over (\alpha_{m})^{n}}\right]
=H_{\infty}^{n,A}+H_{\infty}^{n,B}
\; \; \; \; ,
\eqno (L5.5.3)
$$
where, defining
$$
a_{np}\equiv (p-n)(p-n-2)(p-n-4) \; , \;
b_{np}\equiv 3\Bigr(-p^{3}+(3+2n)p^{2}-(n^{2}+3n+1)p\Bigr) \; ,
\eqno (L5.5.4a)
$$
$$
c_{np}\equiv 3\Bigr(p^{3}-\left(p^{2}+p\right)n-p \Bigr)
\; \; \; \; , \; \; \; \;
d_{np}\equiv -p(p+1)(p+2)
\; \; \; \; ,
\eqno (L5.5.4b)
$$
one has
$$ \eqalignno{
H_{\infty}^{n,A}&\equiv \sum_{p=1}^{l}h_{np}
\sum_{m=0}^{\infty}\left(m^{2}-m\right)
{\left({1\over 2x}{d\over dx}\right)}^{3}\Bigr[\alpha_{m}^{p-n}
(m+\alpha_{m})^{-p}\Bigr]\cr
&=\sum_{p=1}^{l}{h_{np}\over 8}\sum_{m=0}^{\infty}\left(m^{2}-m\right)
\Bigr[a_{np}\alpha_{m}^{p-n-6}(m+\alpha_{m})^{-p}
+b_{np}\alpha_{m}^{p-n-5}
(m+\alpha_{m})^{-p-1}  \cr
& +c_{np}\alpha_{m}^{p-n-4}(m+\alpha_{m})^{-p-2}
+d_{np}\alpha_{m}^{p-n-3}(m+\alpha_{m})^{-p-3}\Bigr]
\; \; \; \; ,
&(L5.5.5)\cr}
$$
$$
H_{\infty}^{n,B}\equiv -{1\over 8}\sum_{r=0}^{2n}k_{nr}(r+n)(r+n+2)
(r+n+4)\left[\sum_{m=0}^{\infty}\left(m^{2}-m\right)m^{r}
\alpha_{m}^{-r-n-6}\right] \; .
\eqno (L5.5.6)
$$
Because we are only interested in understanding the behaviour of (L5.5.5)
as a function of $x$, the application of the Euler-Maclaurin formula
is more useful than the splitting (A.7). In so doing, we find that
the part of Eq. (C.2) of [4]
involving the integral on the left-hand side, when
$n=\infty$, contains the least negative power
of $x$. Thus, if we prove that the conversion of (L5.5.5) into an integral
only contains $x^{-l}$ with $l>6$, $\forall n>3$, we have proved that
$H_{\infty}^{n,A}$ does not contribute to $\zeta(0)$, $\forall n>3$. This is
indeed the case, because in so doing we deal with the integrals defined
in (F.19)-(F.26) of [4], where $I_{1}^{(np)}, I_{3}^{(np)},
I_{5}^{(np)}$ and $I_{7}^{(np)}$
are proportional to $x^{-3-n}$, and $I_{2}^{(np)}, I_{4}^{(np)},
I_{6}^{(np)}$
and $I_{8}^{(np)}$ are proportional to $x^{-4-n}$, where $n>3$.

Finally, in (L5.5.6) we must study the case when $r$ is even and the case when
$r$ is odd. In so doing, defining
$$
\Sigma_{(I)}\equiv \sum_{m=0}^{\infty}m^{2+r}\alpha_{m}^{-r-n-6}
\; \; \; \; , \; \; \; \;
\Sigma_{(II)}\equiv \sum_{m=0}^{\infty}m^{1+r}\alpha_{m}^{-r-n-6}
\; \; \; \; ,
\eqno (L5.5.7)
$$
we find for $r=2k>0$ ($k=1,2,...$)
$$
\Sigma_{(I)}= {x^{-3-n}\over 2}
{\Gamma \left(k+{3\over 2}\right) \Gamma \left({n\over 2}+{3\over 2}\right)
\over
\Gamma \left(3+k+{n\over 2}\right)}
\; \; \; \; ,
\eqno (L5.5.8)
$$
and for $r=2k+1$ ($k=0,1,2,...$)
$$ \eqalignno{
\Sigma_{(I)}&\sim
{x^{-3-n}\over \Gamma \left({1\over 2}\right)}
\sum_{l=0}^{\infty} \left \{{2^{l}\over l!}{{\widetilde B}_{l}\over 2}x^{-l}
\Gamma \left({l\over 2}+{1\over 2}\right)\cos \left({l \pi \over 2}\right)
\right. \cr
&\left. \left[ {\Gamma \left({n\over 2}+{3\over 2}+{l\over 2}\right)
\over \Gamma \left({n\over 2}+{5\over 2}\right)}+...+(-1)^{(1+k)}
{\Gamma \left({n\over 2}+{5\over 2}+k +{l\over 2}\right) \over
\Gamma \left({n\over 2}+{7\over 2}+k \right)}\right]\right \}
\; \; .
&(L5.5.9)\cr}
$$
Moreover, we find for $r=2k>0$ ($k=1,2,...$)
$$ \eqalignno{
\Sigma_{(II)}&\sim
{x^{-4-n}\over \Gamma \left({1\over 2}\right)}\sum_{l=0}^{\infty}
\left \{{2^{l}\over l!}{{\widetilde B}_{l}\over 2}x^{-l}
\Gamma \left({l\over 2}+{1\over 2}\right)\cos \left({l\pi \over 2}\right)
\right. \cr
&\left. \left[ {\Gamma \left({n\over 2}+2+{l\over 2}\right) \over
\Gamma \left({n\over 2}+3 \right)}+... +(-1)^{k}
{\Gamma \left({n\over 2}+2+k+{l\over 2}\right)\over
\Gamma \left({n\over 2}+3+k \right)}\right]\right \} \; \; ,
&(L5.5.10)\cr}
$$
and for $r=2k+1$ ($k=0,1,2,...$)
$$
\Sigma_{(II)}=  {x^{-4-n}\over 2}
{\Gamma \left(k+{3\over 2}\right)\Gamma \left({n\over 2}+2\right) \over
\Gamma \left({7\over 2}+k+{n\over 2}\right)}
\; \; \; \; .
\eqno (L5.5.11)
$$
Once more, in deriving (L5.5.8) and (L5.5.11) we used (A.5), and in
deriving (L5.5.9)-(L5.5.10) we used (A.6).
Thus, also $H_{\infty}^{n,B}$ does not
contribute to $\zeta(0)$, $\forall n>3$, and our proof is completed.
\vskip 0.3cm
\centerline {\bf L5.6 $\zeta(0)$ Value}
\vskip 0.3cm
In light of (L5.3.3), (L5.4.3),
(L5.4.5), and using the result proved in
section L5.5, we conclude that
for the complete massless field $\Bigr(\psi^{A}, \;
{\widetilde \psi}^{A'}\Bigr)$
$$
\zeta(0)= {1\over 30}-{1\over 360}={11\over 360}
\; \; \; \; .
\eqno (L5.6.1)
$$
Remarkably, this coincides with the result obtained in the case of
global boundary conditions [4]. Note that, if we study the classical
boundary-value problem for the massless field obeying the Weyl
equation and subject to homogeneous local boundary conditions
$$
\sqrt{2} \; {_{e}n_{A}^{\; \; A'}}\psi^{A}- \epsilon \;
{\widetilde \psi}^{A'}=\Phi^{A'}=0 \; \; \; \;
{\rm on} \; \; \; \; S^{3}
\; \; \; \; ,
$$
regularity of the solution implies the same conditions
on the modes $m_{np},r_{np}$ as in the spectral case
(section L4.3), even though the conditions on the modes
are quite different if $\Phi^{A'}$ does not vanish, or for
one-loop calculations.
Note also that for our problem the degeneracy
is half the one occurring in the case of global boundary conditions,
since we need twice as many modes to get the same number of eigenvalue
conditions ($J_{n+1}(E)=0, \; \forall n \geq 0$, in the global case, or
(L5.2.1) in the local case). If there are $N$ massless fields,
the full $\zeta(0)$ in (L5.6.1) should be multiplied by $N$.

The boundary conditions studied so far have also been expressed in
terms of $\gamma$-matrices in the literature, as follows [4].
Denoting by $A$ the
Dirac operator, one squares it up and studies the second-order
differential operator $A^{\dagger}A$. Thus, setting
$P_{\pm} \equiv {1\over 2}
\Bigr(1\pm i\gamma_{5}\gamma_{\mu}n^{\mu}\Bigr)$,
one has to impose the local mixed boundary conditions
$$
P_{-}\phi=0
\; \; \; \; ,
\eqno (L5.6.2)
$$
$$
\left(n \cdot \nabla + {(TrK)\over 2}\right)P_{+}\phi=0
\; \; \; \; .
\eqno (L5.6.3)
$$
One can easily check that (L5.6.2) coincides with our local boundary
conditions (L5.1.3).
However, one now deals with the extra condition (L5.6.3) involving the
normal derivatives of the massless field. Condition (L5.6.3) is obtained by
requiring that the image of the first-order operator $C$ defined in (L5.1.64)
also obeys the boundary conditions (L5.1.3), so that on $S^3$
$$
\sqrt{2} \; {_{e}n_{A}^{\; \; A'}}\Bigr(\nabla_{\; \; B'}^{A}
{\widetilde \psi}^{B'}\Bigr)=\epsilon
\Bigr(\nabla_{B}^{\; \; A'}\psi^{B}\Bigr)
\; \; \; \; .
\eqno (L5.6.4)
$$
We then use (L5.1.72),
and we also apply the well-known relations
$$
{ }^{(4)}\nabla_{i}\psi^{A}={ }^{(3)}\nabla_{i}\psi^{A}
- {_{e}n_{\; \; B'}^{A}} \; e^{BB'l}
K_{il} \; \psi_{B}
\; \; \; \; ,
\eqno (L5.6.5)
$$
$$
e_{AB'}^{\; \; \; \; \; \; i} \; e^{BB'l}=
\left[-{1\over 2}h^{il}\epsilon_{A}^{\; \; B}
+ {\rm const.} \; \epsilon^{ilk}\sqrt{h}
\; {_{e}n_{AB'}}e_{\; \; \; \; \; k}^{BB'}
\right]
\; \; \; \; ,
\eqno (L5.6.6)
$$
and of course (L5.1.3). One then finds that the contributions of
${ }^{(3)}\nabla_{i}$ add up to zero, using also the property that
${ }^{(3)}\nabla_{i}$ annihilates ${_{e}n_{AB'}}$. Moreover, the second term
on the right-hand side of (L5.6.6) plays no role because for our
torsion-free model the extrinsic-curvature tensor of the boundary is
symmetric. Thus (L5.6.4) leads to the boundary conditions
$$
\left({_{e}n^{BB'}}\nabla_{BB'}+{(TrK)\over 2}\right)
\Bigr(\sqrt{2} \; {_{e}n_{A}^{\; \; A'}}\psi^{A}+\epsilon \;
{\widetilde \psi}^{A'}\Bigr)=0 \; \;
{\rm on} \; \; S^{3}
\; \; \; \; ,
\eqno (L5.6.7)
$$
which are clearly of the type (L5.6.3).
We hope the reader will find it useful
the translation into two-component spinor language of the approach used
by other groups in the literature [4].
We should emphasize that (L5.6.7) (or equivalently (L5.6.3))
does not change the spectrum determined by (L5.1.3) (or equivalently
(L5.6.2)). In fact, as we said before, (L5.6.7) only shows that
the operator $C$ defined in (L5.1.64) preserves the boundary condition
(L5.1.3). This is automatically guaranteed,
in the approach used in Lecture 4,
by the eigenvalue equations (L5.1.56)-(L5.1.57),
which confirm that $A$ maps
$D(A)$ (cf. (L5.1.87)) to itself.
\vskip 0.3cm
\centerline {\bf L5.7 Summary}
\vskip 0.3cm
The Dirac operator is the naturally occurring first-order elliptic
operator one has to consider in studying massive or massless
spin-${1\over 2}$ particles in Riemannian four-geometries
(here we are not concerned with the Lorentzian framework).
Assuming that a spinor structure exists and is unique, the spinor
fields acted upon by the Dirac operator belong to unprimed or to
primed spin-space. These symplectic spaces are not isomorphic,
and the basic property of the Dirac operator is to interchange
them, in that it turns primed spinor fields into unprimed spinor
fields, and the other way around.

By virtue of the first-order nature of the Dirac operator, only
half of a fermionic field can be fixed on the boundary in the
corresponding elliptic boundary-value problem, providing its
index vanishes. The index is defined as the difference of the
dimensions of the null-spaces of the Dirac operator and its
adjoint, and its vanishing ensures that no topological obstructions
exist to finding a unique, smooth solution of the classical
boundary-value problem. This is well-posed providing spectral
or supersymmetric boundary conditions are imposed. The former
are motivated by the work on spectral asymmetry and Riemannian
geometry by Atiyah, Patodi and Singer. The latter are motivated
by local supersymmetry transformations relating bosonic and
fermionic fields. In the massless case, there is a deep relation
between the classical elliptic problems with local or supersymmetric
boundary conditions. The corresponding quantum theories are also
closely related, and are studied by means of $\eta$- and
$\zeta$-functions. The $\eta$-function combines positive and negative
eigenvalues of first-order elliptic operators. The generalized
$\zeta$-function is applied to regularize quantum amplitudes
involving second-order, positive-definite elliptic operators.
For fermionic fields, providing the asymptotic behaviour of
eigenvalues makes it possible to obtain a good cut-off regularization,
the regularized $\zeta(0)$ value for the Laplace-like operator
on spinor fields is used in the quantum theory.

Positive- and negative-helicity Weyl equations, jointly with spinor
Ricci identities, have also been used to restrict the class of
elliptic boundary-value problems relevant for unified theories
of fundamental interactions. In studying a general Riemannian
four-manifold with boundary, the counterpart of the index-theory
problem is the characterization of self-adjoint extensions of
the Dirac operator.
\vskip 0.3cm
\centerline {\bf Appendix A}
\vskip 0.3cm
The method described in section 7.3 of [4] and in Lecture 5
to perform $\zeta(0)$ calculations relies on the following
properties. Given the zeta-function at large $x$
$$
\zeta(s,x^{2}) \equiv \sum_{n=n_{0}}^{\infty}
{\Bigr(\lambda_{n}+x^{2}\Bigr)}^{-s}
\; \; \; \; ,
\eqno (A.1)
$$
one has in four-dimensions
$$
\Gamma(3) \zeta(3,x^{2})=\int_{0}^{\infty}t^{2}
e^{-x^{2}t}G(t) \; dt \sim
\sum_{n=0}^{\infty}B_{n}\Gamma \left(1+{n\over 2}\right)
x^{-n-2}
\; \; \; \; ,
\eqno (A.2)
$$
where we have used the asymptotic expansion of the heat kernel
for $t \rightarrow 0^{+}$
$$
G(t) \sim \sum_{n=0}^{\infty}B_{n}t^{{n\over 2}-2}
\; \; \; \; .
\eqno (A.3)
$$
Such an expansion does actually exist for boundary conditions
ensuring self-adjointness of the elliptic operators under
consideration (see [4] and references therein). On the other
hand, one also has the identity
$$
\Gamma(3) \zeta(3,x^{2})=\sum_{p=0}^{\infty}N_{p}
{\left(-{1\over 2x}{d\over dx}\right)}^{3}
\log \Bigr((ix)^{-p}F_{p}(ix)\Bigr)
\; \; \; \; ,
\eqno (A.4)
$$
where $N_{p}$ is the degeneracy of the problem and $F_{p}$
is the function expressing the condition obeyed by the
eigenvalues by virtue of boundary conditions. Here we focus
on $F_{p}$ functions which are (linear) combinations of
Bessel functions (and their first derivatives) of order
$p$, but in Lecture 5 a more complicated case is studied.
[Moreover, the method also holds for Bessel functions of
non-integer order, as shown in [4]]. Hence
$\zeta(0)=B_{4}$ is found, by comparison, as half the
coefficient of $x^{-6}$ in the asymptotic expansion of
Eq. (A.4).

The contour formulae used in Lecture 5 to perform $\zeta(0)$
calculations are [4]
$$
\sum_{p=0}^{\infty}p^{2k}\alpha_{p}^{-2k-m}
={\Gamma \left(k+{1\over 2}\right)
\Gamma \left({m\over 2}-{1\over 2}\right) \over
2 \Gamma \left(k+{m\over 2}\right)}x^{1-m}
\; \; \; \; \forall k=1,2,3,...
\; \; \; \; ,
\eqno (A.5)
$$
$$
\sum_{p=0}^{\infty}p \alpha_{p}^{-1-n} \sim
{x^{1-n}\over \sqrt{\pi}}
\sum_{r=0}^{\infty}{2^{r}\over r!}
{\widetilde B}_{r}x^{-r}
\; {\Gamma \left({r\over 2}+{1\over 2}\right)
\Gamma \left({n\over 2}-{1\over 2}+{r\over 2}\right)
\over
2 \Gamma \left({1\over 2}+{n\over 2}\right)}
\; \cos \left({r\pi \over 2}\right)
\; \; \; \; ,
\eqno (A.6)
$$
where $\alpha_{p}(x) \equiv \sqrt{p^{2}+x^{2}}$.
Moreover, we have used the identity [4]
$$
(p+\alpha_{p})^{-3}={(\alpha_{p}-p)^{3}\over x^{6}}
\; \; \; \; .
\eqno (A.7)
$$
\vskip 100cm
\centerline {\bf Further References}
\vskip 1cm
\parindent=0pt
\everypar{\hangindent=20pt \hangafter=1}

Berry M. V. and Mondragon R. J. (1987) {\it Proc. Roy. Soc. London}
{\bf A 412}, 53.

Luckock H. C. (1991) {\it J. Math. Phys.} {\bf 32}, 1755.

Reed M. and Simon B. (1975) {\it Methods of Modern Mathematical
Physics}, Vol. {\bf 2} (New York: Academic Press).

\vskip 100cm
\centerline {\bf Problems for Students}
\vskip 1cm
\noindent
(P.1) Prove regularity at the origin of the $\eta$- and
$\zeta$-functions.
\vskip 0.3cm
\noindent
(P.2) Write an essay on the $\eta$- and $\zeta$-functions,
describing in detail their relevance for Riemannian geometry,
elliptic boundary-value problems and quantum field theory.
\vskip 0.3cm
\noindent
(P.3) Following Carslaw and Jaeger 1959, prove equation
(L3.2.12).
\vskip 0.3cm
\noindent
(P.4) Under which conditions does the (integrated) kernel
admit an asymptotic expansion as $t \rightarrow 0^{+}$ ?
\vskip 0.3cm
\noindent
(P.5) In Riemannian four-geometries with a spinor structure,
unprimed and primed spin-spaces are not isomorphic. Prove
this result, and then give a description of this property
by using the theory of differential forms.
\vskip 0.3cm
\noindent
(P.6) Write down the Infeld-van der Waerden symbols in (flat)
Riemannian four-geometries, bearing in mind their form in
(flat) Lorentzian four-geometries.
\vskip 0.3cm
\noindent
(P.7) Prove that the spinor covariant derivative defined in
section L4.1, following Eq. (L4.1.34), exists and is unique.
\vskip 0.3cm
\noindent
(P.8) Prove that, by virtue of the first-order nature of the
Dirac operator, only half of a fermionic field can be fixed
on the boundary. Then, following sections L4.2-L4.3 and
chapter 4 of [4], use the mode-by-mode expansion of a
massless spin-${1\over 2}$ field on a family of three-spheres
centred on the origin to work out the form of all modes in
the case of a flat Riemannian four-geometry bounded by
$S^{3}$. Which modes are regular ? Which modes are singular at
the origin ? Thus, what does it mean to fix half of the
fermionic field on $S^{3}$ in this case ? What is the
appropriate boundary term in the action functional ?
\vskip 0.3cm
\noindent
(P.9) If we impose spectral or supersymmetric boundary conditions
on a three-sphere for massive Dirac fields, can we find a
relation between the corresponding boundary-value problems
analogous to the one worked out in section L4.3 in the
massless case ?
\vskip 0.3cm
\noindent
(P.10) Repeat the analysis of problems (P.8)-(P.9) in the
case of a flat Riemannian four-geometry bounded by two
concentric three-spheres.
\vskip 0.3cm
\noindent
(P.11) What is the tensor form of the Dirac and Weyl
equations ? For this problem, the reader is referred to
Eqs. (2.5.19)-(2.5.25) of the book by J. M. Stewart: {\it Advanced
General Relativity} (Cambridge University Press, 1990-2).
\vskip 0.3cm
\noindent
(P.12) What is the relation between the independent field
strengths for spin ${3\over 2}$, and the independent
spinor-valued one-forms appearing in the action functional
of simple supergravity in Riemannian four-geometries ?
\vskip 0.3cm
\noindent
(P.13) If you can solve problem (P.12), try to use the Penrose
potentials of section L4.4 to analyze the following boundary
conditions for flat Euclidean four-space bounded by a three-sphere:
$$
\sqrt{2} \; {_{e}}n_{A}^{\; \; A'} \; \psi_{i}^{A}
=\pm {\widetilde \psi}_{i}^{A'}
\; \; \; \; {\rm on}
\; \; \; \; S^{3}
\; \; \; \; .
$$
With our notation, $\Bigr(\psi_{i}^{A},{\widetilde \psi}_{i}^{A'}
\Bigr)$ are the spatial components of
the independent spinor-valued one-forms representing the gravitino
field, and ${_{e}}n_{A}^{\; \; A'} \equiv -i n_{A}^{\; \; A'}$
is the Euclidean normal to the three-sphere. In particular, what is
the role played by gauge-invariance of the spin-${3\over 2}$
potentials, expressed in terms of solutions of positive- and
negative-helicity Weyl equations, for this elliptic boundary-value
problem ?
\vskip 0.3cm
\noindent
(P.14) In Eq. (L5.1.65), we have defined the {\it dagger}
conjugation for spinor fields in flat Riemannian four-geometry.
Is it possible to extend this definition
of conjugation to the curved case ?
\vskip 0.3cm
\noindent
(P.15) In section L5.1 we have proved that the operator $iC$
(see (L5.1.64)) is symmetric and has self-adjoint extensions.
Can one now prove, by using the theory of deficiency
indices, that $iC$ admits an {\it unique} self-adjoint extension ?
Moreover, can one extend the analysis of section L5.1 to
curved Riemannian four-geometries ? Are they all Einstein
manifolds ? [i.e. such that the Ricci tensor is proportional
to the four-metric]
\vskip 0.3cm
\noindent
(P.16) A more powerful way exists to perform the $\zeta(0)$
calculation of sections L5.2-L5.6. For this purpose, read and
try to repeat the analysis by Kamenshchik and Mishakov
appearing in {\it Phys. Rev.} D, {\bf 47}, 1380;
{\it Phys. Rev.} D, {\bf 49}, 816.
\vskip 0.3cm
\noindent
(P.17) Find the $\eta(0)$ value for the boundary-value
problem studied in Lecture 5.
\vskip 0.3cm
\noindent
(P.18) Find the $\eta(0)$ and $\zeta(0)$ values for flat
Riemannian four-geometry bounded by two concentric three-spheres,
when the massless fermionic field is subject to
supersymmetric boundary conditions.
\vskip 0.3cm
\noindent
The solution of problems
(P.9)-(P.10), (P.12)-(P.15), (P.17) may lead to good
research papers. Hence they should be given priority.
\bye